\newcommand{\logg}{$\log\,g$}
\newcommand{\teff}{$T_{\rm eff}$}

\documentclass{aa}
\usepackage{txfonts}
\usepackage{graphicx}
\begin{document}

\title{
New identified ($^{3}$H)4d-($^{3}$H)4f transitions of \ion{Fe}{ii} from UVES spectra of HR\,6000 and 
46\,Aql\thanks{This study is the result of a collaboration with S. Johansson,
who unfortunately left us before this paper started to be written.}
}

\author{
F.\, Castelli
\inst{1}
\and
R.L.\, Kurucz
\inst{2}
\and
S.\,Hubrig
\inst{3}
}

\offprints{F. Castelli}

\institute{
Istituto Nazionale di Astrofisica--
Osservatorio Astronomico di Trieste, Via Tiepolo 11,
I-34131 Trieste, Italy\\
\email{castelli@oats.inaf.it}
\and
Harvard-Smithsonian Center for Astrophysics, 60 Garden Street,
Cambridge, MA 02138, USA
\and
Astrophysical Institute Postsdam, An der Sternwarte 16, D-14482 Potsdam, Germany 
}

\date{}

\abstract
{}
{The analysis of the high-resolution UVES spectra of the CP stars HR\,6000 
and 46\,Aql has revealed the presence of an impressive number of unidentified
lines, in particular in the 5000-5400\,\AA\ region. Because numerous 4d-4f 
transitions of \ion{Fe}{ii} lie in this spectral range, and because both 
stars are iron overabundant, we investigated whether the unidentified lines 
can be due to \ion{Fe}{ii}. 
}
{ATLAS12 model atmospheres with parameters [13450K,4.3] and
[12560K,3.8] were computed for the individual abundances
of HR\,6000 and 46\,Aql, respectively, in order to use the stars as
spectroscopic sources to identify \ion{Fe}{ii} lines and to determine 
\ion{Fe}{ii} gf-values.  
After having identified several unknown lines in the stellar spectra as due  
to ($^{3}$H)4d$-$($^{3}$H)4f  transitions of \ion{Fe}{ii}, we derived 
stellar $\log\,gf$'s for them by comparing observed and computed 
profiles. The energies of the upper levels were assigned on the basis 
of both laboratory iron spectra and predicted energy levels.
}
{
We fixed 21 new levels of \ion{Fe}{ii} with energies 
between 122910.9\,cm$^{-1}$ and 123441.1\,cm$^{-1}$. They
allowed us to add 1700 new lines to the \ion{Fe}{ii}
line list in the range 810 $-$ 15011\,\AA. 
}
{}


\maketitle{}

\section{Introduction}

The analysis of the UVES spectrum of the chemically peculiar star HR\,6000 
performed by Castelli \& Hubrig (2007) has shown 
the presence of a huge number of unidentified lines in the whole 
observed range from 3050\,\AA\ to 9460\,\AA. Most impressive regions 
were those at 4404-4411\,\AA\  and 5000-5400\,\AA.
An attempt to identify unknown lines in the 5130-5136\,\AA\ interval using
the available lists of predicted \ion{Fe}{ii} lines \footnote{http://kurucz.harvard.edu/atoms/2601/gf2601.lines0600} (version 2003)  was very laborious and 
not very successful.

An analysis of the list of the unidentified lines and of the plot of the
observed and computed spectra available at the Castelli 
web-site\footnote{http://wwwuser.oat.ts.astro.it/castelli/hr6000/hr6000.html}
has led S. Johansson to identify about half of them as transitions
between high excitation levels of \ion{Fe}{ii}. The identifications were made 
using unpublished line lists he obtained from laboratory spectra. 
Johansson (2009) was able to fix a group of 13 lines 
concentrated in the 4404-4411\,\AA\ interval as belonging to
the  multiplet 4s($^{7}$S)4d\,$^{8}$D $-$ 4s($^{7}$S)4f\,$^{8}$F. 
The upper terms have energies between 132145\,cm$^{-1}$ and 
132158\,cm$^{-1}$, therefore well above the \ion{Fe}{ii} ionization limit of 
130563\,cm$^{-1}$. Another example of new identifications 
can be found in  Castelli, Johansson \& Hubrig (2008)
where some unknown lines in the 5175-5180\,\AA\ interval of HR\,6000 
were identified as due to ($^{3}$H)4d $-$ ($^{3}$H)4f  transitions of
\ion{Fe}{ii}.
In this case the energy of the upper levels is of the order of 
123000\,cm$^{-1}$, therefore just below the ionization limit.

Laboratory spectroscopic sources show lines in emission and
must populate upper levels to produce a spectrum. Most stars are
observed in absorption so that their line strengths are determined
by the lower level populations; the spectra are stronger.
HR\,6000 and 46\,Aql are bright and can be observed at high
resolution and high signal-to-noise ratio. They have low projected
rotation velocity, 1.5~km\,s$^{-1}$ for HR\,6000 and 1.0~km\,s$^{-1}$
for 46\,Aql. Thus blending is minimal and wavelengths and line
strengths can be well determined by fitting the spectrum.

It is likely that most of the unidentified lines in these stars are \ion{Fe}{ii}.
Some could be due to other overabundant elements, in particular
\ion{P}{ii}, \ion{Mn}{ii} and \ion{Xe}{ii}. However, as far as
\ion{Mn}{ii} is concerned, the unidentified lines are either weaker 
or do not appear at all in the HgMn star HD\,175640 which is iron weak
([Fe/H]= $-$0.25) and manganese overabundant ([Mn/H]= +2.4) 
(Castelli \& Hubrig 2004; Castelli et al. 2008). 
It is also improbable that  so large number of
unidentified lines, mostly concentrated in the 5000-6000\,\AA\ region,
are not also due to \ion{Fe}{ii}. Since these lines are present at
subsolar iron abundance they must be present in all Population\,I
late B-type stars, or in any object with strong \ion{Fe}{ii}
lines, but they are normally smeared out by rotation and
difficult to see.       

Because every new level accounts for hundreds of new lines throughout the 
spectrum from the UV through the IR,  we extended to a larger
range the work already made for the 5175-5181\,\AA\ interval
with the aim to increase the number of the classified \ion{Fe}{ii} levels. 
We used both identifications based on laboratory spectra and
derived from predicted energy levels.
Furthermore, because  numerous high-excitation lines due to the 
3d$^{6}$($^{5}$D)4d $-$ 3d$^{6}$($^{5}$D)4f transitions 
lie in the 5000$-$5400\,\AA\ interval, we examined the 
computed $\log\,gf$'s  that are used for the synthetic spectra computations.  
We compared them with experimental (Johansson, 2002) and stellar 
values, which we derived from both HR\,6000 and  46\,Aql. We used
two stars  in order to check the consistency and estimate the reliability of
the stellar oscillator strengths. Then, we derived stellar $\log$\,gf's for 
the new identified ($^{3}$H)4d $-$ ($^{3}$H)4f lines.
Finally, we show an example of synthetic spectrum computed both with old and
new line lists. 

In order to use HR\,6000 and 46\,Aql as spectroscopic sources to 
identify \ion{Fe}{ii} lines, to determine \ion{Fe}{ii} $\log\,gf$'s, and to
compute synthetic spectra we fixed at first model atmospheres and abundances
for each star. Special care was devoted to derive the iron abundance.

\section{The stars HR\,6000 and 46 Aql}

Both HR\,6000 (HD\,144667) and 46\,Aql (HD\,186122)
are B-type CP stars. 
HR\,6000 was extensively studied by Castelli \& Hubrig (2007) in the 
3050 $-$ 9460\,\AA\ region,
46\,Aql by Sadakane et al. (2001) in the 5100\,$-$\,6400\,\AA\ interval. 
Because model atmospheres are needed for the determination of stellar
$\log\,gf$'s,  we  revised
the model atmosphere and the abundances of HR\,6000 and determined 
the model atmosphere and the abundances for 46\,Aql on the basis 
of our observations.

\subsection{Observations}

The observations of HR\,6000 are described in
Castelli \& Hubrig (2007). Those for 46 Aql were
obtained in the framework of the same observational
program (ESO prg. 076.D-0169(A)). 
They are of the same quality and were reduced
with the same procedures used for HR\,6000.

In this paper we revised our previous analysis of HR\,6000
(Castelli \& Hubrig 2007) because a further investigation has
indicated that Balmer profiles that were used by us to fix the
model parameters are too affected in the UVES spectra 
by  imperfections  related to the echelle orders. As we show in
Appendix\,A, the spectral distorsions for H$_{\delta}$, 
H$_{\gamma}$, and, to a less extent, for H$_{\beta}$ are
so significant that it is not possible to draw a reliable
continuum over the observed profiles. Only H$\alpha$ does
not show manifest problems so that it could be used for the analysis
with some confidence.

\subsection{Model paramaters and abundances}

In Castelli \& Hubrig (2007) the model parameters of HR\,6000 
(\teff=12850\,K, \logg=4.10, $\xi$=0.0\,km\,s$^{-1}$) were
derived from Balmer profiles and from  
\ion{Fe}{i} and \ion{Fe}{ii} ionization equilibrium.
In this paper the model parameters for both HR\,6000 and
46\,Aql were obtained from  
the Str\"omgren photometry, from the requirement of no
correletion of \ion{Fe}{ii} abundances derived from high and low excitation 
lines, and from the requirement of
\ion{Fe}{i} $-$ \ion{Fe}{ii} ionization equilibrium.
This kind of determination has led to a revision of the model parameters
for HR\,6000.

\begin{table*}
\caption{Observed and dereddened Str\"omgren indices for 
HR\,6000 and 46 Aql.}
\centering
\begin{tabular}{rrrrrrrrrrr}
\hline
\hline
\noalign{\smallskip}
\multicolumn{1}{c}{Star} & 
 \multicolumn{1}{c}{b$-$y} &
 \multicolumn{1}{c}{m} &
 \multicolumn{1}{c}{c} &
 \multicolumn{1}{c}{$\beta$} &
 \multicolumn{1}{c}{E(b-y)} &
 \multicolumn{1}{c}{(b-y)$_{0}$} &
 \multicolumn{1}{c}{m$_{0}$} &
 \multicolumn{1}{c}{c$_{0}$} &
 \multicolumn{1}{c}{\teff(K)} &
 \multicolumn{1}{c}{\logg} 
 \\
\hline
\noalign{\smallskip}

HR\,6000&$-$0.030 & 0.116 & 0.521& 2.750& 0.031&$-$0.061& 0.126 & 0.506 &13799$\pm$150& 4.27$\pm$0.05 \\
       &$\pm$0.003& $\pm$0.003&$\pm$0.003\\
46 Aql  &$-$0.019 & 0.094 & 0.641& 2.729& 0.035&$-$0.054& 0.105 & 0.634 &12763$\pm$150& 3.75$\pm$0.05 \\
        &$\pm$0.002& $\pm$0.002 & $\pm$0.002\\
\hline
\noalign{\smallskip}
\end{tabular}
\end{table*}

The parameters of the two stars obtained from the Str\"omgren photometry 
are given in Table\,1. They were derived with the method described in
Castelli \& Hubrig (2007), where the reddening for HR\,6000 is also dicussed.
The observed indices were taken from the Hauck\& Mermilliod (1998) 
Catalogue\footnote{http://obswww.unige.ch/gcpd/gcpd.html}.
ATLAS9 models with parameters [13800K,4.3] for HR\,6000  and [12750K,3.8]
for 46\,Aql were computed for solar abundances and zero microturbulent 
velocity.
The iron abundance was derived from the equivalent 
widths of  a selected sample of high-excitation \ion{Fe}{ii} lines
having experimental $\log\,gf$'s (Johansson, 2002). They are
due to ($^{5}$D)4d $-$ ($^{5}$D)4f transitions and are listed
in Table\,2. 
The equivalent widths of these lines, as well as of the other lines
discussed in this paper, were measured
integrating the residual fluxes over the profiles.
Abundances were obtained
with a Linux version (Castelli 2005) of the WIDTH code (Kurucz 1993).
We note that there are no measured equivalent widths 
for the line at 5100.734\,\AA\ because
it is part of a strong blend  formed by four \ion{Fe}{ii} lines.
The line was included in Table\,2 in order to show the whole set
of  ($^{5}$D)4d $-$ ($^{5}$D)4f transitions having 
experimental $\log\,gf$'s.
The lines from Table\,2  are 
particularly well suited to provide the iron abundance because they
are rather insensitive to the model parameters. In fact, for differences
in \teff\ of 500\,K, they give abundance differences less than 0.05\,dex.
For difference in \logg\ of 0.2\,dex they give abundance differences 
of the order of 0.01\,dex. 
For example, for HR\,6000, the \ion{Fe}{ii} abundance from  ATLAS9 
models computed for \logg=4.3 and \teff= 13300\,K, 13800\,K, and 14300\,K 
is $-$3.66\,dex ,  $-$3.65\,dex (Table\,2), and $-$3.61\,dex, respectively. 
For models computed for \teff=13800\,K and \logg=4.1, 4.3, and 4.5 the iron
abundance is  $-$3.65\,dex, $-$3.65\,dex (Table\,2), and $-$3.64\,dex.
Similar results can be obtained for 46\,Aql.

\begin{table*}[]
\caption{Iron abundance from a selected sample of high excitation (4d-4f) \ion{Fe}{ii} lines 
with experimental $\log$\,gf's  and ATLAS9 models.}
\begin{flushleft}
\begin{tabular}{lrccccccccccccccccc}
\hline\noalign{\smallskip}
\multicolumn{1}{c}{$\lambda$($\AA$)}&
\multicolumn{1}{c}{$\log$\,gf}&
\multicolumn{1}{c}{$\chi_{low}$(cm$^{-1}$)}&
\multicolumn{1}{c}{$\chi_{up}$(cm$^{-1}$)}&
\multicolumn{1}{c}{Config.}&
\multicolumn{1}{c}{W(m\AA)}&
\multicolumn{1}{c}{$\log$$\epsilon$(W)}&
\multicolumn{1}{c}{W(m\AA)}&
\multicolumn{1}{c}{$\log$$\epsilon$(W)}
\\
\hline\noalign{\smallskip}
&exp&&&&\multicolumn{2}{c}{HR\,6000 [13800,4.3]}&
 \multicolumn{2}{c}{46\,Aql [12750,3.8]}\\
\hline\noalign{\smallskip}
4883.292      & $-$0.521 &82853.66 & 103325.93 & ($^{5}$D)4d e$^{6}$F$_{11/2}$$-$ ($^{5}$D)4f\,3[5]$_{11/2}$&14.2 & $-$3.87 & 11.3 & $-$4.10\\ 
4913.295      & 0.016   & 82978.67 & 103325.93 & ($^{5}$D)4d e$^{6}$F$_{9/2}$$-$ ($^{5}$D)4f\,3[5]$_{11/2}$&36.3 & $-$3.67 & 31.1 & $-$3.91 \\
5001.953      & 0.933 & 82853.66 & 102840.27 & ($^{5}$D)4d e$^{6}$F$_{11/2}$$-$ ($^{5}$D)4f\,4[6]$_{13/2}$&76.0 & $-$3.61 & 66.1 & $-$3.86\\
5100.734     &0.671    & 83726.37 & 103325.93 & ($^{5}$D)4d $^{6}$D$_{9/2}$$-$ ($^{5}$D)4f\,3[5]$_{11/2}$ & $--$ & $--$ & $--$ & $--$ \\ 
5227.483     &0.831    & 84296.83 & 103421.16 & ($^{5}$D)4d e$^{6}$G$_{11/2}$$-$ ($^{5}$D)4f\,3[6]$_{13/2}$&66.8 & $-$3.58 & 58.0 & $-$3.83 \\  
5253.647      &$-$0.191 &84296.83 & 103325.93 &($^{5}$D)4d e$^{6}$G$_{11/2}$$-$ ($^{5}$D)4f\,3[5]$_{11/2}$& 30.6 & $-$3.51 & 23.5 & $-$3.83 \\  
5260.254      & 1.090  &84035.12 & 103040.32 & ($^{5}$D)4d e$^{6}$G$_{13/2}$$-$ ($^{5}$D)4f\,4[7]$_{15/2}$&77.9 & $-$3.59 & 66.9 & $-$3.84\\
5316.214      & 0.418  & 84035.12 & 102840.27 & ($^{5}$D)4d e$^{6}$G$_{13/2}$$-$ ($^{5}$D)4f\,4[6]$_{13/2}$&48.7 & $-$3.61 & 39.7 & $-$3.94 \\
5339.592      & 0.568   & 84296.83 & 103019.64 & ($^{5}$D)4d e$^{6}$G$_{11/2}$$-$ ($^{5}$D)4f\,4[7]$_{13/2}$&48.2 & $-$3.74 & 42.8 & $-$3.97\\
5387.063      & 0.593   & 84863.33 & 103421.16 &($^{5}$D)4d e$^{4}$G$_{11/2}$$-$ ($^{5}$D)4f\,3[6]$_{13/2}$& 47.7 & $-$3.75 & 41.3 & $-$4.00\\ 
5414.852      &$-$0.258  & 84863.33 & 103325.93 &($^{5}$D)4d e$^{4}$G$_{11/2}$$-$ ($^{5}$D)4f\,3[5]$_{11/2}$& 25.0 & $-$3.57 & 19.1 & $-$3.87\\
5506.199      & 0.923  & 84863.33 & 103019.64 &($^{5}$D)4d e$^{4}$G$_{11/2}$$-$ ($^{5}$D)4f\,4[7]$_{13/2}$& 73.8 & $-$3.65 & 54.2 & $-$3.94\\
5510.783      & 0.043    & 85184.73 & 103325.93 & ($^{5}$D)4d e$^{4}$G$_{9/2}$$-$ ($^{5}$D)4f\,3[5]$_{11/2}$&32.5 & $-$3.61 & 26.3 & $-$3.89\\
\hline\noalign{\smallskip}
\multicolumn{2}{c}{Average  abundances}                     &     &           &                                                              &       &$-$3.65$\pm$0.09& &$-$3.92$\pm$0.08\\ 
\hline\noalign{\smallskip}
\end{tabular}
\end{flushleft}
\end{table*}

Once the iron abundance was fixed, we checked 
the model parameters  inquiring whether the adopted model gives
the same abundance also  from \ion{Fe}{i} lines and from a large sample 
of \ion{Fe}{ii} lines which includes also low-excitation transitions. 
To this purpose, we measured the 
equivalent widths of the \ion{Fe}{i} and \ion{Fe}{ii} lines
listed in Table\,B.1 and derived the corresponding abundances.
They are  $-$3.68$\pm$0.06\,dex  and $-$3.68$\pm$0.15\,dex,
respectively, for HR\,6000 and 
$-$3.93$\pm$0.06\,dex and $-$3.91$\pm$0.11\,dex,
 respectively, for 46\,Aql.
For both stars  the ionization equilibria are fullfilled 
within the error limits and also the \ion{Fe}{ii} abundance from the large 
sample of lines listed in Appendix\,B  agrees, within the error limits, 
with the abundance yielded 
by the small sample of high-excitation \ion{Fe}{ii} listed in Table\,2.
We conclude that the [13800,4.3] ATLAS9 model for HR\,6000 and the [12750,3.8] 
ATLAS9 model for 46\,Aql reproduce the Str\"omgren colors,  give 
 \ion{Fe}{i}-\ion{Fe}{ii} ionization equilibrium, and also \ion{Fe}{ii} 
abundance independent from the excitation potential. 
Also the H$_{\alpha}$ profile is farly well reproduced by the models in 
both stars.

The ATLAS9 models were used to obtain abundances 
for the other elements different from iron. For \ion{He}{i} we adopted the 
lines listed in Castelli \& Hubrig (2004) and we
analyzed them as described in that paper. For the other elements the lines 
listed in Table\,B.1 were used.  For most lines equivalent widths were
measured, for
weak lines or lines which are blends of transitions belonging to the same
multiplet, as \ion{Mg}{ii} 4481\,\AA\ and most \ion{O}{i} profiles,
we derived the abundance from the line profiles with the synthetic spectrum
method. Synthetic spectra were computed  with a Linux version 
(Sbordone et al. 2004) of the SYNTHE code (Kurucz 2005). When no lines were
observed for a given element an upper abundance limit was fixed by reducing 
the intensity of the computed line at the level of the noise.
To compute synthetic spectra rotational velocities equal to
1.5\,km\,s$^{-1}$ and 1.0\,km\,s$^{-1}$ were adopted for HR\,6000 and 46\,Aql,
respectively. They were derived  from the comparison of the observed
and computed \ion{Mg}{ii} profiles at 4481\,\AA.

Because the abundances found for most elements were far from solar,
 we computed ATLAS12 models (Kurucz 2005) for the individual abundances  
with the same parameters of the ATLAS9 models.  
The structure of the ATLAS12 models  is heavily affected by the large iron 
overabundance,
while the helium underabundance, although large, has a negligible effect, 
in contrast with what we wrongly stated in Castelli \& Hubrig (2007). 
As a consequence,  the \ion{Fe}{i}$-$\ion{Fe}{ii} ionization equilibrium is 
no longer acheived by the ATLAS12 models unless the parameters are changed.  
Keeping fixed the gravity, which reproduces rather well the H$_{\alpha}$ 
profile and affects the wings more than the temperature does,
we modified \teff\ until the \ion{Fe}{i} abundance 
from the lines listed in Appendix\,B agrees with the \ion{Fe}{ii}
abundance obtained from the lines shown in Table\,2.
The final ATLAS12 parameters are \teff=13450\,K, 
\logg=4.3, $\xi$=0.0\,km\,s$^{-1}$
for HR\,6000 and \teff=12560, \logg=3.8, $\xi$=0.0\,km\,s$^{-1}$ for 46\,Aql.
The computed  indices (b-y), m and c for HR\,6000 are $-$0.064, 0.126, 
and 0.535, respectively.  For 46\,Aql they are $-$0.052, 0.116, and 0.662.
For both stars  the observed (b-y) is reproduced by the models
within the observational uncertainties, while the c index is not.
This means
that the models are able to predict the optical spectrum but not
the spectrum shortward of the Balmer discontinuity.     

We note that we could have simply used the  ATLAS9 models computed for 
solar abundances with the parameters given in Table\,2 which reproduce
both Str\"omgren colors and \ion{Fe}{i}-\ion{Fe}{ii} ionization equilibria.
But in this case, the state of the gas in the input model is very different
from that recomputed by the final synthetic spectrum which has
not solar input abundances for several elements, in particular \ion{He},
which heavly affects the state equation results.
We have preferred to use consistent computations in
model and  synthetic spectra 
rather than use different abundances in the two codes, 
although the predictions may appear to be better in this latter case. 

Table \,3  summarizes the final 
stellar abundances used to compute ATLAS12 models and synthetic
spectra.   For HR\,6000,  abundances which  differ from the previous 
determination (Castelli \& Hubrig 2007) by 0.2\,dex  or more are those for 
\ion{He}{i} (+0.20), \ion{Ca}{ii} (+0.2), 
\ion{Ti}{ii} (+0.30), 
\ion{Cr}{ii} (+0.27), 
\ion{Mn}{ii} (+0.42), \ion{Fe}{i} and \ion{Fe}{ii} (+0.21), 
\ion{Y}{ii} (+0.2).
For 46\,Aql, abundances differing by more than 0.2\,dex
from those derived by Sadakane et al. (2001) are those  
for \ion{C}{ii} ($-$0.33), \ion{S}{ii} ($\ge$$+$0.77),
\ion{Ti}{ii} ($-$0.44), \ion{Fe}{i} ($-$0.32). 
The number in parenthesis is the difference
between this paper and the other analyses. 
Sadakane et al. (2001) adopted an ATLAS9 model atmosphere with parameters
\teff=13000\.K, \logg=3.65, and v$_{turb}$=0.3\,km\,s$^{-1}$.

The abundances of 46\,Aql show approximately the same pattern as 
in HR\,6000, but the deviations from solar values are generally lower. 
The most remarkable differences between the two stars are 
the overabundance of copper, zinc, and  arsenic in 46\,Aql, while 
no lines of these elements were observed in HR\,6000. 
The arsenic overabundance can not be quantified owing to the 
lack of $\log\,gf$'s for \ion{As}{ii}. Lines of \ion{As}{ii} observed
in the spectrum are listed in Table\,B.1. We assigned 
to each line the corrisponding transition on the basis of the 
two separate lists for lines and energy levels taken from the NIST 
database\footnote{http://physics.nist.gov/PhysRefData/ADS/lines$_{-}$form.html}$^{,}$ 
\footnote{http://physics.nist.gov/PhysRefData/ADS/levels$_{-}$form.html}. 
Furthermore, \ion{Cr}{ii} is slightly overabundant   
in HR\,6000 ([0.3]) and underabundant  in 46 Aql [$-$1.1],  \ion{Ca}{ii}
is solar in HR\,6000 and slightly underabundant in 46\,Aql [$-$0.3],
and \ion{Y}{ii} and \ion{Hg}{ii} are less overabundant in 
HR\,6000 than in 46\,Aql.

In both HR\,6000 and 46\,Aql the \ion{He}{i} profiles can not be 
reproduced by the same abundance. We adopted the abundance reproducing
the wings of $\lambda\lambda$ 3867, 4026, 4471\,\AA\ at best. The cores of 
all \ion{He}{i} lines would require a lower abundance than that which 
fits the wings.

In both stars the average abundances for phosphorous and manganese show
deviations larger than 0.2\,dex. For phosphorous they are due 
 to a  difference of the order of 0.5\,dex  between the abundance 
 from lines with $\lambda$ $<$ 5000\,\AA\  
and the abundance from lines with $\lambda$ $>$ 5000\,\AA.
For manganese, the large deviation 
is due to a difference of  0.6\,dex in HR\,6000 and 0.4\,dex
in 46\,Aql between the abundances 
from lines lying shortward and longward of the Balmer discontinuity. 
The \ion{Mn}{ii} abundance was derived both from equivalent widths and 
lines profiles. No hyperfine structure was considered in the computations 
in the first case,
while in the second case hyperfine components were taken into account 
for all the lines except for 3917.318\,\AA, in that  no hyperfine constants
are available for its upper energy level. 
The hyperfine components 
were taken from Kurucz\footnote{http://kurucz.harvard.edu/atoms/2501/hyper250155.srt}.  
The difference in the average abundances obtained with the two methods 
are of the order of 0.01\,dex. 

A plausible explanation for the discrepancy in the \ion{Mn}{ii} abundance
is that the  model structure is inadequate to reproduce the ultraviolet
spectrum, as we already deduced from the comparison of 
computed and observed c indices.    
Vertical abundance stratification, which is a consequence of radiative
diffusion acting in CP stars (Michaud 1970), can be invoked 
to explain the anomalous  He\,I line profiles  as well as
the P and Mn non homogeneous abundances. 
A comprehensive discussion on observational evidence for the 
abundance stratification in CP stars is given by Ryabchikova et al. (2003).
It includes the impossibility to fit the wing and the core of strong
spectral lines with the same abundance and the different abundances obtained 
from the lines of the same ions formed at different optical depths as, for
example, longward and shortward of the Balmer discontinuity.
Finally, for a reliable discussion on phosphorous, more studies on the 
oscillator strengths of \ion{P}{ii} and \ion{P}{iii} lines observed in the
optical region  are needed.

In both HR\,6000 and 46\,Aql the \ion{Hg}{ii} line at 3983.890\,\AA\
is mostly due to the heaviest isotope of Hg.
The lines of the \ion{Ca}{ii} infrared triplet at 8498, 8542, and 8662\,\AA\ 
are-red shifted  by 0.14\,\AA\ in HR\,6000 and by 0.13\,\AA\ in 46\,Aql, 
so indicating a nonsolar Ca isotopic composition.
In HR\,6000 emission lines of \ion{Cr}{ii},
\ion{Mn}{ii}, and \ion{Fe}{ii} were observed. Instead, 
in the spectrum of 46\,Aql there are  emission lines of 
\ion{Ti}{ii} and \ion{Mn}{ii}.

\section{The ($^{5}$D)4f and ($^{3}$H)4f states of \ion{Fe}{ii}}

Energy levels of a given atom are most often described by the LS coupling 
in which the total orbital angular momentum {\bf L} of the atom is 
coupled to the
total spin angular momentum {\bf S} to give the total angular momentum  
{\bf J=L+S}. Some high levels, such as the ($^{5}$D)4f  and ($^{3}$H)4f
states of \ion{Fe}{ii}, 
tend to appear in pairs so that they are better described by the jK coupling with the 
notation j$_{c}$[K]$_{J}$, where {\bf j}$_{c}$ is the total angular momentum of
the core and {\bf K}={\bf j}$_{c}$+{\bf l}  is the coupling of 
{\bf j}$_{c}$ with the orbital angular momentum {\bf l}
of the active electron. The level pairs correspond to the two  
values of the total angular momentum {\bf J} that result 
when the spin  {\bf s}=$\pm$1/2 of the active electron is added to {\bf K}.

While the energy levels of the  3d$^{6}$($^{5}$D)4f states are known
and are available for instance at the NIST 
data-base, this is not the case for the levels with the higher parent 
term 3d$^{6}$($^{3}$H)4f. The levels have not been observed in the laboratory 
and are still unclassified.

\begin{table}[]
\begin{center}
\caption{Abundances $\log$(N(elem)/N$_{tot}$) for HR\,6000 and 46\,Aql 
from ATLAS12 models. Solar abundances are from Grevesse \& Sauval (1998).}
\begin{tabular}{lccrccccccccccccccc}
\hline\noalign{\smallskip}
\multicolumn{1}{c}{elem}&
\multicolumn{1}{c}{HR\,6000}&
\multicolumn{1}{c}{46\,Aql}&
\multicolumn{1}{c}{Sun}
\\
& [13450K,4.3]&[12560K,3.8]&\\
\hline\noalign{\smallskip}
\ion{He}{i} & $-$2.10 & $-$2.00 & $-$1.05\\
\ion{Be}{ii} & $-$9.78 & $-$9.91 & $-$10.64\\
\ion{C}{ii}  & $-$5.50 & $-$4.75 & $-$3.52\\ 
\ion{N}{i}  & $\le$$-$5.82&$\le$$-$5.50 & $-$4.12\\
\ion{O}{i}  & $-$3.68$\pm$0.4 & $-$3.51$\pm$0.10 & $-$3.21 \\
\ion{Ne}{i}& $\le$$-$4.86 & $-$4.51 & $-$3.96\\
\ion{Na}{i} &$\le$ $-$5.71 & $-$5.69 & $-$5.71\\
\ion{Mg}{ii} & $-$5.66 & $-$5.45 & $-$4.46\\
\ion{Al}{i} & $\le$$-$7.30 & $-$6.65 & $-$5.57\\
\ion{Al}{ii} & $\le$$-$7.30 &$\le$ $-$7.40 & $-$5.57\\
\ion{Si}{ii} & $-$7.35 & $-$5.61$\pm$0.06 & $-$4.49\\
\ion{P}{ii} & $-$4.44$\pm$0.27 & $-$5.02$\pm$0.30 & $-$6.59\\
\ion{P}{iii}& $-$5.11$\pm$0.26 & $-$5.87          &$-$6.59\\
\ion{S}{ii} &$-$6.26  & $-$5.74 & $-$4.71\\
\ion{Cl}{i} &$\le$$-$7.74 & $\le$$-$7.04 & $-$6.54\\
\ion{Ca}{ii} & $-$5.68 & $-$5.98 & $-$5.68\\
\ion{Sc}{ii} &$\le$ $-$9.50 & $\le$$-$9.50 & $-$8.87\\
\ion{Ti}{ii} & $-$6.47$\pm$0.13 &$-$6.46$\pm$0.06  & $-$7.02\\
\ion{V}{ii}  &$\le$ $-$9.14 & $-$8.94 & $-$8.04\\
\ion{Cr}{ii} & $-$6.10$\pm$0.09 & $-$7.48 & $-$6.37\\
\ion{Mn}{ii} & $-$5.18$\pm$0.32 & $-$5.82$\pm$0.22 & $-$6.65\\
\ion{Fe}{i} & $-$3.65$\pm$0.07 & $-$3.91$\pm$0.06 & $-$4.54\\
\ion{Fe}{ii} & $-$3.65$\pm$0.09 & $-$3.91$\pm$0.08 & $-$4.54\\
\ion{Co}{ii} &$\le$$-$8.42 &$\le$ $-$8.02 & $-$7.12\\
\ion{Ni}{ii} &$-$6.24 & $-$6.47 & $-$5.79\\
\ion{Cu}{ii} &$\le$$-$7.83 & $-$6.24$\pm$0.02 & $-$7.83\\
\ion{Zn}{i}  & $--$        &$-$5.85  & $-$7.44\\
\ion{Zn}{ii} & $--$        &$-$5.76  & $-$7.44\\
\ion{As}{ii} & $--$        & $>$$-$9.67 & $-$9.67\\
\ion{Sr}{ii} &$\le$$-$10.07     & $-$10.67 & $-$9.07\\
\ion{Y}{ii}  &$-$8.60      & $-$8.09$\pm$0.08  & $-$9.80\\
\ion{Xe}{ii} &$-$5.23      & $-$5.81  & $-$9.87\\
\ion{Hg}{ii} & $-$8.20       & $-$7.10  & $-$10.91\\
\hline\noalign{\smallskip}
\end{tabular}
\end{center}
\end{table}

\section{ The 3d$^{6}$($^{5}$D)4d $-$ 3d$^{6}$($^{5}$D)4f transitions of \ion{Fe}{ii}}

The 4d-4f transitions discussed in this paper appear in the 
optical region, mostly between 4800 and 6000\,\AA.
Their presence in stellar spectra  was found long time
ago by Johansson \& Cowley (1984). They are present also
in the iron-rich peculiar stars HR\,6000
and 46\,Aql. We used UVES spectra of these stars to derive stellar
$\log\,gf$'s that we compared with experimental and computed
values.

\subsection{Experimental $\log\,gf$'s}

All the experimental data described in this section have been made available
to F. Castelli by S. Johansson and are briefly described in
Johansson (2002). 
Radiative lifetime measurements of five 3d$^{6}$($^{5}$D)4f levels,
i.e. 4[6]$_{13/2}$, 4[7]$_{13/2}$, 4[7]$_{15/2}$, 3[6]$_{13/2}$,
 3[5]$_{11/2}$, and branching fraction measurements for 13 
transitions 4d-4f with wavelengths in the 4800-5800\,\AA\ region 
were made at Lund.  
Einstein cofficients A, derived by combining the measured branching fractions 
with the lifetime measurements, were converted into experimental
oscillator strengths.
The 4d$-$4f transitions together with the experimental $\log\,gf$'s
are given in Table\,2 and Table\,4.
While the 4f levels with J $>$11/2 decay only to 4d levels,   
the 4f\,3[5]$_{11/2}$ level may decay also to 3d
levels. As a consequence, the $\log\,gf$'s 
for the transitions involving the 
3[5]$_{11/2}$ level  may be less accurate than those 
related with levels with J$>$11/2 which have an estimated error of 0.05\,dex.

\subsection{Computed $\log\,gf$'s}

The computed $\log\,gf$'s  were taken both from
Kurucz's line lists (K09, January, 2009 
version)\footnote{http://kurucz/harvard/edu/atoms/2601/gf2601kjan09.pos} and 
from Raassen \& Uylings (1998) (RU98) data. We note that
Fuhr \& Wiese (2006) (FW06) adopted  
the RU98 data for the few ($^{5}$D)4d $-$ ($^{5}$D)4f transitions that
they listed in their critical compilation.

Both K09 and RU98 results are due to semi-empirical methods,
although different. The K09 results are obtained with the use of the 
Cowan (1981) atomic structure code  while the RU98 rsults are obtained by the
orthogonal operators method. 
The computed $\log\,gf$'s are listed in Table\,4.

\subsection{Stellar $\log\,gf$'s}

In order to derive stellar $\log\,gf$'s we computed 
synthetic profiles for the \ion{Fe}{ii} lines listed in Table\,4.
We used the ATLAS12 models discussed in Sect.\,2, the SYNTHE code (Kurucz 2005)
and line lists based on the Kurucz database   that 
we continually update with new data when available (Castelli \& Hubrig 2004).
Keeping fixed the iron abundance of $-$3.65\,dex for HR\,6000 and 
of $-$3.91\,dex for 46\,Aql (Table\,3) we adjusted 
the $\log\,gf$ values in the calculated
spectrum for the lines listed in Table\,4 until  observed and computed 
profiles agree at best.
All the lines can be well fitted except for the very strong ones
with $\log\,gf$  larger than 0.9. Their observed cores are
stronger than the computed and can never be reproduced by the computed
spectrum because increasing the $\log\,gf$ broadens the wings rather than
increases the core.
An example is the line at 5260.254\,\AA\ that we decided to exclude from
the comparisons. Possibly, the strongest \ion{Fe}{ii} lines are affected by
vertical iron inhomogeneties which do not affect the medium-strong and weak
lines.

\begin{figure}
\centering
\resizebox{4.75in}{!}{\rotatebox{90}{\includegraphics[30,10][400,750]
{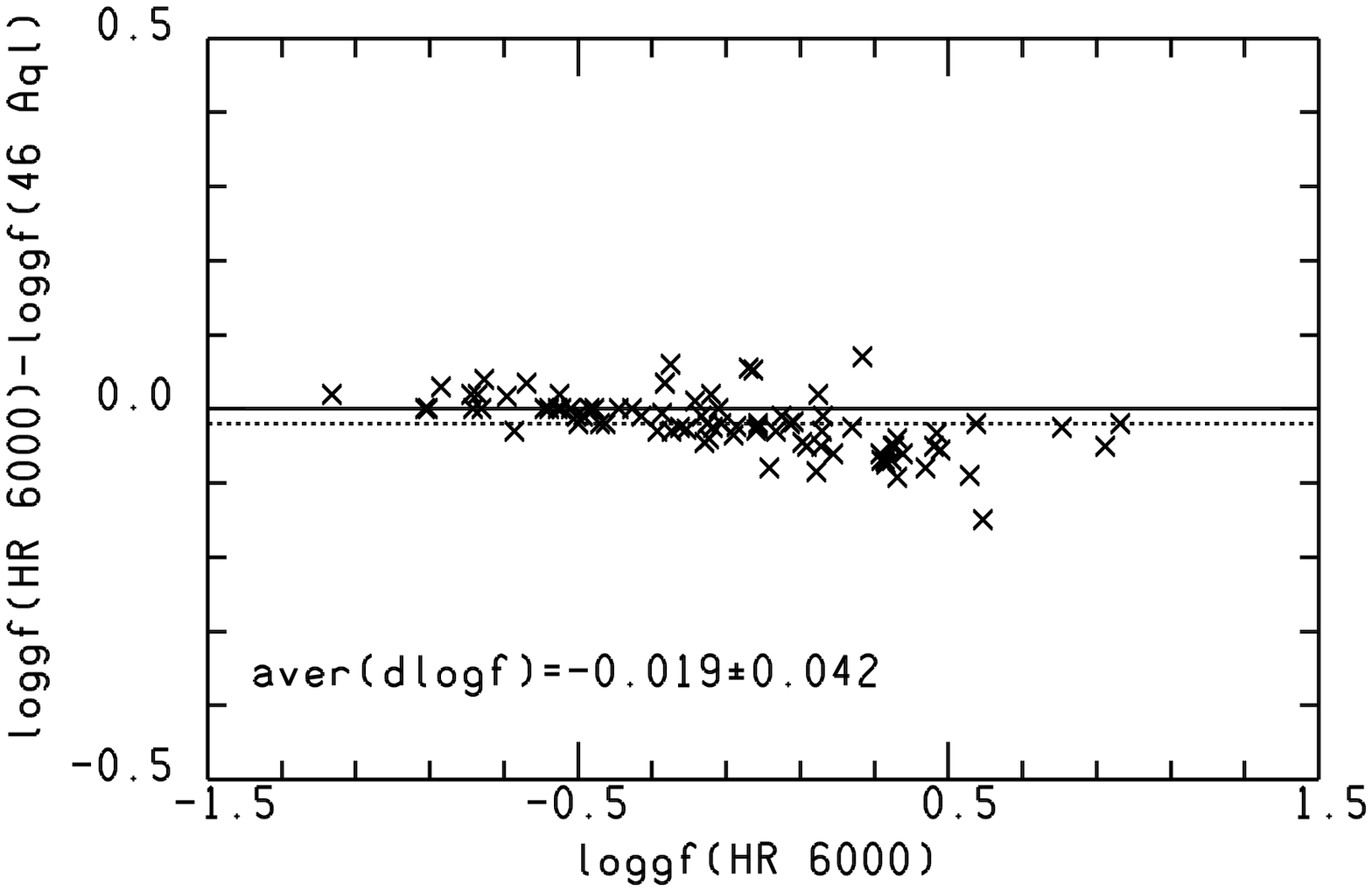}}}
\caption{Comparison of stellar $\log\,gf$'s from HR\,6000 and
46\,Aql for ($^{5}$D)4d-($^{5}$D)4f transitions of \ion{Fe}{ii}.
 The horizontal dashed line indicates the average difference.}
\label{}
\end{figure}

Fig.\,1 shows the difference between the stellar $\log\,gf$'s from
HR\,6000 and from 46 Aql as function of the stellar $\log\,gf$'s from
HR\,6000. The average difference, shown by the dashed line, is 
$-$0.019$\pm$0.042\,dex, but the difference for 
 single lines increases with increasing $\log\,gf$'s from 0.00 
to  0.15\,dex,  in the sense that the values from 46\,Aql become 
larger than those from HR\,6000. 
The largest difference of $-$0.15\,dex occurs for $\lambda$  5961.705\,\AA.
Because each stellar value well reproduces the observed line
in each star we do not have explanation for the discrepancy.

We assumed as stellar $\log\,gf$'s the average of the values obtained
from HR\,6000 and 46\,Aql. Stellar $\log\,gf$'s for the two stars 
and the average are listed in Table\,4.

\subsection{Comparison of $\log\,gf$'s}

The difference between experimental and computed $\log\,gf$'s 
versus the experimental $\log\,gf$'s is shown 
Fig.\,2 and Fig.\,3, where the computed data are those from K09 and
RU98, respectively. The average difference of 0.028$\pm$0.061\,dex 
yielded by the RU98 values is lower than the average difference of  
0.043$\pm$0.071\,dex given by the K09 values, but they agree within the 
error limits. The largest difference of $+$0.130 in Fig.\,2 is
due to the transition at 4883.292\,\AA. However,
the J value of the 4f upper level of this line is 11/2, therefore not 
high enough to assure no additional decays to 3d levels (Johansson, 2002).
This fact could affect the experimental $\log\,gf$'s and the better agreement
with the RU98 data could be fortuitous. In fact, Table\,4 shows that 
the stellar $\log\,gf$ agrees better with the K09 than with the RU98 
value.

\begin{figure}
\centering
\resizebox{4.75in}{!}{\rotatebox{90}{\includegraphics[30,10][400,750]
{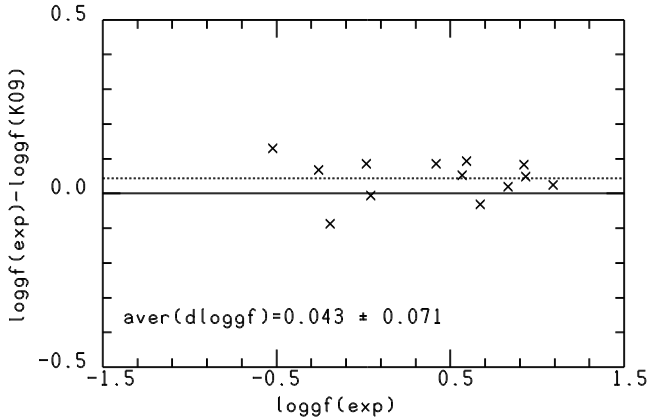}}}
\caption{Calculated $\log\,gf$'s (K09) for ($^{5}$D)4d-($^{5}$D)4f transitions 
of \ion{Fe}{ii} are compared 
with experimental $\log\,gf$'s. The horizontal dashed line indicates the
average difference.}
\label{}
\end{figure}

\begin{figure}
\centering
\resizebox{4.75in}{!}{\rotatebox{90}{\includegraphics[30,10][400,750]
{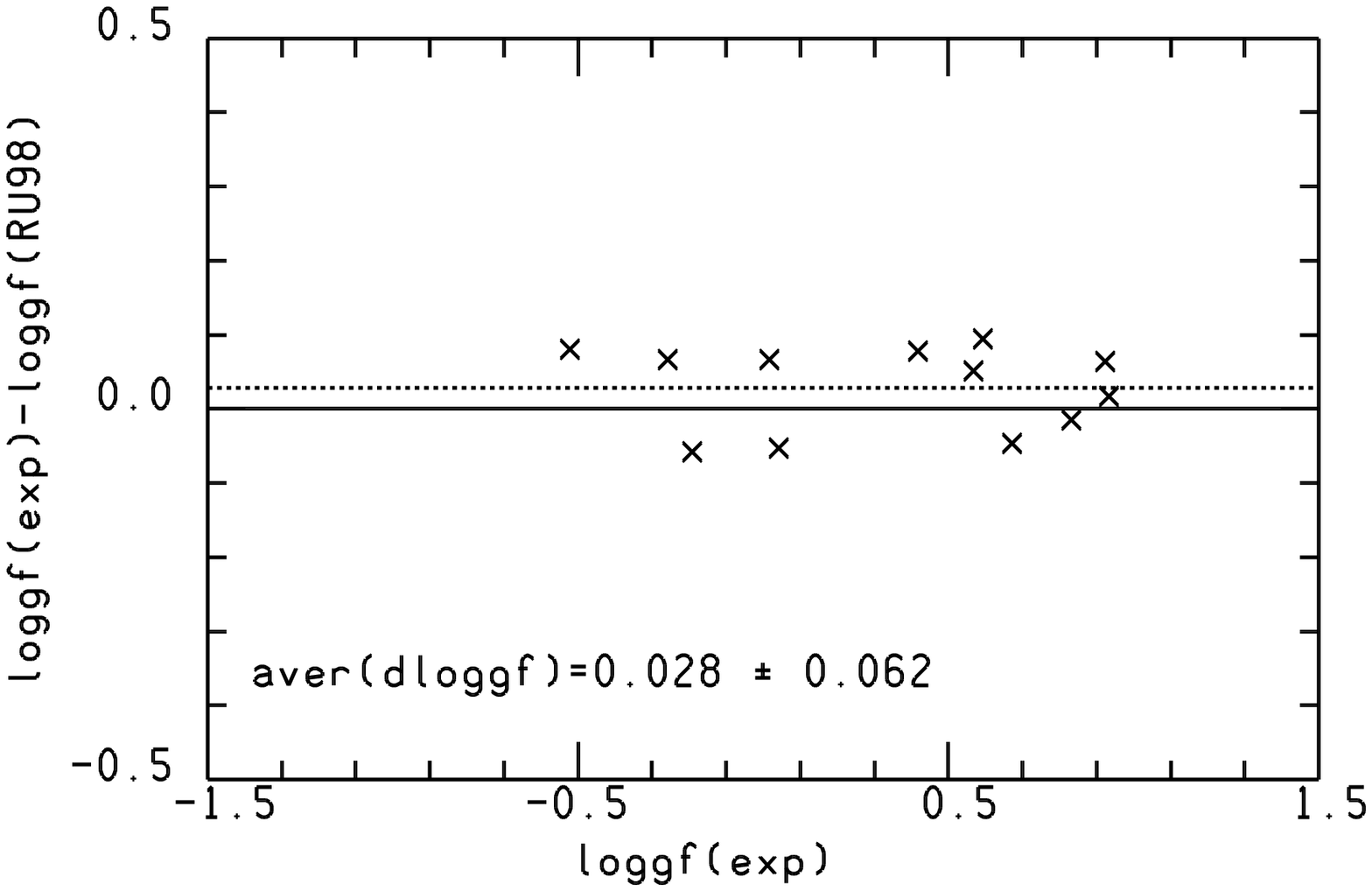}}}
\caption{Calculated $\log\,gf$'s (RU98) for ($^{5}$D)4d-($^{5}$D)4f 
transitions of \ion{Fe}{ii} are compared 
with experimental $\log$gf's. The horizontal dashed line indicates the
average difference}
\label{}
\end{figure}

\begin{figure}
\centering
\resizebox{4.75in}{!}{\rotatebox{90}{\includegraphics[30,10][400,750]
{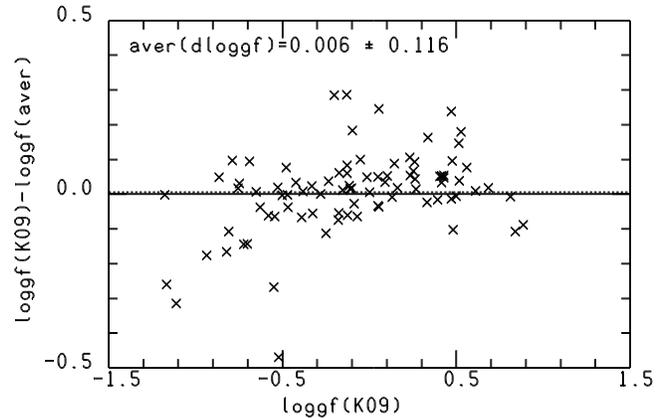}}}
\caption{Calculated $\log\,gf$'s (K09) for ($^{5}$D)4d-($^{5}$D)4f transitions 
of \ion{Fe}{ii} are compared 
with astrophysical $\log\,gf$'s. The horizontal dashed line indicates the
average difference.}
\label{}
\end{figure}

Stellar $\log\,gf$'s are compared with  K09 and RU98  $\log\,gf$'s 
in Fig.\,4 and Fig.\,5, respectively.
The mean difference of 0.006$\pm$0.116\,dex given by the K09 data 
is fully comparable with the average difference of 
$-$0.007$\pm$0.144\,dex yielded by the RU98 $\log\,gf$'s.
In both cases the dispersion around the mean value is 
rather large. The lines giving the largest discrepancies are
different in K09 and RU98. In K09 the lines are those
at  5257.119 ($-$0.47), 
5359.237 ($-$0.314), 5358.872 ($+$0.286), 5355.421 ($+$0.285),
5366.210 ($-$0.267), 5062.927 ($-$0.260)\,\AA. 
In RU98 they are those 
at 5070.583 ($-$0.760), 5140.689 ($-$0.543), 5093.783 ($-$0.399),
5200.798 (+0.287), 5081.898 ($-$0.279) \,\AA.
The values in parenthesis are the difference 
$\log\,gf$(computed)$-$ $\log\,gf$(stellar).
For all these transitions and for both sets of computed $\log\,gf$'s 
the difference between the stellar $\log\,gf$'s from HR\,6000 and
46\,Aql is less than 0.06\,dex, so that 
the cause of the disagreements probably is due to the computed values.

We note that Kurucz updates his calculation whenever new \ion{Fe}{ii} levels
become available. The January 2009 version of the \ion{Fe}{ii}
line list used for the  ($^{5}$D)4d$-$($^{5}$D)4f 
transitions discussed in this paper includes only a few of the new
($^{3}$H)4d$-$($^{3}$H)4f levels presented in Sect.5.
In a near future a new \ion{Fe}{ii} line list with 
all new levels given in Table\,5 and Table\,6  
will be made available on the Kurucz web-site.

\begin{figure}
\centering
\resizebox{4.75in}{!}{\rotatebox{90}{\includegraphics[30,10][460,750]
{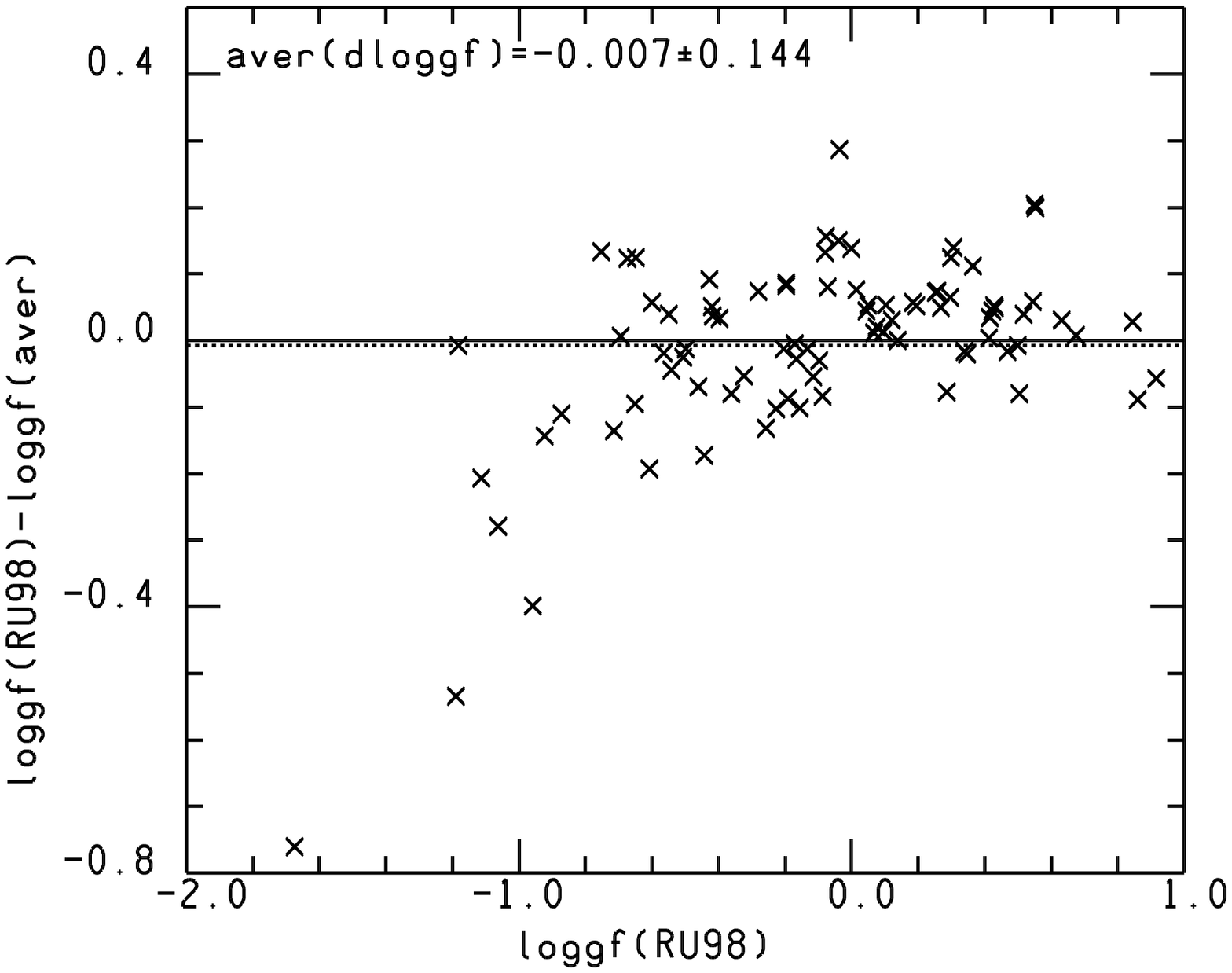}}}
\caption{Calculated $\log\,gf$'s (RU98) for ($^{5}$D)4d-($^{5}$D)4f transitions 
of \ion{Fe}{ii} are compared 
with stellar $\log\,gf$'s. The horizontal dashed line indicates the
average difference.}
\label{}
\end{figure}

\begin{table*}[]
\caption{Stellar $\log\,gf$'s for a sample of ($^{5}$D)4d$-$($^{5}$D)4f lines of \ion{Fe}{ii} observed in HR\,6000 and 46\,Aql. 
$\log\,gf$'s from the two stars are averaged and compared with experimental $\log\,gf$'s from Johansson (2002) 
and calculated $\log\,gf$'s 
from Kurucz (2009)(K09)(footnote\,7) and from Raassen \& Uylings (1998) (RU98). } 

\begin{flushleft}
\begin{tabular}{cllccrrrrrrcccccc}
\hline\noalign{\smallskip}
\multicolumn{1}{c}{$\lambda$($\AA$)}&
\multicolumn{2}{c}{transition}&
\multicolumn{1}{c}{$\chi_{low}$(cm$^{-1}$)}&
\multicolumn{1}{c}{$\chi_{up}$(cm$^{-1}$)}&
\multicolumn{5}{c}{$\log$gf}\\
\hline\noalign{\smallskip}
&&&&& exp & HR\,6000 & 46\,Aql & Aver &K09 &RU98 \\
\hline\noalign{\smallskip}
4883.292 & ($^{5}$D)4d e$^{6}$F$_{11/2}$ & ($^{5}$D)4f\,3[5]$_{11/2}$& 82853.656 & 103325.927&$-$0.521&$-$0.673 &$-$0.643 &$-$0.658&$-$0.651& $-$0.601\\ 
4913.295 & ($^{5}$D)4d e$^{6}$F$_{9/2}$  & ($^{5}$D)4f\,3[5]$_{11/2}$&  82978.668 & 103325.927&  0.016&$-$0.014 &0.006 &$-$0.004&$-$0.069& 
0.050     \\
4990.509 & ($^{5}$D)4d e$^{6}$F$_{5/2}$& ($^{5}$D)4f\,3[4]$_{7/2}$&  83308.195 & 103340.652&  &0.118 &0.168 &0.143&0.161 &0.195     \\
5001.953 & ($^{5}$D)4d \,e$^{6}$F$_{11/2}$ & ($^{5}$D)4f\,4[6]$_{13/2}$& 82853.656 & 102840.269&0.933& 0.963 &0.983&0.973& 0.884&0.916\\
5004.188 & ($^{5}$D)4d \,e$^{6}$F$_{11/2}$ & ($^{5}$D)4f\,4[5]$_{11/2}$& 82853.656 & 102831.344& &0.574 &0.594&0.584& 0.482 & 0.504 \\
5006.840 & ($^{5}$D)4d \,$^{6}$D$_{7/2}$   & ($^{5}$D)4f\,2[4]$_{9/2} $& 83713.534 & 103680.640& &$-$0.265 &$-$0.300&$-$0.282& $-$0.281 &
$-$0.362\\
5007.450 & ($^{5}$D)4d \,$^{6}$D$_{9/2}$ & ($^{5}$D)4f\,2[5]$_{11/2}$& 83726.367 & 103691.045&& $-$0.390 &$-$0.390&$-$0.390& $-$0.382&
$-$0.460 \\
5010.061 & ($^{5}$D)4d \,$^{6}$D$_{9/2}$ & ($^{5}$D)4f\,2[4]$_{9/2}$& 83726.367 & 103680.640&& $-$0.693 &$-$0.710&$-$0.701& $-$0.808&
$-$0.694 \\
5011.026 & ($^{5}$D)4d \,$^{6}$D$_{9/2}$ & ($^{5}$D)4f\,2[3]$_{7/2}$& 83726.367 & 103676.798&& $-$1.165 &$-$1.185&$-$1.175& $-$1.177&
$-$1.183 \\
5022.419 & ($^{5}$D)4d \,e$^{6}$F$_{3/2}$ & ($^{5}$D)4f\,3[3]$_{5/2}$& 83459.674 & 103364.847&& $-$0.142 &$-$0.162&$-$0.152& $-$0.052&
$-$0.072 \\
5026.798 & ($^{5}$D)4d \,e$^{6}$F$_{7/2}$ & ($^{5}$D)4f\,4[2]$_{5/2}$& 83136.462 & 103024.293&& $-$0.287 &$-$0.257&$-$0.272& $-$0.235&
$-$0.444 \\
5030.631 & ($^{5}$D)4d \,e$^{6}$F$_{9/2}$ & ($^{5}$D)4f\,4[5]$_{9/2}$& 82978.668 & 102851.345&& 0.361&0.401&0.381& 0.428&0.431 \\
5032.704 & ($^{5}$D)4d \,$^{6}$D$_{5/2}$ & ($^{5}$D)4f\,2[3]$_{7/2}$& 83812.304 & 103676.798&& 0.050 &0.060&0.055& 0.143&0.076 \\
5035.700 & ($^{5}$D)4d \,e$^{6}$F$_{9/2}$ & ($^{5}$D)4f\,4[5]$_{11/2}$& 82978.668 & 102831.344 &&0.557 &0.647&0.602& 0.611&0.632 \\
5036.713 & ($^{5}$D)4d \,$^{6}$D$_{5/2}$ & ($^{5}$D)4f\,2[2]$_{5/2}$& 83812.304 & 103660.987&& $-$0.546 &$-$0.546&$-$0.546& $-$0.527&
$-$0.565 \\
5045.108 & ($^{5}$D)4d \,e$^{6}$F$_{7/2}$ & ($^{5}$D)4f\,4[3]$_{5/2}$& 83136.462 & 102952.122& &$-$0.150 &$-$0.130&$-$0.140& $-$0.116&
$-$0.002 \\
5060.249 & ($^{5}$D)4d \,$^{6}$P$_{7/2}$ & ($^{5}$D)4f\,0[3]$_{7/2}$& 84266.544 & 104022.912& &$-$0.555 &$-$0.555&$-$0.555& $-$0.479&
$-$0.650 \\
5062.927 & ($^{5}$D)4d \,e$^{6}$F$_{7/2}$ & ($^{5}$D)4f\,4[4]$_{9/2}$& 83136.462 & 102882.375& &$-$0.906 &$-$0.906&$-$0.906& $-$1.166&
$-$1.113 \\
5067.890 & ($^{5}$D)4d \,e$^{6}$F$_{5/2}$ & ($^{5}$D)4f\,4[2]$_{3/2}$& 83308.195 & 103034.770& &$-$0.249 &$-$0.219&$-$0.234& $-$0.173&
$-$0.078 \\
5070.583 & ($^{5}$D)4d \,e$^{6}$F$_{5/2}$ & ($^{5}$D)4f\,4[2]$_{5/2}$& 83308.195 & 103024.293& &$-$0.914 &$-$0.914&$-$0.914& $-$0.865&
$-$1.674 \\
5070.895 & ($^{5}$D)4d \,e$^{6}$F$_{7/2}$ & ($^{5}$D)4f\,4[5]$_{9/2}$& 83136.462 & 102851.345& & 0.189&0.249&0.219& 0.262& 0.268 \\
5075.760 & ($^{5}$D)4d \,$^{6}$P$_{5/2}$ & ($^{5}$D)4f\,0[3]$_{7/2}$& 84326.918 & 104022.912& & 0.105& 0.150&0.127& 0.233& 0.184 \\
5076.597 & ($^{5}$D)4d \,$^{6}$D$_{7/2}$ & ($^{5}$D)4f\,3[2]$_{5/2}$& 83713.534 & 103406.278& & $-$0.770 &$-$0.790&$-$0.780&$-$0.749&
$-$0.924 \\
5081.898 & ($^{5}$D)4d \,$^{6}$D$_{7/2}$ & ($^{5}$D)4f\,3[3]$_{7/2}$& 83713.534 & 103385.735& &$-$0.783&$-$0.783&$-$0.783& $-$0.689&
$-$1.062 \\
5083.503 & ($^{5}$D)4d \,e$^{6}$F$_{3/2}$ & ($^{5}$D)4f\,4[1]$_{1/2}$& 83459.674 & 103125.669& & $-$0.870&$-$0.900&$-$0.885& $-$0.788&
$-$0.752 \\
5086.306 & ($^{5}$D)4d \,$^{6}$D$_{3/2}$ & ($^{5}$D)4f\,2[2]$_{3/2}$& 83990.059 & 103645.211& & $-$0.470 &$-$0.470&$-$0.470&$-$0.472&
$-$0.419 \\
5089.214 & ($^{5}$D)4d \,e$^{6}$F$_{5/2}$ & ($^{5}$D)4f\,4[3]$_{5/2}$& 83308.195 & 102952.122& &$-$0.081&$-$0.046&$-$0.063&$-$0.014&
0.013 \\
5093.783 & ($^{5}$D)4d \,$^{6}$D$_{9/2}$ & ($^{5}$D)4f\,4[5]$_{9/2}$& 83726.367 & 103352.673& &$-$0.550&$-$0.570&$-$0.560&$-$0.703&
$-$0.959 \\
5100.734 & ($^{5}$D)4d \,$^{6}$D$_{9/2}$ & ($^{5}$D)4f\,4[5]$_{11/2}$& 83726.367 & 103352.927&  0.671& & & & 0.703 & 0.718\\
5117.032 & ($^{5}$D)4d \,$^{6}$D$_{1/2}$ & ($^{5}$D)4f\,2[1]$_{3/2}$& 84131.575 & 103668.714& &$-$0.184&$-$0.194&$-$0.189&$-$0.129&$-$0.039 \\
5140.689 & ($^{5}$D)4d \,$^{6}$D$_{3/2}$ & ($^{5}$D)4f\,3[1]$_{1/2}$& 83990.059 & 103437.280& &$-$0.639&$-$0.674&$-$0.656&$-$0.822&$-$1.190 \\
5143.875 & ($^{5}$D)4d \,$^{6}$P$_{7/2}$ & ($^{5}$D)4f\,2[5]$_{9/2}$& 84266.544 & 103701.729& & $-$0.206&$-$0.178&$-$0.192&  0.054 & $-$0.205 \\
5144.352 & ($^{5}$D)4d \,$^{6}$P$_{3/2}$ & ($^{5}$D)4f\,1[2]$_{5/2}$& 84424.376 & 103857.755& & 0.162 &0.172&0.167&  0.260&0.307 \\
5149.465 & ($^{5}$D)4d \,$^{6}$P$_{7/2}$ & ($^{5}$D)4f\,2[4]$_{9/2}$& 84266.544 & 103680.640& & 0.319 &0.389  &0.354&0.405  &0.553 \\
5180.312 & ($^{5}$D)4d \,$^{6}$D$_{5/2}$ & ($^{5}$D)4f\,4[1]$_{3/2}$& 83812.304 & 103110.786& & $-$0.019&0.010&$-$0.004& 0.002& $-$0.088 \\
5199.188 & ($^{5}$D)4d \,$^{6}$D$_{7/2}$ & ($^{5}$D)4f\,4[3]$_{7/2}$& 83713.534 & 102942.208& & 0.082 &0.100&0.091&  0.054 & 0.122 \\
5200.798 & ($^{5}$D)4d \,$^{6}$D$_{5/2}$ & ($^{5}$D)4f\,4[2]$_{3/2}$& 83812.304 & 103034.770& &$-$0.328 &$-$0.318&$-$0.323& $-$0.390 &$-$0.036 \\
5203.634 & ($^{5}$D)4d \,$^{6}$D$_{5/2}$ & ($^{5}$D)4f\,4[2]$_{5/2}$& 83812.304 & 103024.293& &$-$0.073&$-$0.049&$-$0.061&$-$0.088&$-$0.115 \\
5218.841 & ($^{5}$D)4d \,$^{6}$D$_{9/2}$ & ($^{5}$D)4f\,4[4]$_{9/2}$& 83726.367 & 102882.375& & $-$0.160&$-$0.115&$-$0.137& $-$0.250 & $-$0.165\\
5219.920 & ($^{5}$D)4d \,f$^{4}$D$_{5/2}$ & ($^{5}$D)4f\,0[3]$_{7/2}$&84870.864 & 104022.912& & $-$0.590&$-$0.590 &$-$0.590& $-$0.628& $-$0.550 \\
5222.350 & ($^{5}$D)4d \,e$^{6}$G$_{5/2}$ & ($^{5}$D)4f\,1[3]$_{5/2}$&84844.819 & 103987.951& & $-$0.355&$-$0.355&$-$0.355& $-$0.332& $-$0.281 \\
5223.802 & ($^{5}$D)4d \,$^{6}$D$_{7/2}$ & ($^{5}$D)4f\,4[5]$_{9/2}$&83713.534 & 102851.345& & $-$0.486&$-$0.476&$-$0.481& $-$0.546& $-$0.506 \\
5224.404 & ($^{5}$D)4d \,$^{6}$D$_{3/2}$ & ($^{5}$D)4f\,4[1]$_{1/2}$&83990.059 & 103125.669& & $-$0.519&$-$0.519&$-$0.519& $-$0.581& $-$0.428 \\
5227.487 & ($^{5}$D)4d e$^{6}$G$_{11/2}$& ($^{5}$D)4f\,3[6]$_{13/2}$ & 84296.833 & 103421.165&0.831&0.806&0.831 &0.818&0.811&0.846 \\  
5237.949 & ($^{5}$D)4d \,$^{6}$P$_{7/2}$ & ($^{5}$D)4f\,3[5]$_{9/2}$&84266.544 & 103352.673& &  0.035 &0.065&0.050& 0.103 & 0.104\\
5245.455 & ($^{5}$D)4d \,$^{6}$P$_{5/2}$ & ($^{5}$D)4f\,3[3]$_{7/2}$&84326.918 & 103385.735& &  $-$0.504 &$-$0.494&$-$0.499& $-$0.502& $-$0.543 \\
5247.956 & ($^{5}$D)4d \,e$^{6}$G$_{3/2}$ & ($^{5}$D)4f\,1[3]$_{5/2}$&84938.220 & 103987.951& &  0.316  & 0.376&0.346&0.526&0.550 \\
5251.225 & ($^{5}$D)4d e$^{6}$G$_{5/2}$& ($^{5}$D)4f\,1[4]$_{7/2}$  &84844.819 & 103882.692 &  & 0.355 &0.405&0.380&0.475 &0.424 \\  
5253.649 & ($^{5}$D)4d e$^{6}$G$_{11/2}$& ($^{5}$D)4f\,3[5]$_{11/2}$  &84296.833 & 103325.927 &$-$0.191&$-$0.121 &$-$0.121&$-$0.121&$-$0.104 & $-$0.133\\  
5257.119 & ($^{5}$D)4d \,f$^{4}$D$_{7/2}$ & ($^{5}$D)4f\,2[5]$_{9/2}$&84685.198 & 103701.729& &$-$0.028 &$-$0.080&$-$0.054&$-$0.523&0.156 \\
5260.254 & ($^{5}$D)4d \,e$^{6}$G$_{13/2}$ & ($^{5}$D)4f\,4[7]$_{15/2}$&84035.121 & 103040.317 &1.090&$--$ &$--$&$--$& 1.065&$--$ \\
5264.180 & ($^{5}$D)4d \,e$^{6}$G$_{7/2}$ & ($^{5}$D)4f\,2[5]$_{9/2}$&84710.703 & 103701.729 & &0.267&0.197 &0.232&0.470 & 0.297\\
5265.985 & ($^{5}$D)4d \,e$^{6}$G$_{13/2}$ & ($^{5}$D)4f\,4[7]$_{13/2}$&84035.121 & 103019.639& &$-$0.760&$-$0.760&$-$0.760& $-$0.936&
 $-$0.871 \\
5270.029 & ($^{5}$D)4d \,e$^{6}$G$_{7/2}$ & ($^{5}$D)4f\,2[4]$_{9/2}$&84710.703 & 103680.640& &$-$0.250&$-$0.310&$-$0.280&$-$0.097&
$-$0.197 \\
\hline\noalign{\smallskip}
\end{tabular}
\end{flushleft}
\end{table*}

\setcounter{table}{3}

\begin{table*}[]
\caption{cont.} 
\begin{flushleft}
\begin{tabular}{clccrrrrrrrccccc}
\hline\noalign{\smallskip}
\multicolumn{1}{c}{$\lambda$($\AA$)}&
\multicolumn{2}{c}{transition}&
\multicolumn{1}{c}{$\chi_{low}$(cm$^{-1}$)}&
\multicolumn{1}{c}{$\chi_{up}$(cm$^{-1}$)}&
\multicolumn{6}{c}{$\log$gf}\\
\hline\noalign{\smallskip}
&&&&& exp & HR\,6000 & 46\,Aql & Aver&K09&RU98 \\
\hline\noalign{\smallskip}
5291.661 & ($^{5}$D)4d \,e$^{6}$G$_{9/2}$ & ($^{5}$D)4f\,3[6]$_{11/2}$&84527.758 & 103420.158& &0.460 &0.510&0.485& 0.561& 0.544 \\
5306.182 & ($^{5}$D)4d \,f$^{4}$D$_{5/2}$ & ($^{5}$D)4f\,2[4]$_{7/2}$&84870.864 & 103711.562& &$-$0.013 &0.012&$-$0.001 &0.049&0.044 \\
5315.083 & ($^{5}$D)4d \,f$^{4}$D$_{3/2}$ & ($^{5}$D)4f\,1[2]$_{5/2}$&85048.609 & 103857.755& &$-$0.455 &$-$0.455&$-$0.455&$-$0.422&
$-$0.418 \\
5316.214 & ($^{5}$D)4d e$^{6}$G$_{13/2}$& ($^{5}$D)4f\,4[6]$_{13/2}$&  84035.121 & 102840.269& 0.418 &0.378 &0.438 &0.356&0.332&0.340\\
5318.055 & ($^{5}$D)4d \,e$^{6}$G$_{9/2}$ & ($^{5}$D)4f\,3[4]$_{9/2}$&84527.758 & 103326.396& &$-$0.136 &$-$0.111&$-$0.123&$-$0.177&$-$0.226 \\
5339.592 & ($^{5}$D)4d e$^{6}$G$_{11/2}$& ($^{5}$D)4f\,4[7]$_{13/2}$&  84296.833 & 103019.639 & 0.568 &0.438 &0.518 &0.478&0.516&0.517\\
5355.421 & ($^{5}$D)4d \,f$^{4}$D$_{7/2}$ & ($^{5}$D)4f\,3[5]$_{9/2}$&84685.198 & 103352.673& &$-$0.498 &$-$0.478&$-$0.488&$-$0.203&
$-$0.500 \\
5358.872 & ($^{5}$D)4d \,f$^{4}$D$_{7/2}$ & ($^{5}$D)4f\,3[4]$_{7/2}$&84685.198 & 103340.652& &$-$0.426 &$-$0.406&$-$0.416&$-$0.130&
 $-$0.609 \\
5359.237 & ($^{5}$D)4d \,e$^{6}$G$_{7/2}$ & ($^{5}$D)4f\,3[3]$_{5/2}$&84710.703 & 103364.847& &$-$0.788 &$-$0.808&$-$0.798&$-$1.112&
$-$0.675 \\
5366.210 & ($^{5}$D)4d \,e$^{6}$G$_{7/2}$ & ($^{5}$D)4f\,3[4]$_{7/2}$&84710.703 & 103340.652 &&$-$0.265 &$-$0.300&$-$0.282&$-$0.549&
$-$0.196 \\
5387.064 & ($^{5}$D)4d e$^{4}$G$_{11/2}$& ($^{5}$D)4f\,3[6]$_{13/2}$& 84863.334 & 103421.165& 0.593&0.479 &0.534 &0.506&0.500&0.499 \\ 
5395.855 & ($^{5}$D)4d $^{6}$S$_{5/2}$& ($^{5}$D)4f\,0[3]$_{7/2}$& 85495.318 & 104022.912& &0.328&0.398&0.363&0.415 &0.285 \\ 
5402.059 & ($^{5}$D)4d e$^{4}$G$_{9/2}$& ($^{5}$D)4f\,2[5]$_{11/2}$& 85184.725 & 103691.045 & &0.469&0.502 &0.485&0.472 &0.469\\
5414.852 & ($^{5}$D)4d e$^{4}$G$_{11/2}$& ($^{5}$D)4f\,3[5]$_{11/2}$& 84863.334 & 103325.927 & $-$0.258&$-$0.274 &$-$0.268 &$-$0.271&
$-$0.326 & $-$0.324\\
5429.987 & ($^{5}$D)4d \,e$^{4}$G$_{7/2}$ & ($^{5}$D)4f\,1[4]$_{9/2}$&85462.858 & 103873.991& &0.350 &0.400&0.375& 0.429&0.427 \\
5444.386 & ($^{5}$D)4d \,$^{6}$S$_{5/2}$ & ($^{5}$D)4f\,1[2]$_{5/2}$&85495.318 & 103857.755& &$-$0.167 &$-$0.157&$-$0.165& $-$0.153&
$-$0.170 \\
5451.316 & ($^{5}$D)4d \,f$^{4}$D$_{7/2}$ & ($^{5}$D)4f\,4[2]$_{5/2}$&84685.198 & 103024.293& &$-$0.753 &$-$0.793&$-$0.773& $-$0.756&
$-$0.649 \\
5465.932 & ($^{5}$D)4d \,e$^{4}$G$_{5/2}$ & ($^{5}$D)4f\,1[3]$_{7/2}$&85679.709 & 103969.766& &0.331 &0.406&0.368& 0.515&0.348 \\
5472.855 & ($^{5}$D)4d \,f$^{4}$D$_{7/2}$ & ($^{5}$D)4f\,4[3]$_{5/2}$&84685.198 & 102952.122& &$-$0.579 &$-$0.579&$-$0.579&$-$0.723&
$-$0.715 \\
5475.826 & ($^{5}$D)4d \,f$^{4}$D$_{7/2}$ & ($^{5}$D)4f\,4[3]$_{7/2}$&84685.198 & 102942.208& & $-$0.225 &$-$0.200&$-$0.212&$-$0.129&
$-$0.080 \\
5482.307 & ($^{5}$D)4d \,e$^{4}$G$_{9/2}$ & ($^{5}$D)4f\,3[6]$_{11/2}$&85184.725 & 103420.158& &0.363 &0.455 &0.409&0.393 &0.413 \\
5488.776 & ($^{5}$D)4d \,e$^{4}$G$_{7/2}$ & ($^{5}$D)4f\,2[3]$_{7/2}$&85462.858 & 103676.798& & $-$0.440 &$-$0.420&$-$0.430&$-$0.468&
$-$0.397 \\
5492.398 & ($^{5}$D)4d \,f$^{4}$D$_{7/2}$ & ($^{5}$D)4f\,4[4]$_{7/2}$&84685.198 & 102887.124& & $-$0.039 &$-$0.094&$-$0.066&$-$0.127&
$-$0.097 \\
5493.830 & ($^{5}$D)4d \,f$^{4}$D$_{7/2}$ & ($^{5}$D)4f\,4[4]$_{9/2}$&84685.198 & 102882.375& &0.143 &0.228&0.185&0.252&0.259 \\
5502.670 & ($^{5}$D)4d \,e$^{4}$G$_{9/2}$ & ($^{5}$D)4f\,3[5]$_{9/2}$&85184.725 & 103352.673& &$-$0.114 &$-$0.094&$-$0.104&$-$0.179&
$-$0.192 \\
5506.199 & ($^{5}$D)4d e$^{4}$G$_{11/2}$& ($^{5}$D)4f\,4[7]$_{13/2}$&  84863.334 & 103019.639& 0.923&0.923 &0.973 &0.948&0.840&0.859\\
5510.783 & ($^{5}$D)4d e$^{4}$G$_{9/2}$& ($^{5}$D)4f\,3[5]$_{11/2}$& 85184.725 & 103325.927& 0.043&0.073 &0.093 &0.083&0.049&0.096\\
5529.053 & ($^{5}$D)4d f$^{4}$D$_{5/2}$& ($^{5}$D)4f\,3[2]$_{5/2}$& 84870.864 & 102952.122&  &$-$0.146&$-$0.106&$-$0.126&$-$0.110 &$-$0.258\\
5544.763 & ($^{5}$D)4d e$^{4}$G$_{11/2}$& ($^{5}$D)4f\,4[6]$_{11/2}$& 84863.334 & 102893.377&  &0.149 &0.129 &0.139&0.130 &0.139\\
5783.623 & ($^{5}$D)4d e$^{4}$F$_{7/2}$& ($^{5}$D)4f\,2[5]$_{9/2}$& 86416.323 & 103701.729&  &0.241&0.266 &0.253&0.267 &0.365\\
5885.015 & ($^{5}$D)4d e$^{4}$F$_{3/2}$& ($^{5}$D)4f\,2[3]$_{5/2}$& 86710.864 & 103698.466&  &0.158&0.188 &0.173&0.336 &0.298\\
5902.825 & ($^{5}$D)4d e$^{4}$F$_{7/2}$& ($^{5}$D)4f\,3[5]$_{9/2}$& 86416.323 & 103352.673& &0.346&0.416 &0.381&0.415 &0.416\\
5955.698 & ($^{5}$D)4d e$^{4}$F$_{5/2}$& ($^{5}$D)4f\,3[3]$_{7/2}$& 86599.744 & 103385.735&  &0.155&0.205 &0.180&0.234 &0.252\\
5961.705 & ($^{5}$D)4d e$^{4}$F$_{9/2}$& ($^{5}$D)4f\,4[6]$_{11/2}$& 86124.301 & 102893.377& &0.593 &0.742 &0.667&0.685 &0.675\\
5965.622 & ($^{5}$D)4d e$^{4}$F$_{9/2}$& ($^{5}$D)4f\,4[4]$_{9/2}$& 86124.301 & 102882.375&  &0.016 &0.096 &0.056&0.091 &0.068\\
\hline\noalign{\smallskip}
\end{tabular}
\end{flushleft}
\end{table*}

\section{New identified lines in HR\,6000 and 46\,Aql}

The spectrum of HR\,6000 contains a huge number of unidentified lines
mostly concentrated in the 5000-5400\,\AA\ region (Castelli \& Hubrig, 2007).
The same unidentified lines can be observed also in the spectrum
of 46\,Aql.  S. Johansson (2006) remarked that a great number of unidentified 
lines in the plots of HR\,6000 available at the Castelli
web-site (footnote\,2) are also present in the laboratory iron spectra. 
He therefore identified several unknown features in the 4000-5500\,\AA\ 
interval of HR\,6000 as due to iron. These identifications are at different 
levels of completeness.
In a few cases the transition is identified only as \ion{Fe}, 
in most cases as \ion{Fe}{ii}, in some cases as \ion{Fe}{ii} with both levels 
of the transition classified.

\subsection{ The 3d$^{6}$($^{3}$H)4d $-$ 3d$^{6}$($^{3}$H)4f transitions of \ion{Fe}{ii}}

Most of the \ion{Fe}{ii} lines  classified 
by S. Johansson are due to the ($^{3}$H)4d-($^{3}$H)4f
transitions. Their lower excitation potential  is
larger than 103800\,cm$^{-1}$ (12.87 eV) and  upper
excitation potential is of the order of 123000\,cm$^{-1}$ 
(15.25 eV), therefore close to the ionization limit of 130563\,cm$^{-1}$
(16.19\,eV). 

In a preliminary work, Castelli et al. (2008) gave an example
 in HR\,6000 and 46\,Aql of four  lines at 5176.711, 5177.3896,
5177.7762, and 5179.536\,\AA\  identified for the first time as due to 
\ion{Fe}{ii} ($^{3}$H)4d-($^{3}$H)4f transitions.
Only for two of them (5177.3896 and 5179.539\,\AA) 
energies  and  terms were known for both levels, while for the other two 
(5176.711 and 5179.536\,\AA)  the term of the upper level 
was unknown, except for the J quantum number. 
Further lines and energies  indicated by S. Johansson  are those marked 
with a  ``J'' in Table\,5.

In order to complete and extend the number of the identifications we proceeded
as follows.  We started from the ``J'' lines. 
Knowing the term, the energy of the lower level of a ``J'' line  was determined
 either using the NIST database (footnote\,4) or the Kurucz line 
lists (footnote\,1). The  particular ($^{3}$H)4d-($^{3}$H)4f transition was then 
searched among the \ion{Fe}{ii} predicted lines  available
in the Kurucz database (version 2007). For this search
the lower energy level and  the J quantum number of the upper level  
were used as key. Once the predicted line corresponding to the new identified
transition was fixed, the predicted energy was replaced by the  
energy assigned by S. Johansson to the level. This substitution was made
 not only for the given line but also for all the lines having that predicted
level as their upper level. Then we compared the pattern of 
the computed $\log\,gf$ values to the pattern of stellar $\log\,gf$ values
for those lines. If they were similar we accepted the identification.
If not, we tried another match. In this way, thanks to the J lines, we fixed 11 
new energy levels together with all the transitions
having a  new level as upper level.
In addition, we determined another 10 new levels from predicted lines
by searching in the spectrum unidentified lines with corresponding intensity
and wavelength difference as the predicted lines having  
$\log\,gf$ $\ge$ 0.0 or  a negative $\log\,gf$  close to zero and
arising from the same upper predicted level.  The observed
wavelengths and the lower observed level were then used to fix the upper
level of the transitions.

Table\,5 lists both the ($^{3}$H)4d-($^{3}$H)4f transitions originally 
identified by S. Johansson  and those we derived from the above
described procedure. The letter ``J'' is associated to the first group of lines,
the letter ``K''  to the second group. 
Not all the lines listed in table\,5  are observable in 
the spectra, so that  stellar $\log\,gf$'s can not be assigned to all the 
lines. The last column of Table\,5 lists lines observed in HR\,6000 which were
listed as unidentified lines by Castelli \& Hubrig (2007).
 
 Fig.\,6  compares stellar $\log\,gf$s from HR\,6000 and 46\,Aql.
It is the analogous of Fig.\,1, but for the
($^{3}$H)4d-($^{3}$H)4f transitions. The average difference, shown
by the dashed line, is negligible, but the scatter is larger than
that obtained for the ($^{5}$D)4d-($^{5}$D)4f transitions.
The reason is the very low intensity of some 
($^{3}$H)4d-($^{3}$H)4f transitions that makes the fitting of the
profile rather problematic.  
As for the ($^{5}$D)4d-($^{5}$D)4f transitions we assumed as 
stellar $\log\,gf$ the average of the value obtained from HR\,6000
and 46\,Aql, but we excluded the lines with stellar $\log\,gf$'s 
differing more than 0.1\,dex. Stellar $\log\,gf$'s for the two stars and
the average are listed in Table\,5.

\begin{figure}
\centering
\resizebox{4.75in}{!}{\rotatebox{90}{\includegraphics[30,10][400,750]
{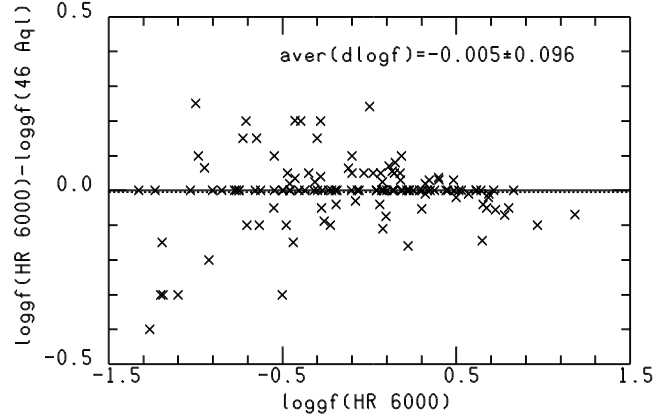}}}
\caption{Comparison of stellar $\log\,gf$'s from HR\,6000 and
46\,Aql for ($^{3}$H)4d-($^{3}$H)4f transitions of \ion{Fe}{ii}.
The horizontal dashed line indicates the average difference.}
\label{}
\end{figure}

The stellar $\log\,gf$'s are compared with the calculated 
$\log\,gf$'s in Fig.\,7. We excluded from this comparison
the lines at $\lambda\lambda$  5177.777 ($-$1.45),
5250.632 ($-$2.61), 5346.098 ($-$1.21), and 5420.234 ($-$1.76)\,\AA,
for which the $\log\,gf$ difference between computed and
astrophysical $\log\,gf$'s, given in parenthesis, 
is larger than 1.0\,dex. Their upper energies of the lines are 122952.73, 
123026.35, 123015.40, and 123251.47\,cm$^{-1}$, respectively.
The most probably explanation for these large discrepancies  
is the presence of some additional unidentified component which
contributes to the absorption, so that the stellar $\log\,gf$ 
is largely overestimated.
The mean difference between the two sets of data is
 $-$0.07 $\pm$ 0.22\,dex, indicating that the stellar $\log\,gf$'s are,
on average, larger than the computed ones.
The differences for the individual lines increase with decreasing
$\log\,gf$'s, namely with decreasing line intensity.
Weak lines are more difficult to be fitted by the synthetic
spectrum owing to the contribution of the noise and the
non-negligible effect of the position for the continuum.

\begin{figure}
\centering
\resizebox{4.75in}{!}{\rotatebox{90}{\includegraphics[30,10][400,750]
{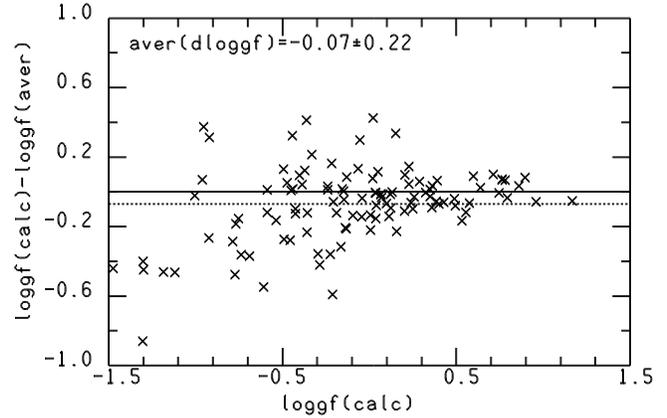}}}
\caption{The calculated $\log\,gf$'s (calc.) for ($^{3}$H)4d-($^{3}$H)4f
transitions of \ion{Fe}{ii} are compared with astrophysical $\log\,gf$'s obtained
averaging stellar from HR\,6000 and 46\,Aql.  The horizontal dashed line 
indicates the average difference.}
\label{}
\end{figure}

\begin{table*}[]
\begin{flushleft}
\caption{Lines due to  ($^{3}$H)4d $-$ ($^{3}$H)4f transitions of \ion{Fe}{ii}} 
\begin{tabular}{lllllllllrl}
\noalign{\smallskip}\hline
\multicolumn{2}{c}{Upper level}&
\multicolumn{2}{c}{Lower level}&
\multicolumn{1}{c}{$\lambda$(calc)}&
\multicolumn{4}{c}{$log$\,gf}&
\multicolumn{1}{c}{}&
\multicolumn{1}{c}{$\lambda$(obs),Notes}
\\
\noalign{\smallskip}\hline
\multicolumn{1}{c}{cm$^{-1}$}&
\multicolumn{1}{c}{} &
\multicolumn{1}{c}{cm$^{-1}$}&
\multicolumn{1}{c}{} &
\multicolumn{1}{c}{\AA} &
\multicolumn{1}{c}{HR6000} &
\multicolumn{1}{c}{46 Aql} &
\multicolumn{1}{c}{Aver} &
\multicolumn{1}{c}{calc.} &
\multicolumn{1}{c}{} &
\multicolumn{1}{c}{\AA} 
\\
\noalign{\smallskip}\hline

122910.92 & ($^{3}$H)4f~~6[ ]$_{15/2}$& 103617.58 & ($^{3}$H)4d $^{4}$H$_{13/2}$  & 5181.693 &$--$& $--$  &$--$  &$-$3.314& K &  \\
          &                           & 103644.80 & ($^{3}$H)4d $^{4}$K$_{17/2}$  & 5189.014 &$--$& $--$  &$--$  &$-$3.086  &K & blend\\
          &                           & 103706.53 & ($^{3}$H)4d $^{4}$K$_{15/2}$  & 5205.693 &$+$0.075& $+$0.050&$+$0.063& $-$0.295  &K &5205.714\\
          &                           & 103832.05 & ($^{3}$H)4d $^{4}$K$_{13/2}$  & 5239.942 &$+$0.225& $+$0.225&$+$0.225& $+$0.006  &K &5239.948\\
          &                           & 103878.37 & ($^{3}$H)4d $^{4}$I$_{15/2}$  & 5252.695 &$+$0.070& $+$0.070 &$+$0.070& $-$0.138  &K &5252.702\\
          &                           & 104064.67 & ($^{3}$H)4d $^{4}$I$_{13/2}$  & 5304.620 &$-$0.314& $-$0.339  &$-$0.327& $-$0.426  &K& 5304.60\\
          &                           & 104119.71 & ($^{3}$H)4d $^{2}$K$_{15/2}$  & 5320.157 &$+$0.078& $+$0.188  & $--$ & $+$0.133 &K &5320.18\\
          &                           & 104315.37 & ($^{3}$H)4d $^{2}$K$_{13/2}$  & 5376.136 &$+$0.253& $+$0.253  &$+$0.253& $+$0.112  &K &5376.12\\
          &                           & 104622.30 & ($^{3}$H)4d $^{2}$I$_{13/2}$  & 5466.362 &$+$0.614& $+$0.614  &$+$0.614& $+$0.713  &K &5466.38\\
\\
122952.73 & ($^{3}$H)4f~~6[9]$_{17/2}$& 103644.80 & ($^{3}$H)4d $^{4}$K$_{17/2}$  & 5177.777 & $+$0.513 &$+$0.513&$+$0.513&$-$0.933& J & 5177.77\\
           &                            & 103706.53 & ($^{3}$H)4d $^{4}$K$_{15/2}$ & 5194.384 & $+$0.724&$+$0.780&$+$0.752& $+$0.745  &J & 5194.387\\
           &                            & 103878.37 & ($^{3}$H)4d $^{4}$I$_{15/2}$ & 5241.181 & $+$0.655&$+$0.695&$+$0.675& $+$0.556  &J & 5241.183\\
           &                            & 104119.71 & ($^{3}$H)4d $^{2}$K$_{15/2}$ & 5308.346 & $+$0.500&$+$0.520&$+$0.510& $+$0.599  &J & 5308.35\\
\\
122954.18 & ($^{3}$H)4f~~6[9]$_{19/2}$& 103644.80 & ($^{3}$H)4d $^{4}$K$_{17/2}$ & 5177.388 & $+$1.183&$+$1.253& $+$1.218& $+$1.166   &J & 5177.394\\
\\
122990.62 & ($^{3}$H)4f~~6[]$_{13/2}$& 103600.43 & ($^{3}$H)4d $^{4}$G$_{11/2}$ & 5155.811 & $--$& $--$ &$--$  &$-$3.353  &K &\\
           &                            & 103617.58 & ($^{3}$H)4d $^{4}$H$_{13/2}$ & 5160.375 &$--$& $--$&$--$ &$-$1.722&K & \\
           &                            & 103706.53 & ($^{3}$H)4d $^{4}$K$_{15/2}$ & 5184.178 &$--$&$--$&$--$&$-$1.029 &K & \\
           &                            & 103751.66 & ($^{3}$H)4d $^{4}$H$_{11/2}$ & 5196.339 & $+$0.114&$+$0.044&$+$0.079&$-$0.134&K & 5196.345\\
           &                            & 103832.05 & ($^{3}$H)4d $^{4}$K$_{13/2}$ & 5218.143 & $+$0.100&$+$0.100&$+$0.100&$-$0.044  &J & 5218.149\\
           &                            & 103878.37 & ($^{3}$H)4d $^{4}$I$_{15/2}$ & 5230.790 &$-$1.200 &$-$0.900&$--$&$-$1.233  &K & very weak\\
           &                            & 103973.78 & ($^{3}$H)4d $^{4}$K$_{11/2}$ & 5257.034 &$--$&$--$&$--$&$-$0.894&K &\\
           &                            & 104064.67 & ($^{3}$H)4d $^{4}$I$_{13/2}$ & 5282.281 & $-$0.704&$-$0.604&$-$0.654&$-$1.118 &K &\\
           &                            & 104119.71 & ($^{3}$H)4d $^{2}$K$_{15/2}$ & 5297.687 &$-$1.327 &$-$1.327&$-$1.327&$-$0.953  &K & very weak\\
           &                            & 104174.27 & ($^{3}$H)4d $^{4}$I$_{11/2}$ & 5313.049 &$--$&$--$&$--$&$-$0.987  &K & \\
           &                            & 104315.37 & ($^{3}$H)4d $^{2}$K$_{13/2}$ & 5353.192 & $+$0.148&$+$0.068&$+$0.108&$+$0.205 &K & blend, 5353.251\\
           &                            & 104622.30 & ($^{3}$H)4d $^{2}$I$_{13/2}$ & 5442.643 & $+$0.130&$+$0.075&$+$0.103&$+$0.070  &J & 5442.65\\
           &                            & 104765.45 & ($^{3}$H)4d $^{2}$I$_{11/2}$ & 5485.393 & $-$0.187&$-$0.187&$-$0.187&$+$0.150  &K & 5485.40\\
           &                            & 106045.69 & ($^{3}$H)4d $^{2}$H$_{11/2}$ & 5899.835 & $+$0.348&$+$0.348&$+$0.348&$+$0.250  &K & 5899.82\\
\\
123002.28 & ($^{3}$H)4f~~6[ ]$_{11/2}$ &  103600.43 & ($^{3}$H)4d $^{4}$G$_{11/2}$ & 5152.712 &$+$0.616 &$+$0.616 &$+$0.616&$+$0.639  & K  &5152.70\\
          &                            & 103617.58 & ($^{3}$H)4d $^{4}$H$_{13/2}$ & 5157.271 & $--$ & $--$ & $--$  & $+$0.383 &K & blend \\
          &                            & 103751.66 & ($^{3}$H)4d $^{4}$H$_{11/2}$ & 5193.192 &$-$0.767 &$-$0.767: &$-$0.767&$-$0.444 &K & weak\\
          &                            & 103771.32 & ($^{3}$H)4d $^{4}$G$_{9/2}$  & 5198.501 &$\le$$-$1.030 &$\le$$-$1.230 &$--$&$-$1.357  &K   &\\
          &                            & 103832.05 & ($^{3}$H)4d $^{4}$K$_{13/2}$  & 5214.970 &$--$ & $--$  & $--$&$-$2.624 &K   &\\
          &                            & 103874.26 & ($^{3}$H)4d $^{4}$H$_{9/2}$  & 5226.478 &$--$ & $--$  & $--$&$-$2.790 &K   &\\
          &                            & 103973.78 & ($^{3}$H)4d $^{4}$K$_{11/2}$ & 5253.813 &$--$ & $--$  & $--$&$-$4.096  &K   &\\
          &                            & 104064.67 & ($^{3}$H)4d $^{4}$I$_{13/2}$  & 5279.028 &$--$ & $--$  & $--$&$-$2.424  &K   &\\
          &                            & 104174.27 & ($^{3}$H)4d $^{4}$I$_{11/2}$ & 5309.759 &$--$ & $--$  & $--$&$-$2.421  &K   &\\
          &                            & 104192.48 & ($^{3}$H)4d $^{4}$I$_{9/2}$  & 5314.899 &$--$ & $--$  & $--$&$-$4.098  &K   &\\
          &                            & 104315.37 & ($^{3}$H)4d $^{2}$K$_{13/2}$  & 5397.852 &$--$ & $--$  & $--$&$-$3.473  &K   &\\
          &                            & 104622.30 & ($^{3}$H)4d $^{2}$I$_{13/2}$  & 5439.191 &$--$ & $--$  & $--$&$-$3.967  &K   &\\
          &                            & 104765.45 & ($^{3}$H)4d $^{2}$I$_{11/2}$ & 5481.886 &$-$1.264 &$-$0.864 &$--$&$-$1.294  &K   &very weak\\
          &                            & 104807.21 & ($^{3}$H)4d $^{2}$G$_{9/2}$  & 5494.468 &$-$0.923 &$-$0.723 &$--$&$-$1.632  &K   &very weak\\
          &                            & 104916.55 & ($^{3}$H)4d $^{4}$F$_{9/2}$  & 5527.868 &$-$1.031 &$-$1.031&$-$1.031&$-$0.962  &K   &very weak\\
          &                            & 105063.55 & ($^{3}$F)4d $^{4}$G$_{11/2}$  & 5572.983 &$-$0.650  &$-$0.550 &$-$0.600  &$-$0.753  &K \\
          &                            & 105155.09 & ($^{3}$F)4d $^{4}$F$_{9/2}$  & 5601.568 &$--$& $--$  & $--$&$-$2.044  &K   &\\
          &                            & 105288.85 & ($^{3}$F)4d $^{4}$H$_{13/2}$  & 5643.867 &$--$& $--$  & $--$&$-$1.416  &K   &\\
          &                            & 105398.85 & ($^{3}$F)4d $^{4}$H$_{11/2}$  & 5679.136 &$--$& $--$  & $--$&$-$2.876  &K   &\\
          &                            & 105524.46 & ($^{3}$F)4d $^{4}$H$_{9/2}$  & 5719.951 &$--$& $--$  & $--$&$-$4.454  &K   &\\
          &                            & 105763.27 & ($^{3}$F)4d $^{2}$H$_{11/2}$  & 5799.189 &$--$& $--$  & $--$&$-$1.904  &K   &\\
          &                            & 106018.64 & ($^{3}$F)4d $^{2}$H$_{9/2}$ & 5886.389 &$--$& $--$  & $--$&$-$6.492  &K   &\\
          &                            & 106045.69 & ($^{3}$H)4d $^{2}$H$_{11/2}$ & 5895.778 &$--$& $--$  & $--$&$-$1.550  &K   &\\
          &                            & 106097.52 & ($^{3}$H)4d $^{2}$H$_{9/2}$  & 5913.855 &$--$& $--$  & $--$&$-$2.902  &K   &\\
\\
123007.91 & ($^{3}$H)4f~~6[8]$_{17/2}$ &  103644.80 & ($^{3}$H)4d $^{4}$K$_{17/2}$ & 5163.021 & $+$0.571 &$+$0.581&$+$0.576&$+$0.495  & J  &5163.00\\
           &                           & 103706.53 & ($^{3}$H)4d $^{4}$K$_{15/2}$ & 5179.534 & $+$0.640 &$+$0.640&$+$0.640&$+$0.574  & J  &5179.54\\
           &                           & 103878.37 & ($^{3}$H)4d $^{4}$I$_{15/2}$ & 5226.062 & $+$0.829 &$+$0.829&$+$0.829&$+$0.794  & J & 5226.07\\
           &                       & 104119.71 & ($^{3}$H)4d $^{2}$K$_{15/2}$ & 5292.838 &$--$ &$--$&$--$&$-$1.288  & K&\\
\hline
\noalign{\smallskip}
\end{tabular}
\end{flushleft}
\end{table*}

\setcounter{table}{4}

\begin{table*}[]
\begin{flushleft}
\caption{cont.} 
\begin{tabular}{lllllllllrl}
\noalign{\smallskip}\hline
\multicolumn{2}{c}{Upper level}&
\multicolumn{2}{c}{Lower level}&
\multicolumn{1}{c}{$\lambda$(calc)}&
\multicolumn{4}{c}{$log$\,gf}&
\multicolumn{1}{c}{}&
\multicolumn{1}{c}{$\lambda$(obs),Notes}
\\
\noalign{\smallskip}\hline
\multicolumn{1}{c}{cm$^{-1}$}&
\multicolumn{1}{c}{} &
\multicolumn{1}{c}{cm$^{-1}$}&
\multicolumn{1}{c}{} &
\multicolumn{1}{c}{\AA} &
\multicolumn{1}{c}{HR6000} &
\multicolumn{1}{c}{46\,Aql} &
\multicolumn{1}{c}{Aver} &
\multicolumn{1}{c}{calc.} &
\multicolumn{1}{c}{} &
\multicolumn{1}{c}{\AA} 
\\ 
\noalign{\smallskip}\hline
123015.40 & ($^{3}$H)4f~~6[]$_{13/2}$ & 103600.43 & ($^{3}$H)4d $^{4}$G$_{11/2}$ & 5149.230    &  $+$ 0.485 & $+$0.455&$+$0.470&$+$0.400& J &5149.25\\
           &                           & 103617.58 & ($^{3}$H)4d $^{4}$H$_{13/2}$ & 5153.783 &  $+$ 0.684& $+$0.704&$+$0.694&$+$0.764& J &  5153.78\\
           &                           & 103706.53 & ($^{3}$H)4d $^{4}$K$_{15/2}$ & 5177.525 &  $--$ &$--$&$--$&$-$0.318& K & \\
           &                           & 103751.66 & ($^{3}$H)4d $^{4}$H$_{11/2}$ & 5189.655 &  $--$ &$--$&$--$&$-$0.576& K & \\
           &                           & 103832.05 & ($^{3}$H)4d $^{4}$K$_{11/2}$ & 5211.403 &$<-$1.238& $--$&$--$&$-$2.671 & K & \\
           &                           & 103878.37 & ($^{3}$H)4d $^{4}$I$_{15/2}$ & 5224.017 & $-$0.102 &$-$0.102&$-$0.102&$-$0.146& J & 5224.025\\
           &                           & 103973.78 & ($^{3}$H)4d $^{4}$K$_{11/2}$ & 5250.193 &$--$ &$--$ & $--$&$-$3.971& K & \\
           &                           & 104064.67 & ($^{3}$H)4d $^{4}$I$_{13/2}$ & 5275.373 &$--$ &$--$& $--$ & $-$1.672& K &\\
           &                           & 104119.71 & ($^{3}$H)4d $^{2}$K$_{15/2}$ & 5290.740 & $-$0.902 &$-$0.902&$-$0.902&$-$1.303& K & 5290.730 \\
           &                           & 104174.27 & ($^{3}$H)4d $^{4}$I$_{11/2}$ & 5306.061 &  $--$ &$--$&$--$&$-$2.648& K &\\
           &                           & 104315.37 & ($^{3}$H)4d $^{2}$K$_{13/2}$ & 5346.098 & $-$0.750 &$-$0.750&$-$0.750&$-$1.957&  K & \\
           &                           & 104622.30 & ($^{3}$H)4d $^{2}$I$_{13/2}$ & 5435.311 &$--$ &$--$&$--$& $-$2.414&  K &  \\
           &                           & 104765.45 & ($^{3}$H)4d $^{2}$I$_{11/2}$ & 5477.945 &$--$  &$--$ &$--$& $-$1.272&  K &\\
           &                           & 106045.69 & ($^{3}$H)4d $^{2}$H$_{11/2}$ & 5891.220 &$--$ &$--$&$--$& $-$1.347&  K & \\
\\
123018.43 & ($^{3}$H)4f~~6[]$_{15/2}$ & 103617.58 & ($^{3}$H)4d $^{4}$H$_{13/2}$ & 5152.978    & $+$0.687  &$+$0.697&$+$0.692& $+$0.763 &J& 5152.98\\
           &                           & 103644.80 & ($^{3}$H)4d $^{4}$K$_{17/2}$ & 5160.218 & $-$0.235  &$-$0.235&$-$0.235&$-$0.357 & K & 5160.2\\
           &                           & 103706.53 & ($^{3}$H)4d $^{4}$K$_{15/2}$ & 5176.713 & $+$0.440  &$+$0.440&$+$0.440&$+$0.385 & J & 5176.72\\ 
           &                           & 103832.05 & ($^{3}$H)4d $^{4}$K$_{13/2}$ & 5210.580 & $--$ &$--$&$--$&$-$1.050 & K & \\
           &                           & 103878.37 & ($^{3}$H)4d $^{4}$I$_{15/2}$ & 5223.190 & $+$0.487 &$+$0.487&$+$0.487& $+$0.427 &K & blend\\
           &                           & 104064.67 & ($^{3}$H)4d $^{4}$I$_{13/2}$ & 5274.530 & $-$1.193 &$-$1.043&$--$&$-$1.138 & K & very weak\\
           &                           & 104119.71 & ($^{3}$H)4d $^{2}$K$_{15/2}$ & 5289.892 & $-$0.656 &$-$0.656&$-$0.656&$-$0.923 & K & 5289.899\\
           &                           & 104315.37 & ($^{3}$H)4d $^{2}$K$_{13/2}$ & 5345.232 &$--$&$--$ &$--$&$-$2.411 &K & \\
           &                           & 104622.30 & ($^{3}$H)4d $^{2}$I$_{13/2}$ & 5434.415 &$--$ &$--$&$--$&$-$1.487 & K &\\
\\
123026.35 & ($^{3}$H)4f~~6[]$_{9/2}$  & 103600.43 & ($^{3}$H)4d $^{4}$G$_{11/2}$ & 5146.326 &$--$ &$--$&$--$&$-$4.447& K &\\
           &                           & 103751.66 & ($^{3}$H)4d $^{4}$H$_{11/2}$ & 5186.706 & $-$0.120 &$-$0.184&$-$0.152& $-$0.149 & J  & 5186.722\\
           &                           & 103771.32 & ($^{3}$H)4d $^{4}$G$_{9/2}$  & 5192.002 & $+$0.080 &$+$0.080&$+$0.080& $+$0.066&  J  & 5192.010\\
           &                           & 103814.55 & ($^{3}$H)4d $^{4}$G$_{7/2}$  & 5203.685 &$--$ & $--$ & $--$& $-$2.154&  K  &\\
           &                           & 103874.26 & ($^{3}$H)4d $^{4}$H$_{9/2}$  & 5219.909 & $-$0.455: &$-$0.455: &$-$0.455&$-$0.440& K   & blend \\
           &                           & 103973.78 & ($^{3}$H)4d $^{4}$K$_{11/2}$ & 5247.175 & $--$& $--$&$--$&$-$3.791 & K& \\
           &                           & 103986.33 & ($^{3}$H)4d $^{4}$G$_{7/2}$  & 5250.632 &$-$0.300 & $-$0.300 ?& $-$0.300&$-$2.907& K &5250.609\\
           &                           & 104174.27 & ($^{3}$H)4d $^{4}$I$_{11/2}$ & 5302.979 &$--$ &$--$&$--$&$-$2.785& K & \\
           &                           & 104192.48 & ($^{3}$H)4d $^{4}$I$_{9/2}$  & 5308.106 &$--$ &$--$&$--$&$-$2.122& K&  \\
           &                           & 104481.59 & ($^{3}$H)4d $^{2}$F$_{7/2}$  & 5390.860 &$--$ &$--$&$--$&$-$1.070& K&\\
           &                           & 104765.45 & ($^{3}$H)4d $^{4}$I$_{11/2}$ & 5474.660 &$--$ &$--$&$--$&$-$1.961& K &\\
           &                           & 104807.21 & ($^{3}$H)4d $^{2}$G$_{9/2}$  & 5487.209 &  $+$0.334  &$+$0.334&$+$0.334&$+$0.348& J & 5497.21\\
           &                           & 104916.55 & ($^{3}$H)4d $^{4}$F$_{9/2}$  & 5520.339 & $-$1.500   &$\le$$-$1.50&$--$&$-$1.382& K &\\
           &                           & 105123.00 & ($^{3}$H)4d $^{2}$G$_{7/2}$  & 5583.996 &$--$ &$--$ & $--$& $-$3.472& K \\
           &                           & 105220.60 & ($^{3}$H)4d $^{4}$F$_{9/2}$  & 5614.605 &$--$ &$--$&  $--$&  $-$1.910& K &\\
           &                           & 106045.69 & ($^{3}$H)4d $^{2}$H$_{11/2}$ & 5887.421 & $-$0.260 &$-$0.172&$-$0.216&$-$0.132& K & 5887.42\\
           &                           & 106097.52 & ($^{3}$H)4d $^{2}$H$_{9/2}$  & 5905.446 & $--$ &$--$&$--$&$-$0.710& K &\\
\\
123037.43 & ($^{3}$H)4f~~6[]$_{11/2}$  & 103600.43 & ($^{3}$H)4d $^{4}$G$_{11/2}$ & 5143.394 &$--$  & $--$& $--$&$-$1.365 & K & \\
           &                        & 103617.58 & ($^{3}$H)4d $^{4}$H$_{13/2}$  & 5147.936 &$--$    & $--$ &$--$& $-$2.244 & K &\\
           &                        & 103751.66 & ($^{3}$H)4d $^{4}$H$_{11/2}$  & 5183.727 &$+$0.303& $+$0.303 & $+$0.303&$+$0.240 & K &5183.713\\
           &                        & 103771.32 & ($^{3}$H)4d $^{4}$G$_{9/2}$  & 5189.016 &$-$0.070 & $-$0.070 & $-$0.070& $-$0.190 & K &5189.013\\
           &                        & 103832.05 & ($^{3}$H)4d $^{4}$K$_{13/2}$  & 5205.425& $-$0.394& $-$0.594 &$--$&$-$0.552 & K &5205.427\\
           &                        & 103874.26 & ($^{3}$H)4d $^{4}$H$_{9/2}$  & 5216.891 &$--$&$--$  &$--$& $-$0.463 & K &\\
           &                        & 103973.78 & ($^{3}$H)4d $^{4}$K$_{11/2}$  & 5244.125 &$--$& $--$& $--$& $-$1.886 & K &\\
           &                        & 104064.67 & ($^{3}$H)4d $^{4}$I$_{13/2}$  & 5269.248 &$-$0.698&noise &$--$&$-$0.828 & K &5269.235\\
           &                        & 104174.27 & ($^{3}$H)4d $^{4}$I$_{11/2}$  & 5299.864 &$--$&$--$ &$--$& $-$1.744 & K &\\
           &                        & 104192.48 & ($^{3}$H)4d $^{4}$I$_{9/2}$  & 5304.985 &$--$ & $--$ & $--$& $-$2.176 & K &\\
           &                        & 104315.37 & ($^{3}$H)4d $^{2}$K$_{13/2}$  & 5339.807 &$-$0.633&$-$0.533 & $-$0.583& $-$0.768 & K &very weak\\
           &                        & 104622.30 & ($^{3}$H)4d $^{2}$I$_{13/2}$  & 5428.808 &$-$0.479&$-$0.379 &$-$0.429& $-$0.387 & K &5288.8\\
           &                        & 104765.45 & ($^{3}$H)4d $^{2}$I$_{11/2}$  & 5471.340 &$-$0.994&$--$ &$--$& $-$0.906 & K &very weak\\
           &                        & 104807.21 & ($^{3}$H)4d $^{2}$G$_{9/2}$  & 5483.874 &$+$0.096 &$+$0.171 &$+$0.134& $+$0.120 & K &5483.85\\
           &                        & 104916.55 & ($^{3}$H)4d $^{4}$F$_{9/2}$  & 5516.963 &$\le$$-$2.00&$\le$$-$2.00 &$--$&$-$1.760 & K &not obs\\
           &                        & 106045.69 & ($^{3}$H)4d $^{2}$H$_{11/2}$  & 5883.582 &$+$0.223&$+$0.383 &$--$& $+$0.259 & K &5883.58\\
           &                        & 106097.52 & ($^{3}$H)4d $^{2}$H$_{9/2}$  & 5901.584 &$--$&$--$ & $--$& $-$0.574 & K &\\
\hline
\noalign{\smallskip}
\end{tabular}
\end{flushleft}
\end{table*}

\setcounter{table}{4}

\begin{table*}[]
\begin{flushleft}
\caption{cont.} 
\begin{tabular}{lllllllllrl}
\noalign{\smallskip}\hline
\multicolumn{2}{c}{Upper level}&
\multicolumn{2}{c}{Lower level}&
\multicolumn{1}{c}{$\lambda$(calc)}&
\multicolumn{4}{c}{$log$\,gf}&
\multicolumn{1}{c}{}&
\multicolumn{1}{c}{$\lambda$(obs),Notes}
\\
\noalign{\smallskip}\hline
\multicolumn{1}{c}{cm$^{-1}$}&
\multicolumn{1}{c}{} &
\multicolumn{1}{c}{cm$^{-1}$}&
\multicolumn{1}{c}{} &
\multicolumn{1}{c}{\AA} &
\multicolumn{1}{c}{HR\,6000} &
\multicolumn{1}{c}{46\,Aql} &
\multicolumn{1}{c}{Aver} &
\multicolumn{1}{c}{calc.} &
\multicolumn{1}{c}{} &
\multicolumn{1}{c}{\AA} 
\\ 
\noalign{\smallskip}\hline
123168.68 & ($^{3}$H)4f~~5[]$_{13/2}$ & 103600.43 & ($^{3}$H)4d $^{4}$G$_{11/2}$ & 5108.895 & $-$1.234 &$-$1.234&$-$1.234& $-$0.921 & K& weak,noise \\
          &                            & 103617.58 & ($^{3}$H)4d $^{4}$H$_{13/2}$ & 5113.377 & $--$ &$--$&$--$& $-$2.499 & K \\
          &                            & 103706.53 & ($^{3}$H)4d $^{4}$K$_{15/2}$ & 5136.747 & $--$ &$--$&$--$& $-$1.281 & K&blend \\
          &                            & 103751.66 & ($^{3}$H)4d $^{4}$H$_{11/2}$ & 5148.687 & $-$0.033 &$-$0.083&$-$0.058& $+$0.020 &K&5148.7\\
          &                            & 103832.05 & ($^{3}$H)4d $^{4}$K$_{11/2}$ & 5170.092 & $--$ &$--$&$--$& $-$1.042 & K& 5170.1,blend \\
          &                            & 103878.37 & ($^{3}$H)4d $^{4}$I$_{15/2}$ & 5182.507 & $--$ &$--$&$--$& $-$1.516 & K \\
          &                            & 103973.78 & ($^{3}$H)4d $^{4}$K$_{11/2}$ & 5208.267 & $+$0.136 &$+$0.136&$+$0.136& $-$0.286 & K&5208.268 \\
          &                            & 104064.67 & ($^{3}$H)4d $^{4}$I$_{13/2}$ & 5233.046 & $+$0.215 &$+$0.215&$+$0.215& $+$0.122 & K & 5233.041 \\
          &                            & 104119.71 & ($^{3}$H)4d $^{2}$K$_{15/2}$ & 5248.167 & $--$ &$--$&$--$& $-$2.450 & K \\
          &                            & 104174.27 & ($^{3}$H)4d $^{4}$I$_{11/2}$ & 5263.242 & $-$0.320 &$-$0.320&$-$0.320& $-$0.689 & K \\
          &                            & 104315.37 & ($^{3}$H)4d $^{2}$K$_{13/2}$ & 5302.633 & $--$ &$--$&$--$& $-$0.553 & K &blend\\
          &                            & 104622.30 & ($^{3}$H)4d $^{2}$I$_{13/2}$ & 5390.389 & $-$0.405 &$-$0.405&$-$0.405& $+$0.021 & K &5390.38\\
          &                            & 104765.45 & ($^{3}$H)4d $^{2}$I$_{11/2}$ & 5432.319 & $+$0.530 &$+$0.530&$+$0.530& $+$0.489 & K&
 5432.31\\
          &                            & 106045.69 & ($^{3}$H)4d $^{2}$H$_{11/2}$ & 5838.483 & $-$0.547 &$-$0.547&$-$0.547& $-$0.332 & K &weak,noise\\
\\

123193.09 & ($^{3}$H)4f~~5[]$_{15/2}$ & 103617.58 & ($^{3}$H)4d $^{4}$H$_{13/2}$ & 5107.001 & $-$0.985 &$-$1.085&$-$1.035& $-$1.475 & K &very weak\\
           &                           & 103644.80 & ($^{3}$H)4d $^{4}$K$_{17/2}$ & 5114.112 & $--$ &$--$&$--$& $-$2.416 & K & \\
           &                           & 103706.53 & ($^{3}$H)4d $^{4}$K$_{15/2}$ & 5130.313 & $-$0.300 &$-$0.450&$--$&$-$0.493   & K & 5130.3\\
           &                           & 103832.05 & ($^{3}$H)4d $^{4}$K$_{13/2}$ & 5163.574 & $+$0.779  &$+$0.850&$+$0.815&$+$0.896& J & 5163.5\\
           &                           & 103878.37 & ($^{3}$H)4d $^{4}$I$_{15/2}$ & 5175.957 & $-$0.429  &$-$0.630&$--$&$-$0.465   & K & 5175.95\\
           &                           & 104064.67 & ($^{3}$H)4d $^{4}$I$_{13/2}$ & 5226.368 & $+$0.397  &$+$0.360&$+$0.379&$-$0.211 & J & 5226.367\\
           &                           & 104119.71 & ($^{3}$H)4d $^{2}$K$_{15/2}$ & 5241.450 & $-$0.100&$-$0.150 &$-$0.125& $-$0.359   & J & 5241.465\\
           &                           & 104315.37 & ($^{3}$H)4d $^{2}$K$_{13/2}$ & 5295.776 &  $-$0.538  &$\le$$-$0.588 &$--$&$-$0.334   & K &5295.773\\
           &                           & 104622.30 & ($^{3}$H)4d $^{2}$I$_{13/2}$ & 5383.304 &  $+$0.401  &$+$0.370 &$+$0.386& $+$0.157   & K & 5383.32\\
\\
123219.199 & ($^{3}$H)4f~~5[8]$_{17/2}$ & 103644.80 & ($^{3}$H)4d $^{4}$K$_{17/2}$ & 5107.290 &$--$  & $--$ & $--$&  $-$0.997 & K & weak,blend\\
           &                           & 103706.53 & ($^{3}$H)4d $^{4}$K$_{15/2}$ &  5123.448 &$+$0.322 & $+$0.332 & $+$0.330&$+$0.390&K &5123.45\\
           &                           & 103878.37 & ($^{3}$H)4d $^{4}$I$_{15/2}$ &  5168.969 &$--$  &$--$ &$--$ & $+$0.201&K & blend\\
           &                           & 104119.71 & ($^{3}$H)4d $^{4}$K$_{15/2}$ &  5234.285 &$+$0.967 &1.067& $+$1.017&$+$0.959&K &5234.283\\
\\
123238.44 & ($^{3}$H)4f~~5[]$_{15/2}$ & 103617.58 & ($^{3}$H)4d $^{4}$H$_{13/2}$ &  5095.196 & $-$0.728 &$-$0.878& $--$ &$-$0.884& K&very weak,blend\\
           &                           & 103644.80 & ($^{3}$H)4d $^{4}$K$_{17/2}$ &  5102.275 & $--$ &$--$ &$--$&$-$2.988&K & \\
           &                           & 103706.53 & ($^{3}$H)4d $^{4}$K$_{15/2}$ &  5118.401 & $-$0.270 &$-$0.270 &$-$0.270& $-$0.239& J & 5118.404\\
           &                           & 103832.05 & ($^{3}$H)4d $^{4}$K$_{13/2}$ &  5151.507 & $-$0.059 &$-$0.059:&$-$0.059&$-$0.608& K & 
5151.52,blend\\
           &                           & 103878.37 & ($^{3}$H)4d $^{4}$I$_{15/2}$ &  5163.831 & $-$0.626 &$-$0.626 &$-$0.626&$-$0.495& K & 5163.82\\
           &                           & 104064.67 & ($^{3}$H)4d $^{4}$I$_{13/2}$ &  5214.007 & $+$0.650 &$+$0.795 &$--$&$+$0.868& K & 5214.96,blend\\
           &                           & 104119.71 & ($^{3}$H)4d $^{2}$K$_{15/2}$ &  5229.017 & $+$0.065 &$+$0.015 &$+$0.040&$-$0.096& J & 5229.038\\
           &                           & 104315.37 & ($^{3}$H)4d $^{2}$K$_{13/2}$ &  5283.085 &  $+$0.343 &$+$0.313 &$+$0.328& $+$0.323& J& 5283.093 \\
           &                           & 104622.30 & ($^{3}$H)4d $^{2}$I$_{13/2}$ &  5370.189 &$--$&$--$ & $--$& $-$1.820& K&\\
\\
123249.65 & ($^{3}$H)4f~~5[]$_{13/2}$ & 103600.43 & ($^{3}$H)4d $^{4}$G$_{11/2}$ &  5087.842 & $-$0.496 &$-$0.496 & $-$0.496& $-$0.402& J & 5087.85\\
           &                           & 103617.58 & ($^{3}$H)4d $^{4}$H$_{13/2}$ &  5092.287 & $-$2.477 &$--$  &$--$& $-$3.143& K & very weak\\
           &                           & 103706.53 & ($^{3}$H)4d $^{4}$K$_{15/2}$ &  5115.465 & $-$0.950 &$-$1.015  &$-$0.983&$-$1.004& K & 5115.5\\
           &                           & 103751.66 & ($^{3}$H)4d $^{4}$H$_{11/2}$ &  5127.305 & $+$0.450:&$+$0.450  &$+$0.450&$+$0.360& J & 5127.32, blend\\
           &                           & 103832.05 & ($^{3}$H)4d $^{4}$K$_{13/2}$ &  5148.533 & $+$0.300 &$+$0.353  &$+$0.326&$+$0.361& J & 5148.52\\
           &                           & 103878.37 & ($^{3}$H)4d $^{4}$I$_{15/2}$ &  5160.844 & $--$ & $--$  & $--$& $-$1.782& K &\\
           &                           & 103973.78 & ($^{3}$H)4d $^{4}$K$_{11/2}$ &  5186.389 & $+$0.060 &$+$0.100  &$+$0.080&$+$0.226& J & 5186.396\\
           &                           & 104064.67 & ($^{3}$H)4d $^{4}$I$_{13/2}$ &  5210.960 & $-$0.280 &$-$0.320  &$-$0.300&$-$0.425& K & 5210.964\\
           &                           & 104119.71 & ($^{3}$H)4d $^{2}$K$_{15/2}$ &  5225.953 & $--$ &$--$  &$--$& $-$0.776& K &\\
           &                           & 104174.27 & ($^{3}$H)4d $^{4}$I$_{11/2}$ &  5240.901 & $-$0.550 &$-$0.500 &$-$0.525&$-$0.474& K & 5240.911\\
           &                           & 104315.37 & ($^{3}$H)4d $^{2}$K$_{13/2}$ &  5279.957 & $-$0.350 &$-$0.400& $-$0.375&$-$0.737& K & blend \\
           &                           & 104622.30 & ($^{3}$H)4d $^{2}$I$_{13/2}$ &  5366.958 & $+$0.143 &$+$0.093  &$+$0.118&$+$0.040& J & 5366.95\\
           &                           & 104765.45 & ($^{3}$H)4d $^{2}$I$_{11/2}$ &  5408.522 &$--$& $--$  &$--$&$-$2.264& K & \\
           &                           & 106045.69 & ($^{3}$H)4d $^{2}$H$_{11/2}$ &  5811.004 & $+$0.050 &bad sp. &$--$&$-$0.189& K  & 5811.00\\

\hline
\noalign{\smallskip}
\end{tabular}
\end{flushleft}
\end{table*}

\setcounter{table}{4}

\begin{table*}[]
\begin{flushleft}
\caption{cont.} 
\begin{tabular}{lllllllllrl}
\noalign{\smallskip}\hline
\multicolumn{2}{c}{Upper level}&
\multicolumn{2}{c}{Lower level}&
\multicolumn{1}{c}{$\lambda$(calc)}&
\multicolumn{4}{c}{$log$\,gf}&
\multicolumn{1}{c}{}&
\multicolumn{1}{c}{$\lambda$(obs),Notes}
\\
\noalign{\smallskip}\hline
\multicolumn{1}{c}{cm$^{-1}$}&
\multicolumn{1}{c}{} &
\multicolumn{1}{c}{cm$^{-1}$}&
\multicolumn{1}{c}{} &
\multicolumn{1}{c}{\AA} &
\multicolumn{1}{c}{HR\,6000} &
\multicolumn{1}{c}{46\,Aql} &
\multicolumn{1}{c}{Aver} &
\multicolumn{1}{c}{calc.} &
\multicolumn{1}{c}{} &
\multicolumn{1}{c}{\AA} 
\\ 
\noalign{\smallskip}\hline
123251.47 &($^{3}$H)4f~~5[]$_{11/2}$  & 103600.43 & ($^{3}$H)4d $^{4}$G$_{11/2}$ &  5087.371 & $--$ & $--$  & $--$& $-$1.402 & K &  \\
           &                           & 103617.58 & ($^{3}$H)4d $^{4}$H$_{13/2}$ &  5091.815 &$--$ &$-$1.000 &$--$&$-$2.881 & K & blend\\
           &                           & 103751.66 & ($^{3}$H)4d $^{4}$H$_{11/2}$ &  5126.827 & $+$0.185 &$+$0.085&$+$0.135&$-$0.224 & J & 5126.84, bl \\
           &                           & 103771.32 & ($^{3}$H)4d $^{4}$G$_{9/2}$  &  5132.001 & $+$0.180 &$+$0.150&$+$0.165&$+$0.097 & J & 5132.0\\
           &                           & 103832.05 & ($^{3}$H)4d $^{4}$K$_{13/2}$ &  5148.050 &$--$ &$--$&$--$&$-$2.123 & K & very weak \\
           &                           & 103874.26 & ($^{3}$H)4d $^{4}$H$_{9/2}$  &  5159.265 & $+$0.140 &$+$0.140&$+$0.140&$+$0.005 & J & 5159.29\\
           &                           & 103973.78 & ($^{3}$H)4d $^{4}$K$_{11/2}$ &  5185.899 & $+$0.080 &$+$0.080&$+$0.080&$+$0.065 & J & 5185.901\\
           &                           & 104064.67 & ($^{3}$H)4d $^{4}$I$_{13/2}$ &  5210.466 & $--$&$--$&$--$&$-$0.582& K & 5210.546,blend\\  
           &                           & 104174.27 & ($^{3}$H)4d $^{4}$I$_{11/2}$ &  5240.401 & $-$0.100 &$-$0.200&$-$0.150&$-$0.207 & J & 5240.405\\
           &                           & 104192.48 & ($^{3}$H)4d $^{4}$I$_{9/2}$  &  5245.408 & $--$ &$--$ &$--$&$-$1.056 & K & \\
           &                           & 104315.37 & ($^{3}$H)4d $^{2}$K$_{13/2}$ &  5279.449 & $-$0.850 &$-$0.850&$-$0.850&$-$1.299 & K & weak\\
           &                           & 104622.30 & ($^{3}$H)4d $^{2}$I$_{13/2}$ &  5366.433 & $-$1.000 &$-$1.250 &$--$&$-$3.032 & K & very weak\\
           &                           & 104765.45 & ($^{3}$H)4d $^{2}$I$_{11/2}$ &  5407.990 & $+$0.040 &$+$0.040&$+$0.040&$+$0.035 & J & 5407.99\\
           &                           & 104807.21 & ($^{3}$H)4d $^{2}$G$_{9/2}$  &  5420.234 & $-$0.750 &$-$0.750&$-$0.750&$-$2.508 & K & blend\\
           &                           & 104916.55 & ($^{3}$H)4d $^{4}$F$_{9/2}$  &  5452.558 & $-$0.500 &$-$0.500 &$-$0.500&$-$0.785 & K& 5452.55\\
           &                           & 106045.69 & ($^{3}$H)4d $^{2}$H$_{11/2}$ &  5810.389 &bad sp.&$-$0.500 &$--$&$-$1.336 & K& blend\\
           &                           & 106097.52 & ($^{3}$H)4d $^{2}$H$_{9/2}$  &  5827.945 & $+$0.190 &$+$0.190&$+$0.190&$+$0.037 & K & 5827.95\\
\\
123258.99 &($^{3}$H)4f~~5[]$_{9/2}$  & 103600.43 & ($^{3}$H)4d $^{4}$G$_{11/2}$ &  5085.425 & $-$1.369 & $--$  & $--$& $-$1.127 & K &weak,noise  \\
          &                          & 103751.66 & ($^{3}$H)4d $^{4}$H$_{11/2}$ &  5124.850 & $+$0.043 & $+$0.043  & $+$0.043& $+$0.037 & K &5124.82  \\
          &                          & 103771.32 & ($^{3}$H)4d $^{4}$G$_{9/2}$  &  5130.020 & $+$0.287 & $+$0.287  & $+$0.287& $+$0.257 & K &5130.0  \\
          &                          & 103814.55 & ($^{3}$H)4d $^{4}$G$_{7/2}$  &  5141.426 & $--$ & $--$  & $--$& $-$3.748 & K &  \\
          &                          & 103874.26 & ($^{3}$H)4d $^{4}$H$_{9/2}$  &  5157.263 & $--$ & $--$  & $--$& $-$0.606 & K & blend \\
          &                          & 103973.78 & ($^{3}$H)4d $^{4}$K$_{11/2}$ &  5183.877 & $--$ & $--$  & $--$& $-$2.156 & K &  \\
          &                          & 103986.33 & ($^{3}$H)4d $^{4}$H$_{7/2}$  &  5187.253 & $--$ & $--$  & $--$& $-$4.544 & K &  \\
          &                          & 104174.27 & ($^{3}$H)4d $^{4}$I$_{11/2}$ &  5238.336 & $--$ & $--$  & $--$& $-$4.551 & K &  \\
          &                          & 104192.48 & ($^{3}$H)4d $^{4}$I$_{9/2}$  &  5243.339 & $--$ & $--$  & $--$& $-$3.145 & K &  \\
          &                          & 104481.59 & ($^{3}$H)4d $^{2}$F$_{7/2}$  &  5324.070 & $-$0.548 & $-$0.648& $-$0.598& $-$0.588 & K & weak,blend \\
          &                          & 104765.45 & ($^{3}$H)4d $^{2}$I$_{11/2}$ &  5405.791 & $--$ & $--$  & $--$& $-$2.078 & K &  \\
          &                          & 104807.21 & ($^{3}$H)4d $^{2}$G$_{9/2}$  &  5418.025 & $+$0.001 & $-$0.241  & $--$& $-$0.061 & K &blend  \\
          &                          & 104916.55 & ($^{3}$H)4d $^{4}$F$_{9/2}$  &  5450.323 & $-$0.471 & $-$0.521  & $-$0.496& $-$0.373 & K &  \\
          &                          & 105123.00 & ($^{3}$H)4d $^{2}$G$_{7/2}$  &  5512.367 & $-$0.917 & $<$$-$0.917  & $--$& $-$1.134 & K &  \\
          &                          & 105220.60 & ($^{3}$H)4d $^{4}$F$_{7/2}$  &  5542.193 & $--$ & $--$  & $--$& $-$1.725 & K &  \\
          &                          & 106045.69 & ($^{3}$H)4d $^{2}$H$_{11/2}$ &  5807.851 & $-$0.226 & $-$0.126  & $-$0.176& $-$0.455 & K &  \\
          &                          & 106097.52 & ($^{3}$H)4d $^{2}$H$_{9/2}$  &  5825.392 & $--$ & $--$  & $--$& $-$0.790 & K &  \\
\\
123270.34 & ($^{3}$H)4f~~5[]$_{11/2}$ & 103600.43 & ($^{3}$H)4d $^{4}$G$_{11/2}$ & 5082.492 &$-$1.101 &$-$0.801 &$--$& $-$0.654 & K& blend \\
          &                           & 103617.58 & ($^{3}$H)4d $^{4}$H$_{13/2}$ & 5086.927 &$--$ &$--$ &$--$& $-$2.184 & K \\
          &                           & 103751.66 & ($^{3}$H)4d $^{4}$H$_{11/2}$ & 5121.871 &$+$0.375 &$+$0.375 &$+$0.375& $+$0.348 & K&5121.89 \\
          &                           & 103771.32 & ($^{3}$H)4d $^{4}$G$_{9/2}$  & 5127.035 &$-$0.458 &$-$0.478 &$-$0.468& $-$0.587 & K \\
          &                           & 103832.05 & ($^{3}$H)4d $^{4}$K$_{11/2}$ & 5143.054 &$-$0.459 &$-$0.459 &$-$0.459& $-$0.450 & K \\
          &                           & 103874.26 & ($^{3}$H)4d $^{4}$H$_{9/2}$  & 5154.246 &$+$0.133 &$+$0.133 &$+$0.133& $+$0.130 & K& 5154.25 \\
          &                           & 103973.78 & ($^{3}$H)4d $^{4}$K$_{11/2}$ & 5180.829 &$-$0.217 &$-$0.217 &$-$0.217& $-$0.491 & K \\
          &                           & 104064.67 & ($^{3}$H)4d $^{4}$I$_{13/2}$ & 5205.347 &$-$1.188 &$-$0.888 &$--$& $-$0.854 & K \\
          &                           & 104174.27 & ($^{3}$H)4d $^{4}$I$_{11/2}$ & 5235.223 &$-$0.372 &$-$0.372 &$-$0.372& $-$0.537 & K&5235.225 \\
          &                           & 104192.48 & ($^{3}$H)4d $^{4}$I$_{9/2}$ & 5240.220 &$--$ &$--$ &$--$& $-$1.237 & K \\
          &                           & 104315.37 & ($^{3}$H)4d $^{2}$K$_{13/2}$ & 5274.195 &$--$ &$--$ &$--$& $-$1.382 & K \\
          &                           & 104622.30 & ($^{3}$H)4d $^{2}$I$_{13/2}$ & 5361.004 &$-$0.438 &$-$0.288 &$--$& $-$0.419 & K \\
          &                           & 104765.45 & ($^{3}$H)4d $^{2}$I$_{11/2}$ & 5402.476 &$--$ &$--$ &$--$& $-$1.619 & K \\
          &                           & 104807.21 & ($^{3}$H)4d $^{2}$G$_{9/2}$ & 5414.696 &$-$0.192 &$-$0.152 &$-$0.172& $-$0.156 & K & blend\\
          &                           & 104916.55 & ($^{3}$H)4d $^{4}$F$_{9/2}$ & 5446.953 &$-$0.709 &$-$0.909 &$--$& $-$0.733 & K \\
          &                           & 106045.69 & ($^{3}$H)4d $^{2}$H$_{11/2}$ & 5804.025 &$+$0.022 &$-$0.028 &$-$0.003& $-$0.041 & K&5804.02 \\
          &                           & 106097.52 & ($^{3}$H)4d $^{2}$H$_{9/2}$ & 5821.543 &$--$ &$--$ &$--$& $-$3.123 & K \\
\\
123355.49 & ($^{3}$H)4f~~4[]$_{13/2}$ & 103600.43 & ($^{3}$H)4d $^{4}$G$_{11/2}$ & 5060.583 &$--$ &$--$ &$--$& $-$1.129 & K&\\
          &                           & 103617.58 & ($^{3}$H)4d $^{4}$H$_{13/2}$ & 5064.980 &$--$ &$--$ &$--$& $-$3.032 & K&\\
          &                           & 103706.53 & ($^{3}$H)4d $^{4}$K$_{15/2}$ & 5087.910 &$--$ &$--$ &$--$& $-$1.855 & K&\\
          &                           & 103751.66 & ($^{3}$H)4d $^{4}$H$_{11/2}$ & 5099.623 &$-$0.281 &$-$0.481 &$-$0.381& $-$0.218 & K&5099.6\\
\hline
\noalign{\smallskip}
\end{tabular}
\end{flushleft}
\end{table*}

\setcounter{table}{4}

\begin{table*}[]
\begin{flushleft}
\caption{cont.} 
\begin{tabular}{lllllllllrl}
\noalign{\smallskip}\hline
\multicolumn{2}{c}{Upper level}&
\multicolumn{2}{c}{Lower level}&
\multicolumn{1}{c}{$\lambda$(calc)}&
\multicolumn{4}{c}{$log$\,gf}&
\multicolumn{1}{c}{}&
\multicolumn{1}{c}{$\lambda$(obs),Notes}
\\
\noalign{\smallskip}\hline
\multicolumn{1}{c}{cm$^{-1}$}&
\multicolumn{1}{c}{} &
\multicolumn{1}{c}{cm$^{-1}$}&
\multicolumn{1}{c}{} &
\multicolumn{1}{c}{\AA} &
\multicolumn{1}{c}{HR\,6000} &
\multicolumn{1}{c}{46\,Aql} &
\multicolumn{1}{c}{Aver} &
\multicolumn{1}{c}{calc.} &
\multicolumn{1}{c}{} &
\multicolumn{1}{c}{\AA} 
\\ 
\noalign{\smallskip}\hline
          &                           & 103832.05 & ($^{3}$H)4d $^{4}$K$_{13/2}$ & 5120.621 &$-$0.650 &$-$0.800 &$-$0.725& $-$1.186 & K& 5120.62\\
          &                           & 103878.37 & ($^{3}$H)4d $^{4}$I$_{15/2}$ & 5132.799 &$--$ &$--$ &$--$& $-$3.301 & K&\\
          &                           & 103973.78 & ($^{3}$H)4d $^{4}$K$_{11/2}$ & 5158.067 &$+$0.715 &$+$0.715 &$+$0.715& $+$0.781 & K&5158.05\\
          &                           & 104064.67 & ($^{3}$H)4d $^{4}$I$_{13/2}$ & 5182.370 &$-$0.081 &$-$0.051 &$-$0.066& $+$0.049 & K&5182.37\\
          &                           & 104119.71 & ($^{3}$H)4d $^{2}$K$_{15/2}$ & 5197.198 &$--$ &$--$ &$--$& $-$1.494 & K&\\
          &                           & 104174.27 & ($^{3}$H)4d $^{4}$I$_{11/2}$ & 5211.982 &$--$ &$--$ &$--$& $-$1.774 & K&\\
          &                           & 104315.37 & ($^{3}$H)4d $^{2}$K$_{13/2}$ & 5250.606 &$-$0.298 &$-$0.298 &$-$0.298& $-$0.774 & K& 5250.609\\
          &                           & 104622.30 & ($^{3}$H)4d $^{2}$I$_{13/2}$ & 5336.635 &$-$0.276 &$-$0.226 &$-$0.251& $-$0.241 & K&5336.62\\
          &                           & 104765.45 & ($^{3}$H)4d $^{2}$I$_{11/2}$ & 5377.729 &$+$0.176 &$+$0.126 &$+$0.151& $-$0.165 & K&5377.71\\
          &                           & 106045.69 & ($^{3}$H)4d $^{2}$H$_{11/2}$ & 5775.473 &$--$ &$--$ &$--$& $-$0.703 & K&no spect.\\
\\
123396.25 & ($^{3}$H)4f~~4[]$_{15/2}$ & 103617.58 & ($^{3}$H)4d $^{4}$H$_{13/2}$ & 5054.542 &$--$ &$--$ &$--$& $-$2.432 & K \\
          &                            & 103644.80 & ($^{3}$H)4d $^{4}$K$_{17/2}$ & 5061.508 &$--$ &$--$ &$--$& $-$5.014 & K \\
          &                            & 103706.53 & ($^{3}$H)4d $^{4}$K$_{15/2}$ & 5077.377 &$--$ &$--$ &$--$& $-$1.424 & K \\
          &                            & 103832.05 & ($^{3}$H)4d $^{4}$K$_{13/2}$ & 5109.953 &$-$0.197 &$-$0.197 &$-$0.197& $-$0.064 & K&5109.29 \\
          &                            & 103878.37 & ($^{3}$H)4d $^{4}$I$_{15/2}$ & 5122.080 &$--$ &$--$ &$--$& $-$1.735 & K \\
          &                            & 104064.67 & ($^{3}$H)4d $^{4}$I$_{13/2}$ & 5171.443 &$+$0.230 &$+$0.230 &$+$0.230& $+$0.289 & K & 5171.45\\
          &                            & 104119.71 & ($^{3}$H)4d $^{2}$K$_{15/2}$ & 5186.209 &$--$ &$--$ &$--$& $-$2.259 & K \\
          &                            & 104315.37 & ($^{3}$H)4d $^{2}$K$_{13/2}$ & 5239.390 &$+$0.800 &$+$0.850 &$+$0.825& $+$0.860 & K & 5239.394\\
          &                            & 104622.30 & ($^{3}$H)4d $^{2}$I$_{13/2}$ & 5325.048 &$+$0.322 &$+$0.302 &$+$0.312& $+$0.201 & K \\
\\
123441.10 & ($^{3}$H)4f~~4[]$_{11/2}$ & 103600.43 & ($^{3}$H)4d $^{4}$G$_{11/2}$ & 5038.747 &$--$ &$--$ &$--$& $-$3.261 & K \\
          &                           & 103617.58 & ($^{3}$H)4d $^{4}$H$_{13/2}$ & 5043.107 &$--$ &$--$ &$--$& $-$4.974 & K \\        
          &                           & 103751.66 & ($^{3}$H)4d $^{4}$H$_{11/2}$ & 5077.449 &$--$ &$--$ &$--$& $-$2.014 & K \\        
          &                           & 103771.32 & ($^{3}$H)4d $^{4}$G$_{9/2}$  & 5082.524 &$-$0.776 &$-$0.776 &$-$0.776& $-$0.363 & K&blend \\        
          &                           & 103832.05 & ($^{3}$H)4d $^{4}$K$_{13/2}$ & 5098.265 &$--$ &$--$ &$--$& $-$1.713 & K \\        
          &                           & 103874.26 & ($^{3}$H)4d $^{4}$H$_{9/2}$  & 5109.263 &$-$0.501 &$-$0.201 &$-$0.351& $+$0.054 & K \\        
          &                           & 103973.78 & ($^{3}$H)4d $^{4}$K$_{11/2}$ & 5135.383 &$-$0.428 &$-$0.463 &$-$0.445& $-$1.305 & K \\        
          &                           & 104064.67 & ($^{3}$H)4d $^{4}$I$_{13/2}$ & 5159.472 &$--$ &$--$ &$--$& $-$2.203 & K \\        
          &                           & 104174.27 & ($^{3}$H)4d $^{4}$I$_{11/2}$ & 5188.822 &$+$0.184 &$+$0.184 &$+$0.184& $+$0.228 & K & 5188.831\\        
          &                           & 104192.48 & ($^{3}$H)4d $^{4}$I$_{9/2}$ & 5193.731 &$+$0.675 &$+$0.725 &$+$0.700& $+$0.533 & K &5193.74\\        
          &                           & 104315.37 & ($^{3}$H)4d $^{2}$K$_{13/2}$ & 5227.103 &$--$ &$--$ &$--$& $-$1.337 & K \\        
          &                           & 104622.30 & ($^{3}$H)4d $^{2}$I$_{13/2}$ & 5312.357 &$--$ &$--$ &$--$& $-$1.799 & K \\        
          &                           & 104765.45 & ($^{3}$H)4d $^{2}$I$_{11/2}$ & 5353.077 &$--$ &$--$ &$--$& $-$0.252 & K& blend \\        
          &                           & 104807.21 & ($^{3}$H)4d $^{2}$G$_{9/2}$ & 5365.074 &$--$ &$--$ &$--$& $-$1.303 & K \\        
          &                           & 104916.55 & ($^{3}$H)4d $^{4}$F$_{9/2}$ & 5396.741 &$--$ &$--$ &$--$& $-$3.494 & K \\        
          &                           & 106045.69 & ($^{3}$H)4d $^{2}$H$_{11/2}$ & 5747.049 &$--$ &$--$ &$--$& $-$1.647 & K \\        
          &                           & 106097.52 & ($^{3}$H)4d $^{2}$H$_{9/2}$ & 5764.224 &$--$ &$--$ &$--$& $-$0.340 & K \\        
\hline
\noalign{\smallskip}
\end{tabular}
\end{flushleft}
\end{table*}

\begin{figure*}
\centering
\resizebox{5.00in}{!}{\rotatebox{90}{\includegraphics[30,100][600,700]
{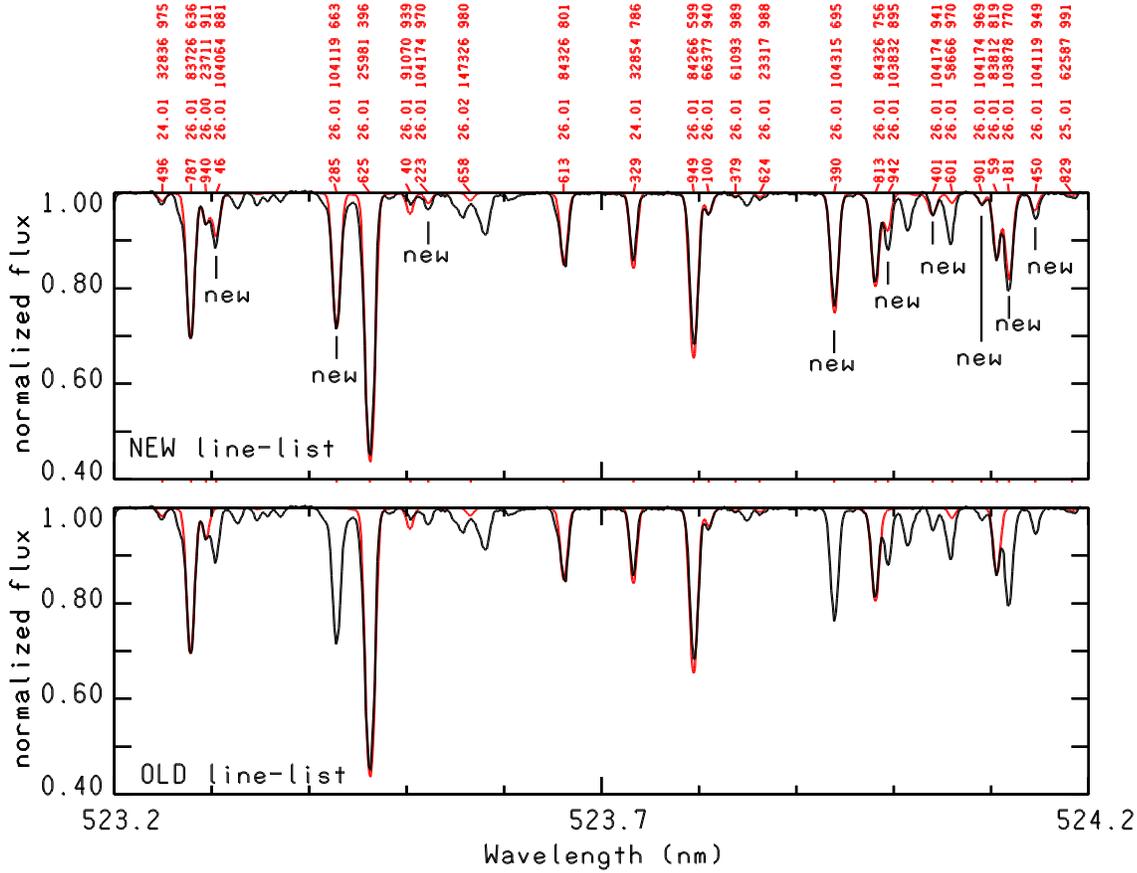}}}
\vskip 0.5cm
\caption{Comparison of the UVES spectrum of HR\,6000 with a synthetic
spectrum computed after this study (upper plot) and before this 
study (lower plot). The black line is the observed spectrum,
the red line is the synthetic spectrum. Nine \ion{Fe}{ii} lines corresponding
to ($^{3}$H)4d $-$ ($^{3}$H)4f new identified transitions are marked in the
upper plot. Their calculated $\log\,gf$'s were used for the synthetic spectrum. }
\end{figure*}

\begin{table*}[]
\begin{flushleft}
\caption{More new \ion{Fe}{ii} identified lines} 
\begin{tabular}{ll|c|lr|lr|lllrr}
\noalign{\smallskip}\hline
\multicolumn{1}{c}{$\lambda$(obs)}&
\multicolumn{1}{c}{$\lambda$(calc)}&
&
\multicolumn{2}{c}{Lower level}&
\multicolumn{2}{c}{Upper level}&
\multicolumn{2}{c}{$\log\,gf$}&
\\
\noalign{\smallskip}\hline
\multicolumn{1}{c}{\AA} & 
\multicolumn{1}{c}{\AA} &
&
\multicolumn{1}{c}{} &
\multicolumn{1}{c}{cm$^{-1}$}&
\multicolumn{1}{c}{} &
\multicolumn{1}{c}{cm$^{-1}$}&
\multicolumn{1}{c}{stellar} &
\multicolumn{1}{c}{calc.} 
\\ 
\noalign{\smallskip}\hline
4042.72 & 4042.741& K &($^{4}$P)4s4p($^{3}$P)\,$^{6}$P$_{5/2}$& 89444.458 & ($^{5}$D)6d\,$^{6}$D$_{7/2}$ &114173.163&$-$1.035 & $-$1.243 \\
4123.65 & 4123.667& K &($^{5}$D)5p\,$^{6}$D$_{9/2}$ & 88723.400 & ($^{5}$D)6d\,$^{6}$P$_{7/2}$  & 112966.820 & $-$1.453 & $-$1.996 \\
4145.98 & 4145.922& K &($^{5}$D)5p\,$^{6}$D$_{7/2}$ & 88853.533 & ($^{5}$D)6d\,$^{6}$P$_{7/2}$  & 112966.820 & $-$1.741 & $-$1.847\\
4187.72 & 4187.733& K &($^{5}$D)5p\,$^{6}$F$_{7/2}$& 90300.625 & ($^{5}$D)6d\,$^{6}$D$_{7/2}$ &114173.163&$-$1.177 & $-$1.505\\
$--$    & 4192.154& K &($^{5}$D)5p\,$^{6}$D$_{5/2}$ & 89119.457 & ($^{5}$D)6d\,$^{6}$P$_{7/2}$  & 112966.820 & $-$1.882 & $-$2.153 \\
5039.704& 5039.700& J  & b($^{3}$F)4p\,$^{4}$D$_{7/2}$ & 93129.900 &($^{5}$D)6d\,$^{6}$P$_{7/2}$  & 112966.820 & $-$0.559 &$-$5.578 \\
5129.929 & 5129.915 &J &($^{4}$D)4s4p\,$^{6}$P$_{5/2}$& 94685.090 & ($^{5}$D)6d\,$^{6}$D$_{7/2}$ &114173.163&$-$0.285 & $-$4.251\\
5131.450 & 5131.445 & J &b($^{3}$F)4p\,u$^{4}$F$_{9/2}$& 93484.580 &  ($^{5}$D)6d\,$^{6}$P$_{7/2}$& 112966.820 &$-$1.112 & $-$7.130  \\
5134.070 & 5134.014 & K &($^{3}$P)4p\,$^{2}$D$_{5/2}$ & 94700.660 & ($^{5}$D)6d $^{6}$D$_{7/2}$ & 114173.163 & $-$0.225 & $-$5.483\\
\hline
\noalign{\smallskip}
\end{tabular}
\end{flushleft}
\end{table*}

\subsection{More new \ion{Fe}{ii}  identified lines}

There are a few other lines not belonging to the
 ($^{3}$H)4d-($^{3}$H)4f transitions that were identified by 
S. Johansson in HR\,6000. They are indicated
with a ``J'' in Table\,6.  
The  lines  marked with a ``K'' were then obtained from
predicted lines \footnote{http://kurucz.harvard.edu/atmos/2601/gf2601.lines0500}$^{,}$
\footnote{http://kurucz.harvard.edu/atmos/2601/gf2601.lines0600}
thanks to the coincidence of  the term of the upper predicted level with 
the term of a ``J'' line.
Table\,6 shows that we fixed 2 new ($^{5}$D)6d levels.
Some  computed $\log\,gf$'s  are much weaker than the stellar ones.
Probably, some unknown transition is the main component of the
observed lines so that the stellar $\log\,gf$'s are not reliable values.

\section{Conclusions}

Figure\,8 compares synthetic spectra for HR\,6000 
computed before and after the study presented in this paper.
The interval plotted in the figure is a significant example of the
whole 5100-5400\,\AA\ region. It shows that as many as
nine new \ion{Fe}{ii} lines have been identified in 10\,\AA\ range. 
Nevertheless, there are still several unidentified absorptions
that are probably due to \ion{Fe}{ii} whose levels 
have still to be fixed. A very similar plot was obtained for 46\,Aql.
We can  infer that a 
large part of the unidentified lines observed in the spectra of
B-type stars are due to unknown high-excitation \ion{Fe}{ii} transitions. 

The new lines identified in this paper correspond to high excitation 
transitions of \ion{Fe}{ii} with upper level just below the ionization limit. 
We have fixed 21 new levels of \ion{Fe}{ii} with energies between
 122910.9\,cm$^{-1}$ and 123441.1\,cm$^{-1}$ and we have added 1700 lines
to the \ion{Fe}{ii} linelist in the range 810 $-$15011\,\AA.
Furthermore, Johansson (2009) identified in the spectrum of HR\,6000 other 
\ion{Fe}{ii} lines with lower level near the
ionization limit and upper level above it, the  
multiplet 4s4d$^{8}$D $-$ 4s4f$^{8}$F at 4410\,\AA. 

Among the new identified  high-excitation \ion{Fe}{ii} lines, several
have residual flux in HR\,6000 and 46\,Aql as deep as 0.7 
and numerous other can be observed as weak absorptions or part of blends.
The two stars are iron overabundant stars, but
these lines are present with lower intensity also in the UVES spectrum of   
HD\,175640, a B-type peculiar star with an  iron underabundance
of $-$0.25\,dex in respect to the sun (Castelli \& Hubrig 2004).
 This fact implies that \ion{Fe}{ii} lines from the new high excitation
levels contribute to the spectrum of all Population\,I late B-type stars
even when their abundance is less than solar. 
The lines are clearly observable in high resolution, high signal-to-noise
spectra of slow rotating stars, while they contribute to the broad observed
features in B-type stars with high rotational velocities. In general,
they would appear in any object with strong
\ion{Fe}{ii} lines.

We conclude that we have clarified the nature of several unidentified lines
observed in the optical spectra of B-type stars  and mostly concentrated
in the 5000-5400\,\AA\ region (see also Wahlgren et al.\, 2000), but that
a large amount of work has still to be done to well 
reproduce stellar observations. More than 1000 energy levels of 
\ion{Fe}{ii} are known,  
but we have seen that they are not enough.
Ignorance  of them and of the involved transitions 
is still a  surviving  shortcoming affecting the  
model atmosphere and synthetic spectra computations.


\begin{appendix}

\section{The Balmer lines of HR\,6000 observed on the UVES spectra}

Figure\,A.1 shows the UVES spectra of HR\,6000 reduced 
by the UVES pipeline\footnote{http://www.eso.org/ \\
observing/dfo/quality/UVES/pipeline/pipe$_{-}$reduc.html}
 Data  Reduction Software (version 2.5; 
Ballester et al. 2000) that were used by Castelli \& Hubrig (2007) and
also in this paper. All spectra are FLUXCAL$_{-}$SCIENCE products. Those at 
$\lambda\lambda$ 3290-4520\,\AA\ and 4780-5650\,\AA\ are flux-calibrated 
spectra in 10$^{-16}$\,erg\,s$^{-1}$\,cm$^{-2}$\,A$^{-1}$ corrected for  terrestrial 
extinction.  The red spectrum at 5730-7560\,\AA\ is in non-physical units
'quasi-ADU' in that the flux calibration
procedure is not implemented in the reduction software for the REDL and REDU
data taken with the  red  mosaic CCD's. 
The rather impressive distorsions of the UVES spectra make 
evident the difficulty in drawing a true
continuum over H$_{\gamma}$ and H$_{\delta}$. Also the use of 
H$_{\beta}$ 
produces troubles due to the position of this line at the left end
of the spectrum order. Only H$_{\alpha}$ does not show manifest problems.

\begin{figure*}
\centering
\resizebox{3.25in}{!}{\rotatebox{90}{\includegraphics[30,100][600,700]
{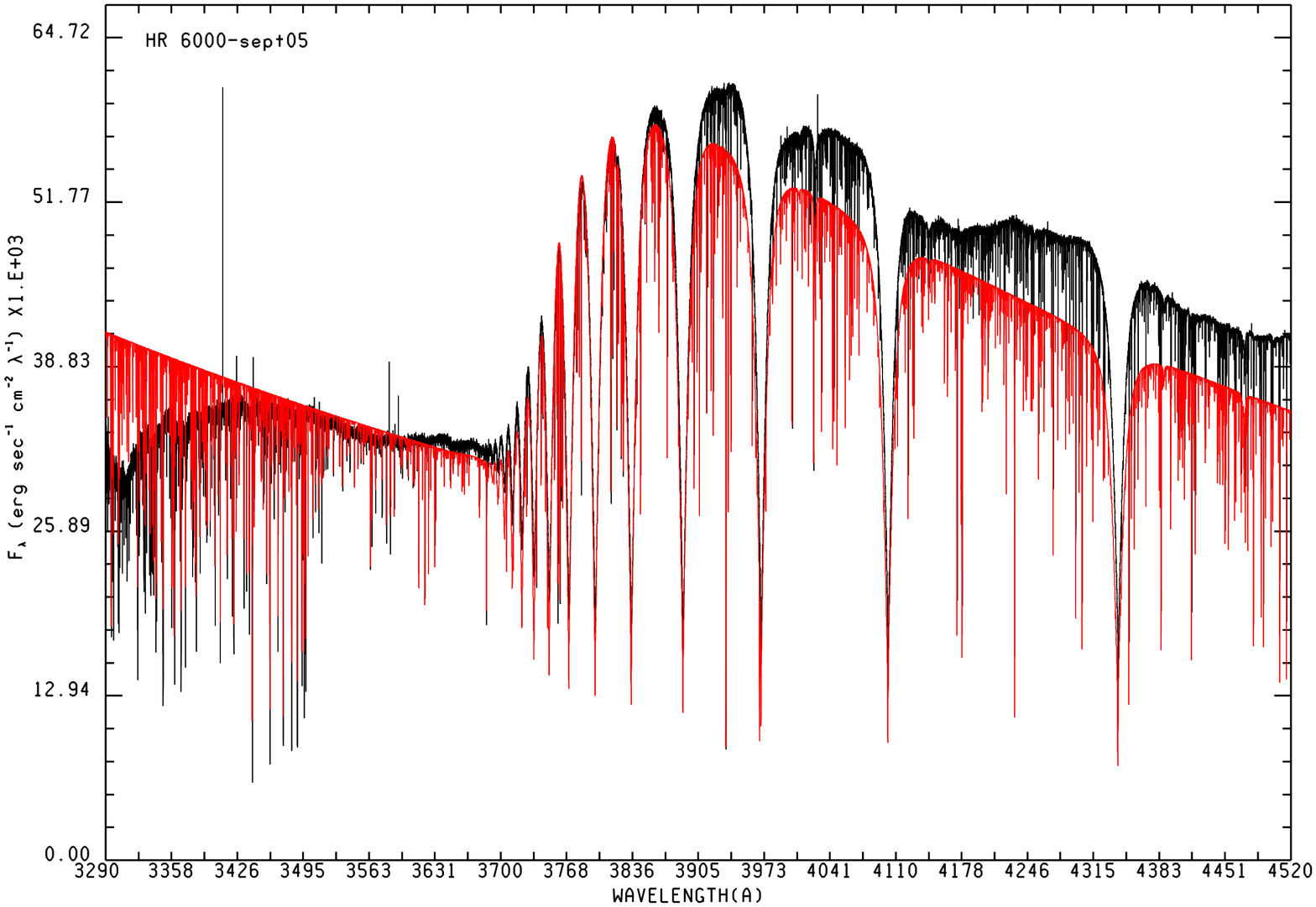}}}
\vskip 0.3cm
\resizebox{3.25in}{!}{\rotatebox{90}{\includegraphics[30,100][600,700]
{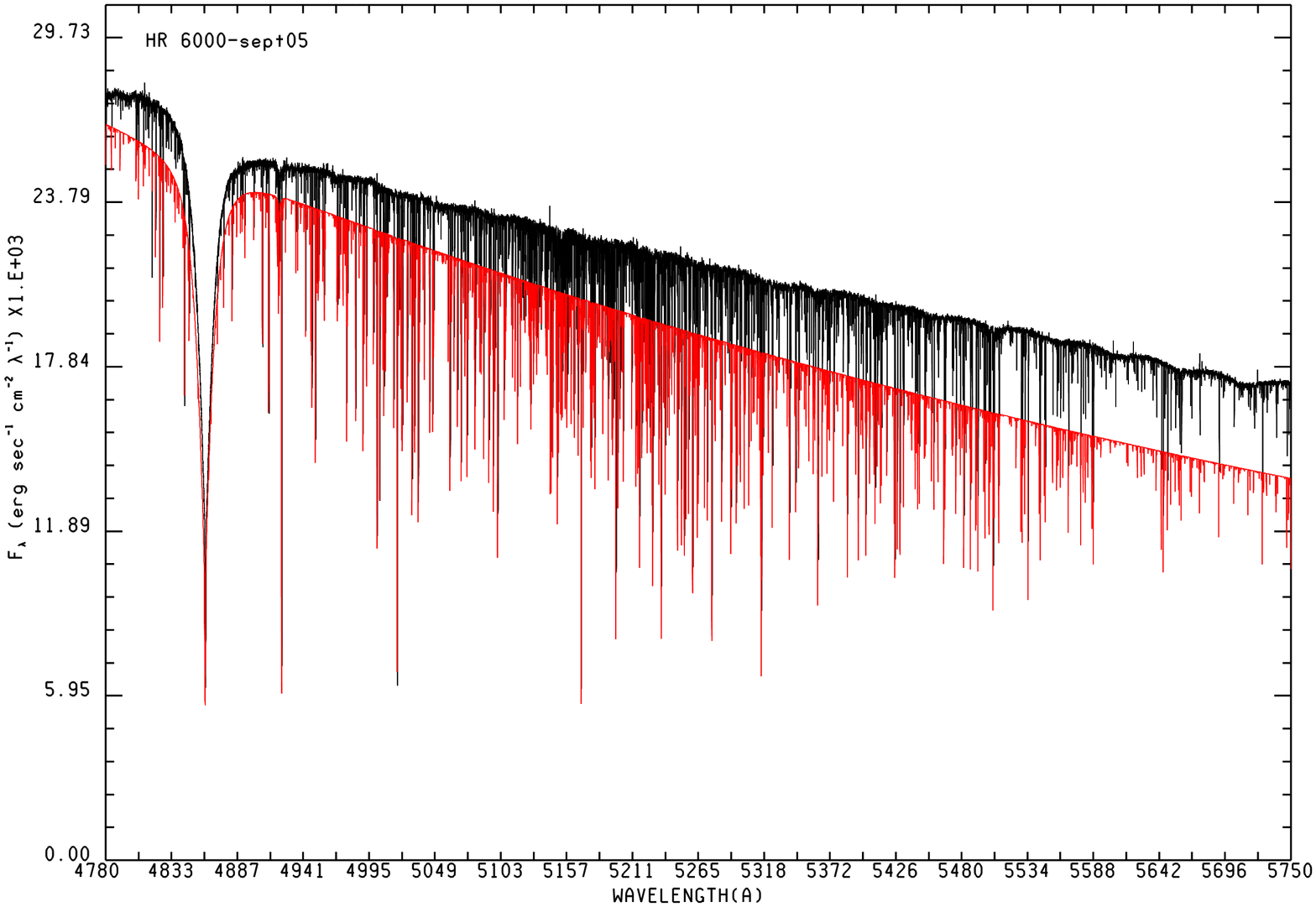}}}
\vskip 0.3cm
\resizebox{3.25in}{!}{\rotatebox{90}{\includegraphics[30,100][600,700]
{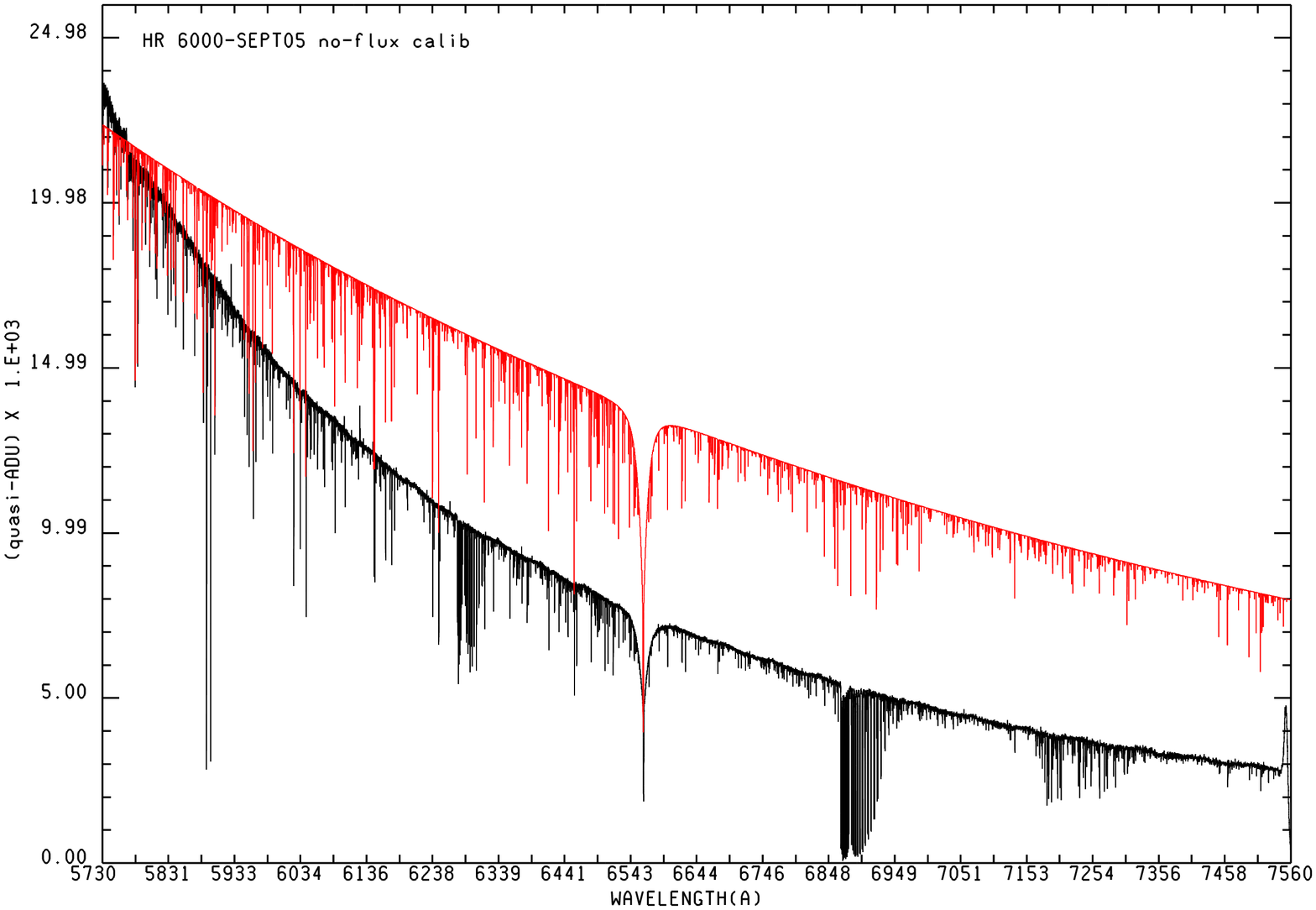}}}
\caption{The observed UVES spectra of HR\,6000 (black line)
are plotted together the computed spectra (red line) in order to show
the different slopes of the observed and computed continua. The computed
fluxes are scaled by a given arbitrary quantity to be roughly overimposed
on the UVES spectra. The ATLAS12
final model with parameters \teff=13450\,K, \logg=4.3 was used for the
computations.} 
\label{}
\end{figure*}

Computed spectra from the final ATLAS12 model ([13450,4.3],Sect.\,2.2)
are also plotted in Fig.\,A.1 in order to show the different
slopes of the observed and computed continua. The computed fluxes
were scaled  by a given arbitrary quantity to be roughly overimposed
on the UVES spectra.

\section{Lines used for the abundance analysis}

Table\,B.1 lists the lines that were examined in the spectra
of HR\,6000 and 46\,Aql in order to derive the stellar abundances.
The wording ``not obs'' is given for
lines not present in the spectra, while the wordings ``profile''
and ``blend'' are given for lines well observed in the spectra,
but which do not have measurable equivalent widths either because they
are too weak to be measurable or because other components affects
the line. These wording also indicate lines for which 
adequate equivalent widths can not be computed as in the cases of
\ion{Mg}{ii} at 4481\,\AA\ and of most \ion{O}{i} lines which are
blends of transitions belonging to the same multiplet. 
The abundances from the final ATLAS12 models derived from the equivalent
widths or from the profiles are given in the Table, as well as
upper abundance limits from lines not observed, 
but predicted  for solar abundance by the synthetic spectrum.
For \ion{Fe}{i} and \ion{Fe}{ii}, $\log\,gf$'s were taken 
from Fuhr \& Wiese (2006) (FW06) when available. Otherwise Kurucz's 
last detrmination was adopted (Kurucz, 2009), except for \ion{Fe}{ii}
at 5257.119\,\AA. In this case the previous values (Kurucz, 2007)
produces synthetic profiles in better agreement with  
the observations.

\begin{table*}[!hbp]
\caption[ ]{Analyzed lines in the stellar spectra, measured equivalent widths
in m\AA, and relative abundances.\ion{He}{i} is not included.} 
\font\grande=cmr7
\grande
\begin{flushleft}
\begin{tabular}{llrrrccccrr}
\hline\noalign{\smallskip}
 & & & & &
\multicolumn{2}{c}{HR\,6000[13450,4.3,AT12]}&
\multicolumn{2}{c}{46\,Aql[12560,3.8,AT12]}
\\
\hline\noalign{\smallskip}
\multicolumn{1}{c}{Species}&
\multicolumn{1}{c}{$\lambda$($\AA$)}&
\multicolumn{1}{c}{$\log\,gf$}&
\multicolumn{1}{c}{Ref.$^{a}$}&
\multicolumn{1}{c}{$\chi_{low}$}&
\multicolumn{1}{c}{W(m$\AA$)}&
\multicolumn{1}{c}{$log(N_{elem}/N_{\rm tot}$)}&
\multicolumn{1}{c}{W(m$\AA$)}&
\multicolumn{1}{c}{$log(N_{elem}/N_{\rm tot}$)}
\\
\hline\noalign{\smallskip}
\ion{Be}{ii} & 3130.420 & $-$0.170 & NIST3 &    0.00  & 26.7   & $-$9.78 & 31.0  & $-$9.91\\  
\ion{C}{ii}  & 4267.001 & $+$0.562 &NIST3&145549.270  &profile & $-$5.50  &profile& $-$4.75 \\
\ion{C}{ii}  & 4267.261 & $+$0.716 &NIST3&145550.700  &profile & $-$5.50  &profile& $-$4.75 \\
\ion{N}{i}   &8680.282  & $+$0.347 &NIST3& 83364.620  & not obs& $\le$$-$5.80 &not obs& $\le$$-$5.50\\
\ion{N}{i}   &8683.403  & $+$0.087 &NIST3& 88317.830  & not obs& $\le$$-$5.80 &not obs& $\le$$-$5.50\\
\ion{O}{i}   &3947.295  & $-$2.095 &NIST3&73768.200   &profile & $-$3.71 & profile & $-$3.71\\
\ion{O}{i}   &3947.481  & $-$2.244 &NIST3&73768.200   &profile & $-$3.71 & profile & $-$3.71\\
\ion{O}{i}   &3947.586  & $-$2.467 &NIST3&73768.200   &profile & $-$3.71 & profile & $-$3.71\\
\ion{O}{i}   &4368.193  & $-$2.665 &NIST3&76794.978   &profile & $-$3.71 & profile & $-$3.51\\
\ion{O}{i}   &4368.242  & $-$1.964 &NIST3&76794.978   &profile & $-$3.71 & profile & $-$3.51\\
\ion{O}{i}   &4368.258  & $-$2.186 &NIST3&76794.978   &profile & $-$3.71 & profile & $-$3.51\\
\ion{O}{i}   &5329.096  & $-$1.938 &NIST3&86625.757   &profile & $-$3.71 & profile & $-$3.46\\
\ion{O}{i}   &5329.099  & $-$1.586 &NIST3&86625.757   &profile & $-$3.71 & profile & $-$3.46\\
\ion{O}{i}   &5329.107  & $-$1.695 &NIST3&86625.757   &profile & $-$3.71 & profile & $-$3.46\\
\ion{O}{i}   &6155.961  & $-$1.363 &NIST3&86625.757   &profile & $-$3.64 & profile & $-$3.46\\
\ion{O}{i}   &6155.971  & $-$1.011 &NIST3&86625.757   &profile & $-$3.64 & profile & $-$3.46\\
\ion{O}{i}   &6155.989  & $-$1.120 &NIST3&86625.757   &profile & $-$3.64 & profile & $-$3.46\\
\ion{O}{i}   &6156.737  & $-$1.487 &NIST3&86627.778   &profile & $-$3.64 & profile & $-$3.43\\
\ion{O}{i}   &6156.755  & $-$0.898 &NIST3&86627.778   &profile & $-$3.64 & profile & $-$3.43\\
\ion{O}{i}   &6156.778  & $-$0.694 &NIST3&86627.778   &profile & $-$3.64 & profile & $-$3.43\\
\ion{O}{i}   &6455.977  & $-$0.920 &NIST3&86631.454  &profile  & $-$3.59 & profile & $-$3.43\\
\ion{O}{i}   &7002.196  & $-$1.489 &NIST3&88631.146 &profile & $-$3.71 &  H$_{2}$O &$--$\\
\ion{O}{i}   &7002.230  & $-$0.741 &NIST3&88631.146 &profile & $-$3.71 & H$_{2}$O  & $--$\\
\ion{O}{i}   &7002.250  & $-$1.364 &NIST3&88631.303 &profile & $-$3.71 & H$_{2}$O  & $--$\\
\ion{Ne}{i}  &7032.413  & $-$0.249 &NIST3&134041.840& noise? & $-$4.86 & noise ? &$-$4.51 \\
\ion{Na}{i}  &5688.205  & $-$0.452 &NIST3&16973.368 &not obs & $\le$$-$5.71& profile & $-$5.65\\
\ion{Na}{i}  &5889.950  & $+$0.108 &NIST3& 0.00     &inters. & $--$      & profile & $-$5.67\\
\ion{Na}{i}  &5895.924  & $-$0.194 &NIST3& 0.00     &inters. & $--$      & profile & $-$5.72\\
\ion{Mg}{ii} &4481.126  & $+$0.749 &NIST3&71490.190 &profile &$-$5.66    & profile & $-$5.45\\
\ion{Mg}{ii} &4481.150  & $-$0.553 &NIST3&71490.190 &profile &$-$5.66    & profile & $-$5.45\\
\ion{Mg}{ii} &4481.325  & $+$0.594 &NIST3&71491.063 &profile &$-$5.66    & profile & $-$5.45\\
\ion{Al}{i}  &3944.006  & $-$0.638 &NIST3&    0.000 &not obs &$\le$$-$7.30&not obs & $\le$$-$6.65\\
\ion{Al}{i}  &3961.520  & $-$0.336 &NIST3&  112.061 &not obs &$\le$$-$7.30&profile &   $-$6.65 \\
\ion{Al}{ii} &7042.083  & $+$0.340 &NIST3  &91274.500 &noise ? &$\le$$-$7.30&noise ? &$\le$$-$7.40\\
\ion{Al}{ii} &7056.712  & $+$0.110 &NIST3  &91274.500 &noise ? &$\le$$-$7.30&noise ? &$\le$$-$7.00\\
\ion{Si}{ii} &3853.665  & $-$1.341 &NIST3  &55309.350&not obs &$--$ & 31.8& $-$5.62\\
\ion{Si}{ii} &3862.595  & $-$0.406 &NIST3  &55325.180&not obs &$--$ & 71.8 & $-$5.68\\
\ion{Si}{ii} &3862.595  & $-$0.757 &NIST3  &55309.350&profile &$-$7.35 & 62.6 & $-$5.53\\
\ion{P}{ii}& 4044.576 & $+$0.481 &K,MRB & 107360.250 & 45.3 & $-$4.42 & 19.9 & $-$5.16\\
\ion{P}{ii} & 4062.149 &   $-$0.585 &K,MRB & 103339.140 & 8.6 & $-$4.79 & 2.4 & $-$5.45\\
\ion{P}{ii}  & 4127.559 &   $-$0.110 &K,KP & 103667.860 & 25.2 & $-$4.54 &9.3 & $-$5.21\\
\ion{P}{ii}  & 4160.623 &   $-$0.410 &K,KP & 103339.140 & 9.4 & $-$4.90 & 7.6 & $-$5.03\\
\ion{P}{ii}  & 4244.622 &   $-$0.308 &K,MRB & 107360.250 &9.6 & $-$4.78 &2.6 & $-$5.46\\
\ion{P}{ii}  & 4288.606 &   $-$0.630 &K,MRB & 101635.690 & 7.0 & $-$4.88&blend& $--$\\
\ion{P}{ii}  & 4420.712 &   $-$0.329 &NIST3 & 88893.220  & 40.1 & $-$4.51&19.4& $-$5.18\\
\ion{P}{ii}  & 4452.472 &   $-$0.194 &K,MRB & 105302.170  & 19.9 & $-$4.48 &6.9 & $-$5.13\\
\ion{P}{ii}  & 4463.027 &   $+$0.026 &K,MRB & 105549.670  & 25.6 & $-$4.52 &9.1 & $-$5.19\\
\ion{P}{ii}  & 4466.140 &   $-$0.560 &NIST3 & 105549.670  & 9.8 & $-$4.57 & 2.2 & $-$5.32\\
\ion{P}{ii}  & 4468.000 &   $-$0.208 &K,MRB & 105244.060  & 24.0 & $-$4.35 &blend&  $--$\\
\ion{P}{ii}  & 4475.270 &   $+$0.440 &NIST3 & 105549.670   & 36.8 & $-$4.63 & 16.2 & $-$5.25 \\
\ion{P}{ii}  & 4483.693 &   $-$0.431 &NIST3 & 105302.370   & 10.0 & $-$4.69 & 2.3 & $-$5.43\\
\ion{P}{ii}  & 4499.230 &   $+$0.470 &NIST3 & 107922.930   &38.0 & $-$4.44 & blend &$--$\\
\ion{P}{ii}  & 5296.077 &   $-$0.160 &NIST3 & 87124.600    &56.3 & $-$4.15 & 33.9 & $-$4.74\\
\ion{P}{ii}  & 5316.055 &   $-$0.294 &NIST3 & 86743.960    & 47.9 & $-$4.31 & 28.3 & $-$4.82\\
\ion{P}{ii}  & 5344.729 &   $-$0.390 &NIST3 & 86597.550    & 45.6 & $-$4.28 &25.7 & $-$4.81\\
\ion{P}{ii}  & 5386.895 &   $-$0.520 &NIST3 & 86743.960    & 51.8 & $-$3.98 &25.6 & $-$4.67\\
\ion{P}{ii}  & 5425.880 &   $+$0.180 &NIST3 & 87124.600    & 71.8 & $-$4.22 & 47.4 & $-$4.71\\
\hline
\noalign{\smallskip}
\end{tabular}
\end{flushleft}
\end{table*}

\setcounter{table}{0}

\begin{table*}[!hbp]
\caption[ ]{cont.}
\font\grande=cmr7
\grande
\begin{flushleft}
\begin{tabular}{llrrrccccrr}
\hline\noalign{\smallskip}
\multicolumn{5}{c}{}&
\multicolumn{2}{c}{HR\,6000[13450,4.3,AT12]}&
\multicolumn{2}{c}{46\,Aql[12560,3.8,AT12]}
\\
\hline\noalign{\smallskip}
\multicolumn{1}{c}{Species}&
\multicolumn{1}{c}{$\lambda$($\AA$)}&
\multicolumn{1}{c}{$\log\,gf$}&
\multicolumn{1}{c}{Ref.$^{a}$}&
\multicolumn{1}{c}{$\chi_{low}$}&
\multicolumn{1}{c}{W(m$\AA$)}&
\multicolumn{1}{c}{$log(N_{elem}/N_{\rm tot}$)}&
\multicolumn{1}{c}{W(m\AA)}&
\multicolumn{1}{c}{$log(N_{elem}/N_{\rm tot}$)}
\\
\hline\noalign{\smallskip}

\ion{P}{ii}  & 6024.178 &   $+$0.137 &NIST3 & 86743.960    & 69.8 & $-$4.04 & 45.7 & $-$4.57\\
\ion{P}{ii}  & 6034.039 &   $-$0.220 &NIST3 & 86597.550    & 52.6 & $-$4.10 & 29.4 & $-$4.70\\
\ion{P}{ii}  & 6043.084 &   $+$0.416 &NIST3 & 87124.600    & 81.1 & $-$4.05 & 55.5 & $-$4.57\\
\ion{P}{iii}& 4059.312 &   $-$0.051 &NIST3 &116885.870     &11.1 & $-$4.97  & 2.5 & $-$5.87\\
\ion{P}{iii} & 4080.089 &   $-$0.306 &NIST3 &116874.560     & 5.5 & $-$5.21 & not ob & $--$\\
\ion{P}{iii} & 4222.198 &   $+$0.210 &NIST3 &117835.950     &19.5 & $-$5.45 & not obs &$--$\\
\ion{P}{iii} & 4246.720 &   $-$0.120 &NIST3 &117835.950     & 10.6 & $-$4.78 & not obs & $--$\\
\ion{S}{ii}  & 4162.665 &   $+$0.777 &NIST3 &128599.160      & 2.0 & $-$6.26 & 5.3 & $-$5.74\\
\ion{Cl}{i}  & 4794.556 &   $+$0.400 &NIST3 &107879.660   & not obs &$\le$$-$7.24& not obs& $\le$$-$7.04\\
\ion{Ca}{i}  & 4226.728 &   $+$0.244 &NIST3 &     0.000    &profile &$>$$-$5.68 & profile & $>$$-$5.98 \\       
\ion{Ca}{ii} & 3933.663 &   $+$0.135 &NIST3 &     0.000    & profile &$-$5.68 & profile &$-$5.98\\       
\ion{Ca}{ii} & 3968.469 &   $-$0.180 &NIST3 &     0.000    & profile &$-$5.68 & profile & $-$5.98\\       
\ion{Sc}{ii} & 4246.822 &   $+$0.242 &NIST3 &   2540.950    & blend & $--$ & blend & $--$  \\       
\ion{Sc}{ii} & 4314.083 &   $-$0.100 &NIST3 &   4987.790    & not obs &$\le$ $-$9.50 & not obs &$\le$$-$9.50\\       
\ion{Ti}{ii} & 4053.821 &   $-$1.130 &PTP   &  15265.620 & 6.4 & $-$6.40  & 11.1 & $-$6.46\\       
\ion{Ti}{ii} & 4163.644 &   $-$0.130 &PTP   &  20891.660 & 19.3 & $-$6.51 & 30.4 & $-$6.50 \\       
\ion{Ti}{ii} & 4287.873 &   $-$1.790 &PTP   &  8710.440  & 3.5 & $-$6.38  &  6.1 & $-$6.49\\       
\ion{Ti}{ii} & 4290.215 &   $-$0.850 &PTP   &  9395.710  & 15.6 &$-$6.55  & 28.3 & $-$6.52\\       
\ion{Ti}{ii} & 4294.094 &   $-$0.930 &PTP   &  9744.250  & 16.5 & $-$6.47 & 28.3 & $-$6.48\\       
\ion{Ti}{ii} & 4300.042 &   $-$0.440 &PTP   &  9518.060  & 30.2 & $-$6.52 & 45.2 & $-$6.49\\       
\ion{Ti}{ii} & 4301.922 &   $-$1.150 &PTP   &  9363.620  & 8.6 &  $-$6.56 & 17.1 & $-$6.55\\       
\ion{Ti}{ii} & 4367.652 &   $-$0.860 &PTP   & 20891.660  & 5.7 &  $-$6.42 &  9.9 & $-$6.46\\       
\ion{Ti}{ii} & 4395.031 &   $-$0.540 &PTP   &  8744.250  &29.2 &  $-$6.49 & 45.2 & $-$6.43\\       
\ion{Ti}{ii} & 4399.765 &   $-$1.190 &PTP   &  9975.920  &3.3 &   $-$6.94 & 17.2 & $-$6.48\\       
\ion{Ti}{ii} & 4411.072 &   $-$0.670 &PTP   & 24961.030  &5.0 &   $-$6.45 & 10.1 & $-$6.40\\       
\ion{Ti}{ii} & 4417.714 &   $-$1.190 &PTP   &  9395.710  &8.8 &   $-$6.51 & 16.4 & $-$6.54\\       
\ion{Ti}{ii} & 4418.331 &   $-$1.970 &PTP   &  9975.920  &2.4 &   $-$6.31 &  3.5 & $-$6.50\\       
\ion{Ti}{ii} & 4443.810 &   $-$0.720 &PTP   &  8710.440  &22.3 &  $-$6.50 & 37.1 & $-$6.47\\       
\ion{Ti}{ii} & 4450.482 &   $-$1.520 &PTP   &  8744.250  &5.5 &   $-$6.45 & 11.5 & $-$6.44\\       
\ion{Ti}{ii} & 4464.448 &   $-$1.810 &PTP   &  9363.620  &3.1 &   $-$6.40 &  6.8 & $-$6.38\\       
\ion{Ti}{ii} & 4468.492 &   $-$0.620 &NIST3 &  9118.260  &25.6 &  $-$6.50 & 39.6 & $-$6.50\\       
\ion{Ti}{ii} & 4488.325 &   $-$0.510 &PTP   & 25192.710  &7.0 &   $-$6.44 & 12.4 & $-$6.44\\       
\ion{Ti}{ii} & 4501.270 &   $-$0.770 &PTP   &  8997.710  &23.0 &  $-$6.41 & 37.6 & $-$6.39\\       
\ion{Ti}{ii} & 4805.085 &   $-$1.120 &NIST3 & 16625.110  &7.2 &   $-$6.28 & 14.5 & $-$6.27\\       
\ion{Ti}{ii} & 4911.195 &   $-$0.610 &PTP   & 25192.790  &5.4 &   $-$6.45 & 10.7 & $-$6.42\\       
\ion{V}{ii}  & 3093.105 &   $+$0.530 &K,NBS   &  3162.800  &blend &$\le$$-$9.14 & noise ?& $\le$$-$8.94   \\       
\ion{V}{ii}  & 3102.289 &   $+$0.410 &K,NBS   &  2968.220  &not obs & $\le$$-$9.14  &noise &$\le$$-$8.94 \\       
\ion{V}{ii}  & 3110.695 &   $+$0.300 &K,NBS   &  2808.720  &blend &$\le$$-$9.14  &blend  & $\le$$-$8.94\\       
\ion{V}{ii}  & 3113.376 &   $+$0.180 &K,NBS   &  2687.010  &not obs & $\le$$-$9.14 &not obs &$\le$$-$8.94 \\       
\ion{V}{ii}  & 3125.276 &   $+$0.070 &K,NBS   &  2604.820  &blend &   $\le$$-$9.14 &blend & $-$8.94 \\       
\ion{Cr}{ii} & 4812.337 &   $-$1.800 &NIST3   &  31168.580  &3.7 &$-$6.21& not obs & $--$ \\       
\ion{Cr}{ii} & 4824.127 &   $-$0.970 &SL      &  31219.350  &26.4 &$-$5.98 & blend & $--$\\       
\ion{Cr}{ii} & 4836.229 &   $-$2.000 &SL      &  31117.390  &3.9 &$-$5.99  &not obs  & $--$\\       
\ion{Cr}{ii} & 5237.329 &   $-$1.160 &NIST3   &  32854.310  &12.5 &$-$6.16  & 1.5 & $-$7.48 \\       
\ion{Cr}{ii} & 5246.768 &   $-$2.460 &NIST3   &  29951.880  &1.2 &$-$6.14  &not obs  & $--$\\       
\ion{Mn}{ii} & 3441.988 &   $-$0.270 &NIST3   &  14325.860  &85.3 &$-$4.92 ($-$5.00)$^{b}$  &74.5  & $-$5.48 ($-$5.35)$^{b}$\\       
\ion{Mn}{ii} & 3460.316 &   $-$0.540 &NIST3   &  14593.820  &74.5 &$-$4.91 ($-$5.00)  &60.2  & $-$5.65 ($-$5.65)\\       
\ion{Mn}{ii} & 3482.905 &   $-$0.740 &NIST3   &  14781.190  &65.3 &$-$4.98 ($-$5.00)  &53.0  & $-$5.70 ($-$5.65)\\       
\ion{Mn}{ii} & 3488.677 &   $-$0.860 &NIST3   &  14901.180  &65.7 &$-$4.85 ($-$4.85)  &52.5  & $-$5.60 ($-$5.60)\\       
\ion{Mn}{ii} & 3495.833 &   $-$1.220 &NIST3   &  14959.840  &55.3 &$-$4.85 ($-$4.85)  &42.2  & $-$5.65 ($-$5.65) \\       
\ion{Mn}{ii} & 3496.809 &   $-$1.690 &NIST3   &  14781.190  &38.1 &$-$5.09 ($-$5.00)  &27.0  & $-$5.87 ($-$5.90)\\       
\ion{Mn}{ii} & 3497.526 &   $-$1.330 &NIST3   &  14901.180  &47.7 &$-$5.05 ($-$4.90)  &35.5  & $-$5.84 ($-$5.85)\\       
\ion{Mn}{ii} & 3917.318 &   $-$1.169 &K03Mn   &  55759.270  &4.9 &$-$5.61 ($-$5.65)   &2.8  & $-$6.09  ($-$6.17)\\       
\ion{Mn}{ii} & 4363.258 &   $-$1.929 &K03Mn   &  44899.820  &4.2 &$-$5.49 ($-$5.58)  & 2.3  & $-$6.03 ($-$6.15)\\       
\ion{Mn}{ii} & 4365.219 &   $-$1.339 &K03Mn   &  44899.820  &4.7 &$-$5.59 ($-$5.63)  & 3.1  & $-$6.02 ($-$6.07)\\       
\ion{Mn}{ii} & 4478.635 &   $-$0.942 &K03Mn   &  53597.130  &8.2 &$-$5.68 ($-$5.63)  & 5.0  & $-$6.16 ($-$6.15)\\       
\ion{Fe}{i}  & 3581.193 &   $+$0.406 &FW06 & 6928.27  & 17.3 & $-$3.67 & 19.2& $-$3.86\\
\ion{Fe}{i}   & 3618.768 &   $-$0.003 &FW06 & 7985.78  & 8.0 & $-$3.73  & 10.3& $-$3.88\\
\ion{Fe}{i}   & 4005.242 &   $-$0.610 &FW06 &12560.93  & 6.6 & $-$3.59  &  6.4& $-$3.94\\
\hline
\noalign{\smallskip}
\end{tabular}
\end{flushleft}
\end{table*}

\setcounter{table}{0}

\begin{table*}[!hbp]
\caption[ ]{cont.}
\font\grande=cmr7
\grande
\begin{flushleft}
\begin{tabular}{rrrrrrcrc}
\hline\noalign{\smallskip}
\multicolumn{5}{c}{}&
\multicolumn{2}{c}{HR\,6000[13450,4.3,AT12]}&
\multicolumn{2}{c}{46\,Aql[12560,3.8,AT12]}
\\
\hline\noalign{\smallskip}
\multicolumn{1}{c}{Species}&
\multicolumn{1}{c}{$\lambda$($\AA$)}&
\multicolumn{1}{c}{$\log\,gf$}&
\multicolumn{1}{c}{Ref.$^{a}$}&
\multicolumn{1}{c}{$\chi_{low}$}&
\multicolumn{1}{c}{W(m$\AA$)}&
\multicolumn{1}{c}{$log(N_{elem}/N_{\rm tot}$)}&
\multicolumn{1}{c}{W(m$\AA$)}&
\multicolumn{1}{c}{$log(N_{elem}/N_{\rm tot}$)}
\\
\hline\noalign{\smallskip}
\ion{Fe}{i}   & 4071.738 &   $-$0.022 &FW06 &12698.55  &15.4 &  $-$3.71 & 17.0& $-$3.99\\
\ion{Fe}{i}   & 4202.029 &   $-$0.708 &FW06 &11976.24  & 4.5 &  $-$3.71 &  5.7& $-$3.93\\
\ion{Fe}{i}   & 4219.360 &   $+$0.000 &FW06 &28819.95  & 4.0 &  $-$3.55 &  4.4& $-$3.78\\
\ion{Fe}{i}   & 4235.936 &   $-$0.341 &FW06 &19562.44  & 4.9 &  $-$3.62 &  5.9& $-$3.84\\
\ion{Fe}{i}   & 4271.760 &   $-$0.164 &FW06 &11976.24  &14.7 &  $-$3.66 & 16.4& $-$3.93\\
\ion{Fe}{i}   & 4383.545 &   $+$0.200 &FW06 &11976.24  &25.2 &  $-$3.67 & 27.6& $-$3.95\\
\ion{Fe}{i}   & 4404.750 &   $-$0.142 &FW06 &12560.93  &18.6 &  $-$3.51 & 18.0& $-$3.87\\
\ion{Fe}{i}   & 4415.122 &   $-$0.615 &FW06 &12968.55  & 6.2 &  $-$3.61 &  8.0& $-$3.81\\
\ion{Fe}{i}   & 5324.179 &   $-$0.103 &FW06 &25899.99  & blend&  $--$    &  3.3& $-$3.98\\
\ion{Fe}{i}   & 5364.871 &   $+$0.228 &FW06 &35856.40  &2.4  &  $-$3.61  &  2.2& $-$3.92\\
\ion{Fe}{i}   & 5369.962 &   $+$0.536 &FW06 &35257.32  &5.0  &  $-$3.62  &  4.3& $-$3.95\\
\ion{Fe}{i}   & 5410.910 &   $+$0.398 &FW06 &36079.37  &2.6  &  $-$3.74  &  2.9& $-$3.94\\
\ion{Fe}{i}   & 5424.068 &   $+$0.580 &FW06 &34843.95  &blend& $--$      &  5.0& $-$3.94\\
\ion{Fe}{i}   & 5615.644 &   $+$0.050 &FW06 &26874.55  &blend &$--$      &  3.9& $-$3.99\\ 
\ion{Fe}{ii}  & 3906.035 &   $-$1.700 &FW06 &44929.55  &42.8 & $-$3.82  & 43.7 & $-$4.02\\
\ion{Fe}{ii}   & 3914.503 &   $-$4.370 &FW06 &13473.41  &18.1 &  $-$3.60 & 22.4 & $-$3.80\\
\ion{Fe}{ii}   & 3935.962 &   $-$1.720 &FW06 &44915.05  &43.3 & $-$3.79 & 44.1 & $-$3.99\\
\ion{Fe}{ii}   & 3938.290 &   $-$4.070 &FW06 &13471.41  &37.1 &  $-$3.35 & 31.7 & $-$3.83\\
\ion{Fe}{ii}   & 3945.210 &   $-$4.440 &FW06 &13673.18  &14.7 &  $-$3.65 & 18.1 & $-$3.85\\
\ion{Fe}{ii}   & 4122.668 &   $-$3.300 &FW06 &20830.58 &33.4 &  $-$3.82  & 36.8 & $-$4.03\\
\ion{Fe}{ii}   & 4128.748 &   $-$3.580 &FW06 &20830.58 &23.6 &   $-$3.82 & 27.9 & $-$4.00\\
\ion{Fe}{ii}   & 4178.862 &   $-$2.440 &FW06 &20830.58 &58.3 &  $-$3.93  & 61.7 & $-$4.11\\
\ion{Fe}{ii}   & 4258.154 &   $-$3.480 &FW06 &21812.05 &26.6 &   $-$3.78 & 29.7 &$-$3.99\\
\ion{Fe}{ii}   & 4273.326 &   $-$3.300 &FW06 &21812.05 &32.6 &  $-$3.79  & 39.6 &$-$3.90\\
\ion{Fe}{ii}   & 4296.572 &   $-$2.930 &FW06 &21812.05 &44.5 &  $-$3.82  & 50.1 &$-$3.95\\
\ion{Fe}{ii}   & 4303.176 &   $-$2.610 &FW06 &21812.05 &57.6 &  $-$3.74  & 63.1 &$-$3.86\\
\ion{Fe}{ii}   & 4369.411 &   $-$3.580 &FW06 &22409.85 &21.5 &   $-$3.80 & 25.5 &$-$3.98\\
\ion{Fe}{ii}   & 4413.601 &   $-$4.190 &FW06 &21581.64 &10.4&   $-$3.66 & 13.7 &$-$3.81\\
\ion{Fe}{ii}   & 4416.830 &   $-$2.600 &FW06 &22409.85 &56.7 &  $-$3.75  & 59.3 &$-$3.96\\
\ion{Fe}{ii}   & 4491.405 &   $-$2.640 &FW06 &23031.30 &49.8 &  $-$3.88  & 53.5 &$-$4.06\\
\ion{Fe}{ii}   & 4508.288 &   $-$2.350 &FW06 &23031.30 &66.2 &  $-$3.68  & 70.1 &$-$3.84\\
\ion{Fe}{ii}   & 4515.339 &   $-$2.360 &FW06 &23939.36 &57.8 &  $-$3.86  & 61.3 &$-$4.04\\
\ion{Fe}{ii}   & 4520.224 &   $-$2.620 &FW06 &22637.21 &54.4 &  $-$3.78 &no spectra&$--$ \\
\ion{Fe}{ii}   & 4522.634 &   $-$1.990 &FW06 &22939.36 &68.7 &  $-$3.97   &no spectra&$--$\\
\ion{Fe}{ii}   & 4923.927 &   $-$1.210 &FW06 &23317.63 &100.8 &  $-$3.91  &blend&$--$\\
\ion{Fe}{ii}   & 4993.358 &   $-$3.680 &FW06 &22637.20 &19.6 &   $-$3.74 & 23.2 &$-$3.93\\
\ion{Fe}{ii}   & 5004.188 &   $+$0.482 &K09&82853.66 &      &           & 47.6 &$-$3.88\\
\ion{Fe}{ii}   & 5006.840 &   $-$0.281 &K09&83713.53 &      &           & 20.4 &$-$3.90\\
\ion{Fe}{ii}   & 5007.450 &   $-$0.382 &K09&83726.37 &      &           & 16.4 &$-$3.94\\
\ion{Fe}{ii}   & 5010.060 &   $-$0.808 &K09&83726.37 &      &           & 10.1 &$-$3.80\\
\ion{Fe}{ii}   & 5011.026 &   $-$1.177 &K09&83726.37 &      &           &  3.6 &$-$3.95\\ 
\ion{Fe}{ii}   & 5018.440 &   $-$1.350 &FW06 &23317.63 &111.7 &  $-$3.57 &116.7  &$-$3.77\\
\ion{Fe}{ii}   & 5022.419 &   $-$0.052 &K09&83459.67 &      &           & 23.6 &$-$4.03\\
\ion{Fe}{ii}   & 5026.798 &   $-$0.235 &K09&83136.46 &      &           & 22.0 &$-$3.92\\
\ion{Fe}{ii}   & 5030.631 &   $+$0.431 &FW06 &82978.68 &50.7  &   $-$3.69 & 43.2 &$-$3.95\\
\ion{Fe}{ii}   & 5032.704 &   $+$0.143 &K09&83812.30 &      &           & 30.9 &$-$3.97\\
\ion{Fe}{ii}   & 5035.700 &   $+$0.630 &FW06 &82978.68 &59.7  &   $-$3.64 & 51.6 &$-$3.90\\
\ion{Fe}{ii}   & 5036.713 &   $-$0.527 &K09&83812.30 &      &           & 13.0 &$-$3.94\\
\ion{Fe}{ii}   & 5045.108 &   $-$0.116 &K09&83136.46 &      &           & 26.8 &$-$3.88\\
\ion{Fe}{ii}   & 5060.249 &   $-$0.479 &K09&84266.54 &      &           & 12.7 &$-$3.97\\
\ion{Fe}{ii}   & 5062.927 &   $-$1.166 &K09&83136.46 &      &           &  7.0 &$-$3.68\\
\ion{Fe}{ii}   & 5067.890 &   $-$0.173 &K09&83308.19 &      &           & 22.4 &$-$3.95\\
\ion{Fe}{ii}   & 5070.583 &   $-$0.865 &K09&83308.19 &      &           &  6.6 &$-$3.99\\
\ion{Fe}{ii}   & 5070.895 &   $+$0.262 &K09&83136.46 &      &           & 37.9 &$-$3.92\\
\ion{Fe}{ii}   & 5075.760 &   $+$0.233 &K09&84326.92 &      &           & 33.3 & $-$3.95\\
\ion{Fe}{ii}      & 5076.597 &   $-$0.749 &K09&83713.53 &      &           &  8.8 & $-$3.95\\
\ion{Fe}{ii}      & 5081.898 &   $-$0.689 &K09&83713.53 &      &           &  8.3 & $-$4.02\\
\hline
\noalign{\smallskip}
\end{tabular}
\end{flushleft}
\end{table*}

\setcounter{table}{0}

\begin{table*}[!hbp]
\caption[ ]{cont.}
\font\grande=cmr7
\grande
\begin{flushleft}
\begin{tabular}{rrrrrrcrc}
\hline\noalign{\smallskip}
\multicolumn{5}{c}{}&
\multicolumn{2}{c}{HR\,6000[13450,4.3,AT12]}&
\multicolumn{2}{c}{46\,Aql[12560,3.8,AT12]}
\\
\hline\noalign{\smallskip}
\multicolumn{1}{c}{Species}&
\multicolumn{1}{c}{$\lambda$($\AA$)}&
\multicolumn{1}{c}{$\log\,gf$}&
\multicolumn{1}{c}{Ref.$^{a}$}&
\multicolumn{1}{c}{$\chi_{low}$}&
\multicolumn{1}{c}{W(m$\AA$)}&
\multicolumn{1}{c}{$log(N_{elem}/N_{\rm tot}$)}&
\multicolumn{1}{c}{W(m$\AA$)}&
\multicolumn{1}{c}{$log(N_{elem}/N_{\rm tot}$)}
\\
\hline\noalign{\smallskip}

\ion{Fe}{ii}      & 5086.306 &   $-$0.472 &K09&83990.06 &      &           & 14.7 & $-$3.90\\
\ion{Fe}{ii}      & 5089.214 &   $-$0.014 &K09&83308.19 &      &           & 27.0 & $-$3.96\\
\ion{Fe}{ii}      & 5093.783 &   $-$0.703 &K09&83726.37 &      &           & 12.7 & $-$3.77\\
\ion{Fe}{ii}      & 5117.032 &   $-$0.129 &K09&84131.58 &      &           & 22.0 & $-$3.96\\    
\ion{Fe}{ii}      & 5132.669 &   $-$4.090 &FW06 &22637.20 &17.5 &   $-$3.40 & 14.0 & $-$3.83\\
\ion{Fe}{ii}      & 5140.692 &   $-$0.822 &K09&83990.06 &      &           &  9.9 & $-$3.78\\
\ion{Fe}{ii}      & 5143.875 &   $+$0.054 &K09&84266.54 &      &           & 21.6 & $-$4.14\\
\ion{Fe}{ii}      & 5144.352 &   $+$0.307 &FW06 &84424.37 &40.1 &   $-$3.73 & 32.8 & $-$4.03\\
\ion{Fe}{ii}      & 5169.033 &   $-$0.870 &FW06 &23317.63 &115.1 &  $-$4.01 & 121.0 & $-$4.19\\
\ion{Fe}{ii}      & 5180.312 &   $+$0.002 &K09&83812.30 &       &          & 26.9 & $-$3.94\\
\ion{Fe}{ii}      & 5199.118 &   $+$0.054 &K09&83713.53 &       &          & 30.3 & $-$3.87\\
\ion{Fe}{ii}      & 5200.798 &   $-$0.390 &K09&83812.30 &       &          & 17.4 & $-$3.87\\
\ion{Fe}{ii}      & 5203.634 &   $-$0.088 &K09&83812.30 &       &          & 25.9 & $-$3.88\\
\ion{Fe}{ii}      & 5218.841 &   $-$0.250 &K09&83726.37 &       &          & 23.6 & $-$3.85\\
\ion{Fe}{ii}      & 5219.920 &   $-$0.628 &K09&84870.86 &       &          & 10.3 & $-$3.89\\
\ion{Fe}{ii}      & 5222.350 &   $-$0.332 &K09&84844.82 &       &          & 15.0 & $-$3.96\\
\ion{Fe}{ii}      & 5223.802 &   $-$0.546 &K09&83713.53 &       &          & 14.6 & $-$3.83\\
\ion{Fe}{ii}      & 5224.404 &   $-$0.581 &K09&83990.06 &       &          & 12.5 & $-$3.87\\    
\ion{Fe}{ii}      & 5234.625 &   $-$2.210 &FW06&25981.63 &69.6 &  $-$3.54  & 71.0 & $-$3.76\\
\ion{Fe}{ii}      & 5237.949 &   $+$0.103 &K09&84266.54 &       &          & 29.8 & $-$3.90\\
\ion{Fe}{ii}      & 5245.455 &   $-$0.502 &K09&84326.92 &       &          & 13.6 & $-$3.88\\
\ion{Fe}{ii}      & 5247.956 &   $+$0.550 &FW06 &84938.18 &46.2 &  $-$3.74 & 39.7 & $-$4.01\\
\ion{Fe}{ii}      &  5257.119 &  $+$0.115 &K07 &84685.20 &      &          & 26.7 & $-$4.00\\
\ion{Fe}{ii}      &  5265.985 &  $-$0.936 &K09 &84035.12 &      &          &  9.3 & $-$3.68\\
\ion{Fe}{ii}      &  5270.029 &  $-$0.097 &K09 &84710.70 &      &          & 18.0 & $-$4.08\\
\ion{Fe}{ii}      & 5272.397 &   $-$2.010 &FW06 &48039.09 &35.0 &   $-$3.51  & 35.4 & $-$3.71\\
\ion{Fe}{ii}      & 5276.002 &   $-$1.900 &FW06 &25805.33 &67.0 &  $-$3.93   & 71.7 & $-$4.07\\
\ion{Fe}{ii}      & 5284.109 &   $-$3.200 &FW06 &23317.63 &38.2 &  $-$3.63   & 40.6 & $-$3.86\\
\ion{Fe}{ii}      & 5291.661 &   $+$0.561 &K09&84527.76 &      &            & 42.8 & $-$3.95\\ 
\ion{Fe}{ii}      & 5306.182 &   $+$0.044 &FW06 &84870.87 &30.4  &  $-$3.70   & 26.7 & $-$3.91\\
\ion{Fe}{ii}      & 5315.083 &   $-$0.422 &K09&85048.61 &      &            & 13.2 & $-$3.93\\
\ion{Fe}{ii}      & 5316.615 &   $-$1.780 &FW06 &25428.78 &79.9 &  $-$3.71   & 81.9 & $-$3.93\\
\ion{Fe}{ii}      & 5318.055 &   $-$0.177 &K09&84527.76 &      &            & 22.3 & $-$3.85\\
\ion{Fe}{ii}      & 5355.421 &   $-$0.203 &K09&84685.20 &      &            & 13.2 & $-$4.17\\
\ion{Fe}{ii}      & 5358.872 &   $-$0.130 &K09&84685.20 &      &            & 14.8 & $-$4.17\\
\ion{Fe}{ii}      & 5359.237 &   $-$1.112 &K09&84710.70 &      &            &  7.2 & $-$3.59\\
\ion{Fe}{ii}      & 5366.210 &   $-$0.549 &K09&84710.70 &      &            & 16.4 & $-$3.68\\ 
\ion{Fe}{ii}      & 5425.257 &   $-$3.390 &FW06 &25805.33 &27.8  &  $-$3.59   & 30.2 & $-$3.81\\
\ion{Fe}{ii}      & 5430.003  &   $+$0.427&FW06 &85462.86 &       &            & 36.7 & $-$3.92\\
\ion{Fe}{ii}      & 5444.386  &  $-$0.153 &K09 &85495.32 &      &            & 19.5 & $-$3.90\\
\ion{Fe}{ii}      & 5451.316  &  $-$0.756 &K09 &84685.20 &      &            & 7.1  & $-$3.95\\ 
\ion{Fe}{ii}      & 5465.932  &  $+$0.348 &FW06 &85679.70 &42.3   &  $-$3.58   & 35.6  & $-$3.86\\
\ion{Fe}{ii}      & 5472.855  &  $-$0.723 &K09 &84685.20 &      &            & 10.5 & $-$3.77 \\
\ion{Fe}{ii}      & 5475.826  &  $-$0.129 &K09 &84685.20 &      &            & 19.7 & $-$3.96\\
\ion{Fe}{ii}      & 5488.776  &  $-$0.468 &K09 &85462.86 &      &            & 13.3 & $-$3.84\\
\ion{Fe}{ii}      & 5492.398  &  $-$0.127 &K09 &84685.20 &      &            & 25.1 & $-$3.78\\
\ion{Fe}{ii}      &  5493.830   &  $+$0.259 &FW06 &84685.20 &37.7   &  $-$3.68 & 32.5 & $-$3.93\\
\ion{Fe}{ii}      & 5502.670  &  $-$0.179 &K09 &85184.73 &      &            & 21.2 & $-$3.83\\
\ion{Fe}{ii}      & 5534.847 & $-$2.860   &FW06 &26170.18&50.5 &  $-$3.44   & 51.1 & $-$3.70\\
\ion{Co}{ii}      & 3501.708 & $-$1.111 & K06 &17771.506 &not obs& $--$ & not obs& $\le$ $-$8.02\\
\ion{Co}{ii}      & 4160.657 & $-$1.751 & K06 &27484.371 &blend&$--$ & blend & $--$\\
\ion{Ni}{ii}      & 4067.031 & $-$1.834 & K03Ni &32499.530 & 9.9 & $-$6.24 & 9.8 & $-$6.47\\
\ion{Cu}{ii}      & 4909.734 & $+$0.790 & K03Cu  &115568.985& not obs & $--$& 3.1 & $-$6.26\\
\ion{Cu}{ii}      & 4931.698 & $+$0.704 & K03Cu  &115662.550& not obs & $--$& 2.7 & $-$6.22\\
\ion{Zn}{i}       & 4810.528 & $-$0.137 &K,WAR & 32890.352 &not obs & $--$& 5.1 & $-$5.85\\
\ion{Zn}{ii}      & 4911.625 & $+$0.540 &NIST3 &96909.740  &not obs & $--$&23.0 & $-$5.76\\
\hline
\noalign{\smallskip}
\end{tabular}
\end{flushleft}
\end{table*}

\setcounter{table}{0}

\begin{table*}[!hbp]
\caption[ ]{cont.}
\font\grande=cmr7
\grande
\begin{flushleft}
\begin{tabular}{rlrrrrcrc}
\hline\noalign{\smallskip}
\multicolumn{5}{c}{}&
\multicolumn{2}{c}{HR\,6000[13450,4.3,AT12]}&
\multicolumn{2}{c}{46\,Aql[12560,3.8,AT12]}
\\
\hline\noalign{\smallskip}
\multicolumn{1}{c}{Species}&
\multicolumn{1}{c}{$\lambda$($\AA$)}&
\multicolumn{1}{c}{$\log\,gf$}&
\multicolumn{1}{c}{Ref.$^{a}$}&
\multicolumn{1}{c}{$\chi_{low}$}&
\multicolumn{1}{c}{W(m$\AA$)}&
\multicolumn{1}{c}{$log(N_{elem}/N_{\rm tot}$)}&
\multicolumn{1}{c}{W(m$\AA$)}&
\multicolumn{1}{c}{$log(N_{elem}/N_{\rm tot}$)}
\\
\hline\noalign{\smallskip}
\ion{Ga}{ii}      & 4255.722 & $+$0.634 &RS94 & 113842.190&not obs & $--$ & not obs & $--$\\
\ion{Ga}{ii}      & 6334.069 & $+$1.000 &RS94 & 102944.550&not obs & $--$ & not obs & $--$\\
\ion{As}{ii}      & 5105.58  & $--$ & $--$    & 81508.925 &not obs & $--$ & 4.0     & $--$\\
\ion{As}{ii}      & 5107.55  & $--$ & $--$    & 82819.214 &not obs & $--$ & blend  & $--$\\
\ion{As}{ii}      & 5231.38  & $--$ & $--$    & 79128.330 &not obs & $--$ &  3.2 & $--$ \\
\ion{As}{ii}      & 5331.23  & $--$ & $--$    & 81508.925 &not obs & $--$ &  9.4 & $--$\\
\ion{As}{ii}      & 5497.73  & $--$ & $--$    & 78730.893 &not obs & $--$ & blend & $--$\\
\ion{As}{ii}      & 5558.09  & $--$ & $--$    & 79128.330 &not obs & $--$ &  9.4 & $--$\\
\ion{As}{ii}      & 5651.32  & $--$ & $--$    & 81508.925 &not obs & $--$ &  11.0 & $--$\\
\ion{As}{ii}      & 6110.07  & $--$ & $--$    & 83819.214 &not obs & $--$ & 3.3 & $--$\\
\ion{As}{ii}      & 6170.27  & $--$ & $--$    & 79128.330 &not obs & $--$ & blend & $--$\\
\ion{Sr}{ii}      & 4077.709 & $+$0.151 &NIST3 &     0.000&not obs & $--$ & not obs & $--$\\
\ion{Y}{ii}       & 3950.349 & $-$0.490 &K,HL &840.213&profile&$-$8.60& 4.2 & $-$8.20\\
\ion{Y}{ii}       & 4883.682 & $+$0.070 &K,HL &8743.316&profile&$-$8.60 & 7.3 & $-$8.03 \\
\ion{Y}{ii}       & 4900.120 & $-$0.090 &K,HL &8328.041&profile&$-$8.60 & 5.5 & $-$8.04\\
\ion{Xe}{ii}       & 4844.330 & $+$0.491 &NIST3 &93068.440&28.8&$-$5.23 & 17.5 & $-$5.81\\
\ion{Hg}{ii}      & 3983.890 & $-$1.510 &NIST3 &35514.000&profile&$-$8.20 & profile & $-$7.10\\
\hline
\noalign{\smallskip}
\end{tabular}
\end{flushleft}
$^{a}$ (NIST3) NIST Atomic Spectra Database, version 3 at http://physics.nist.gov;\\
(FW06) Fuhr \& Wiese (2006);\\
(PTP) Pickering et al. (2002);\\
(RS94) Ryabchikova \& Smirnov (1994);\\ 
(SL) Sigut \& Landstreet (1990);\\
K03Mn: http://kurucz.harvard.edu/atoms/2501/gf2501.pos;\\
K03Ni: http://kurucz.harvard.edu/atoms/2801/gf2801.pos;\\
K03Cu: http://kurucz.harvard.edu/atoms/2901/gf2901.pos;\\
K06: http://kurucz.harvard.edu/atoms/2701/gf2701.pos;\\
K07: http://kurucz.harvard.edu/atoms/2601/gf2601.pos, 2007 version, no longer available.
The $\log\,gf$ gives better agreement with the observations than the K09 new determination;\\
K09: http://kurucz.harvard.edu/atoms/2601/gf2601.pos,January 2009 version;\\
``K'' before another $\log\,gf$ source means that the $\log\,gf$ is from Kurucz files
available at http://kurucz.harvard.edu/linelists/gf100/; 
(HL) Hannaford et al. (1982);(KP) Kurucz \& Peytremann (1975); (MRB) Miller et al. (1971);
(NBS) Younger et al. (1979); (WAR) Warner (1968).\\
$^b$ The abundances given in parenthesis were derived from line profiles. All computed line profiles include
hyperfine components, except for that at 3917.318\,$\AA$.    

\end{table*}

\end{appendix}


\begin{thebibliography}{}

\bibitem[Ballester et al.(2000)]{2000SPIE.4010..246B} Ballester, P., 
Grosbol, P., Banse, K., Disaro, A., Dorigo, D., Modigliani, A., Pizarro de 
la Iglesia, J.~A., \& Boitquin, O.\ 2000, \procspie, 4010, 246 

\bibitem[2005]{Castelli2005}
Castelli, F.\ 2005,
MSAIS, 8, 44

\bibitem[Castelli \& Hubrig(2004)]{2004A&A...425..263C} 
Castelli, F., \& Hubrig, S.\ 2004, \aap, 425, 263 


\bibitem[Castelli \& Hubrig(2007)]{2007A&A...475.1041C} 
Castelli, F., \& Hubrig, S.\ 2007, \aap, 475, 1041 


\bibitem[Castelli et al.(2008)]{2008JPhCS.130a2003C}
Castelli, F., Johansson, S., \& Hubrig, S.\ 2008, 
Journal of Physics Conference Series, 130, 012003 

\bibitem[1981]{Cowan1981}
Cowan, R.~D.\ 1981,
The Theory of Atomic Structure and Spectra (Berkeley: Univ. California Press)

\bibitem[Fuhr \& Wiese(2006)]{2006JPCRD..35.1669F} 
Fuhr, J.~R., \& Wiese, W.~L.\ 2006, Journal of Physical and Chemical 
Reference Data, 35, 1669 

\bibitem[Grevesse \& Sauval(1998)]{1998SSRv...85..161G} 
Grevesse, N., \& Sauval, A.~J.\ 1998, Space Science Reviews, 85, 161 

\bibitem[Hannaford et al.(1982)]{1982ApJ...261..736H} Hannaford, P., Lowe, 
R.~M., Grevesse, N., Biemont, E., \& Whaling, W.\ 1982, \apj, 261, 736 

\bibitem[Hauck \& Mermilliod(1998)]{1998A&AS..129..431H} Hauck, B., \& 
Mermilliod, M.\ 1998, \aaps, 129, 431 

\bibitem[Johansson(2002)]{2002HiA....12...84J} Johansson, S.\ 2002, 
Highlights of Astronomy, 12, 84 

\bibitem[Johansson(2009)]{ } Johansson, S.\ 2009, 
Physica Scripta, T134, 014013

\bibitem[Johansson \& Cowley(1984)]{1984A&A...139..243J} 
Johansson, S., \& Cowley, C.~R.\ 1984, \aap, 139, 243 

\bibitem[1993]{Kurucz1993}
Kurucz, R.~L. \ 1993,
SYNTHE Spectrum Synthesis Programs and Line Data, CD-ROM, No. \ 18

\bibitem[Kurucz(2005)]{2005MSAIS...8...14K} Kurucz, R.~L.\ 2005, Memorie 
della Societa' Astronomica Italiana Supplement, 8, 14 

\bibitem[Kurucz \& Peytremann(1975)]{1975SAOSR.362.....K}
Kurucz, R.~L., \& Peytremann, E.\ 1975, 
SAO Special Report, 362,  

\bibitem[Michaud(1970)]{1970ApJ...160..641M} Michaud, G.\ 1970, \apj, 160, 
641 
\bibitem[Miller et al.(1971)]{1971PhRvA...4.1709M} Miller, M.~H., Roig, 
R.~A., \& Bengtson, R.~D.\ 1971, \pra, 4, 1709 


\bibitem[Pickering et al.(2002)]{2002ApJS..138..247P} Pickering, J.~C., 
Thorne, A.~P., \& Perez, R.\ 2002, \apjs, 138, 247 


\bibitem[Raassen \& Uylings(1998)]{1998A&A...340..300R} 
Raassen, A.~J.~J., \& Uylings, P.~H.~M.\ 1998, \aap, 340, 300 


\bibitem[Ryabchikova 
\& Smirnov(1994)]{1994ARep...38...70R} Ryabchikova, T.~A., \& Smirnov, Y.~M.\ 1994, 
Astronomy Reports, 38, 70 

\bibitem[Ryabchikova et al.(2003)]{2003IAUS..210..301R} Ryabchikova, T., 
Wade, G.~A., 
\& LeBlanc, F.\ 2003, Modelling of Stellar Atmospheres, 210, 301 

\bibitem[Sadakane et al.(2001)]{2001PASJ...53.1223S} Sadakane, K., et al.\ 
2001, \pasj, 53, 1223 

\bibitem[Sbordone et al.(2004)]{2004MSAIS...5...93S} Sbordone, L., 
Bonifacio, P., Castelli, F., \& Kurucz, R.~L.\ 2004, 
Memorie della Societa Astronomica Italiana Supplement, 5, 93 


\bibitem[Sigut \& Landstreet(1990)]{1990MNRAS.247..611S}
Sigut, T.~A.~A., \& Landstreet, J.~D.\ 
1990, \mnras, 247, 611 


\bibitem[NBS]{younger et al.}
Younger, S.~M., Fuhr, Y.~R., Martin, G.~A., \& Wiese, W.~L.\
1978, Journal of Physical and Chemical 
Reference Data, 7, 495 

\bibitem[Wahlgren et al.(2000)]{2000ApJ...539..908W} Wahlgren, G.~M., Dolk, 
L., Kalus, G., Johansson, S., Litz{\'e}n, U., 
\& Leckrone, D.~S.\ 2000, \apj, 539, 908 

\bibitem[WAR]{warner}
Warner, B.,\ 1968,
\mnras,  140, 53

\end{thebibliography}
\end{document}